\documentclass[5p, twocolumn]{elsarticle}
\usepackage{graphicx} 
\usepackage{blindtext}
\usepackage{appendix}
\usepackage{url}

\title{\textbf{Safety-Critical Control for Autonomous Systems:
\\ Control Barrier Functions via Reduced-Order Models}}
\author[1]{Max H. Cohen}
\ead{maxcohen@caltech.edu}
\author[2]{Tamas G. Molnar}
\author[1]{Aaron D. Ames}

\affiliation[1]{
    organization={Department of Mechanical and Civil Engineering, California Institute of Technology},
    city={Pasadena}, 
    state={CA}, 
    postcode={91125},
    country={USA},
}

\affiliation[2]{
    organization={Department of Mechanical Engineering, Wichita State University},
    city={Wichita}, 
    state={KS}, 
    postcode={67260},
    country={USA},
}

\newtheorem{theorem}{Theorem} 
\newtheorem{lemma}{Lemma}

\newtheorem{corollary}{Corollary}
\newdefinition{definition}{Definition} 
\newdefinition{remark}{Remark}
\newdefinition{example}{Example}
\newdefinition{property}{Property}
\newproof{proof}{Proof}

\usepackage{amsmath}
\usepackage{amssymb}
\usepackage{mathtools}
\usepackage{bm}
\usepackage{derivative}

\newcommand{\argmin}{\operatornamewithlimits{arg\,min}}

\newcommand{\R}{\mathbb{R}}

\newcommand{\C}{\mathcal{C}}

\newcommand{\T}{^\top}

\newcommand{\Kinf}{\mathcal{K}_{\infty}}
\newcommand{\relu}{\operatorname{ReLU}}

\newcommand{\bzero}{\mathbf{0}}

\newcommand{\bd}{\mathbf{d}}

\renewcommand{\bf}{\mathbf{f}} 
\newcommand{\bg}{\mathbf{g}}

\newcommand{\bk}{\mathbf{k}}

\newcommand{\bn}{\mathbf{n}}

\newcommand{\bp}{\mathbf{p}}
\newcommand{\bq}{\mathbf{q}}
\newcommand{\br}{\mathbf{r}}

\newcommand{\bu}{\mathbf{u}}
\newcommand{\bv}{\mathbf{v}}

\newcommand{\bx}{\mathbf{x}}
\newcommand{\by}{\mathbf{y}}

\newcommand{\bB}{\mathbf{B}}
\newcommand{\bC}{\mathbf{C}}
\newcommand{\bD}{\mathbf{D}}

\newcommand{\bG}{\mathbf{G}}
\newcommand{\bH}{\mathbf{H}}
\newcommand{\bI}{\mathbf{I}}
\newcommand{\bJ}{\mathbf{J}}

\newcommand{\bxi}{\bm{\xi}}

\newcommand{\bzeta}{\bm{\zeta}}

\newcommand{\Q}{\mathcal{Q}}
\newcommand{\Tq}{T_{\bq}\Q}
\newcommand{\TQ}{T\Q}

\usepackage{xcolor}
\definecolor{myblue}{RGB}{49, 114, 174}
\definecolor{myred}{rgb}{0.796, 0.235, 0.2}
\definecolor{mygreen}{rgb}{0.22, 0.596, 0.149}
\definecolor{mypurple}{rgb}{0.584,0.345,0.698}

\begin{document}
\begin{abstract}
    Modern autonomous systems, such as flying, legged, and wheeled robots, are generally characterized by high-dimensional nonlinear dynamics, which presents challenges for model-based safety-critical control design. 
    Motivated by the success of reduced-order models in robotics, this paper presents a tutorial on constructive safety-critical control via reduced-order models and control barrier functions (CBFs). 
    To this end, we provide a unified formulation of techniques in the literature that share a common foundation of constructing CBFs for complex systems from CBFs for much simpler systems. 
    Such ideas are illustrated through formal results, simple numerical examples, and case studies of real-world systems to which these techniques have been experimentally applied.
\end{abstract}

\maketitle

\section{Introduction}
The control stack for modern autonomous systems – from legged robots to self-driving vehicles – typically
consists of a complex interconnection of decision-making, planning, and control modules, all of which may
leverage different model representations to strike a balance between computational efficiency, model uncertainty,
and satisfaction of system-level specifications. Among the various specifications that such autonomous
systems must satisfy, safety – informally thought of as requiring a system never to do anything “bad” – is
often given precedence, as the violation of specifications deemed to be safety-critical could result in undesirable
behavior. Over the past decade, control barrier functions (CBFs) \cite{AmesCDC14,AmesADHS15,AmesTAC17,AmesECC19} have emerged as a powerful tool
for designing controllers that ensure the safety of autonomous systems. Despite their success, constructing
CBFs for high-dimensional autonomous systems remains an open challenge since their dynamics may be
nontrivial or not even known. 

To address these challenges, there has been recent
interest in constructing CBFs for complex autonomous systems based on reduced-order models (ROMs)
– lower-dimensional representations that are rich enough to capture the high-level behavior of the full-order system but that are simple enough to synthesize safety-critical controllers \cite{TamasRAL22,TamasACC23,DrewIROS21}. This approach has demonstrated success in controlling seemingly complex systems, such as underactuated and dynamic robotic systems, in a computationally efficient
manner, and naturally integrates into the existing control stack present in many autonomous systems. 

In this paper, we provide a self-contained introduction and detailed overview of CBF techniques based on ROMs. Here, we highlight the theoretical foundations of this approach and illustrate its applications across different domains through a collection of case studies. Before diving into this discussion, however, we first
review current state-of-the-art techniques in the field of safety-critical control and motivate the techniques covered and perspective taken in this tutorial.

\subsection{The Different Flavors of Control Barrier Functions}
The property of safety is often formalized using the framework of set invariance \cite{Blanchini} in which a system is said to be safe if its trajectories remain within a desirable set of the state space \cite{AmesECC19}. 
That is, a closed-loop system is safe if there exists an invariant set that does not intersect with a set of states deemed by the user to be dangerous. 
Such an invariant set is referred to as a \emph{safe set}.

By moving from invariant sets to controlled invariant sets -- those that can be rendered forward invariant through the application of a feedback controller -- this notion of safety may also be applied to systems with control inputs. 
Control designs in which safety is a high-priority requirement are often referred to as \emph{safety-critical} controllers.
Among the various tools that have emerged to address safety-critical control, including, but not limited to, model predictive control (MPC) \cite{Borelli,ZeilingerARCAS20}, reachability analysis \cite{TomlinTAC05,BansalCDC17}, and symbolic control \cite{Tabuada,Belta}, CBFs \cite{AmesTAC17,AmesECC19} have demonstrated success in synthesizing safety-critical controllers for high-dimensional nonlinear systems.

Since the introduction of CBFs \cite{AmesCDC14,AmesTAC17} (see \cite{AmesECC19} for a more in-depth survey on the history of CBFs), there has been a large body of work in developing various types of CBFs for different classes of systems and control objectives. 
Given that CBFs are a model-based tool, and that most models are coarse representations of the underlying system, many of these developments have been motivated by controlling systems subject to uncertainty \cite{AmesCSM23}. 
These include, for example, robust CBFs for systems with unstructured uncertainty \cite{jankovic2018robust,AmesLCSS19,AnilLCSS22,AnilTCST23}, adaptive CBFs for systems with parametric uncertainty \cite{AndrewACC20,LopezLCSS21,Cohen}, data-driven CBFs for systems with unknown dynamics \cite{AndrewL4DC20,SchoelligL4DC22,DhimanTAC21}, and stochastic CBFs for systems with stochastic dynamics \cite{SantoyoAutomatica21,CosnerRSS23}. 

Other lines of work have developed classes of CBFs to account for different assumptions on systems' actuation and sensing capabilities. 
For example, measurement-robust \cite{DeanCoRL20} and observer-based CBFs \cite{DevLCSS23,XuACC22} have been developed to design safety-critical controllers for systems with measurement uncertainty, whereas event-triggered \cite{GuangACC19,AndrewLCSS21,WeiTAC23,LongSCL22} and sampled-data CBFs \cite{KrsticACC18,BreedenLCSS21,AndrewCDC22-sampled-data} have been developed to enforce safety when one may only update control inputs at discrete instances in time. 
Variants of CBFs have also been developed to address more nuanced notions of safety including input-to-state safety (ISSf) \cite{AmesLCSS19,AnilLCSS22} and finite/fixed/prescribed-time safety \cite{GargACC21,PolyakovTAC23,AbelTAC23}, whereas others have been used to enforce satisfaction of more general temporal logic specifications \cite{LarsLCSS19,CooganTRO21,CohenNAHS23}.

\subsection{Constructive Methods for Control Barrier Functions}
Although much attention has been given to defining different classes of CBFs for various systems and control objectives of interest, relatively less attention has been given to the construction of such CBFs. As a result, there exists a plethora of different types of CBFs, but a lack of constructive techniques required to obtain such CBFs in the first place. This lack of constructive techniques often limits the applicability of CBFs to relatively simple low-dimensional systems. Motivated by these limitations, researchers have begun to investigate constructive techniques for safety-critical control and CBFs. 

A central challenge in constructing a CBF is finding a scalar function whose time derivative directly depends on the system's control input and whose zero superlevel set defines a controlled invariant subset of the state space. This challenge highlights the crucial distinction between a \emph{safe} set and a \emph{constraint} set. The former is a controlled invariant set that does not intersect with the set of failure states. The latter is simply the set of states deemed by the user to not be in violation of a given safety constraint. These sets need not coincide and, in general, they do not. For example, in robot motion planning problems, the ``distance to the obstacle" function -- depending only on the robot's position -- defines the obstacle-free space (constraint set) but is not a CBF (i.e., it does not yield a safe set) unless the derivatives of the position directly depend on the control inputs. 

The challenges mentioned above are related to the \emph{relative degree} of the function -- the number of times it must be differentiated along the system dynamics until the input appears -- defining the safety constraint.
A popular approach to address such challenges is through the use of \emph{extended}, also called exponential \cite{SreenathACC16} or high-order \cite{WeiTAC22}, CBFs, which have roots in work on non-overshooting control \cite{KrsticTAC06}. Here, one differentiates a high relative degree constraint function until the control input appears and then enforces CBF-like conditions upon its highest-order derivative. Such an approach has demonstrated success in safety-critical control of high-dimensional systems \cite{Wei}, but also faces challenges in verifying the satisfaction of CBF-like conditions \cite{TanTAC22}.

Some limitations of extended CBFs have been addressed by leveraging the structure present in certain classes of systems. For example, constructive CBF techniques have been developed for robotic systems \cite{DrewLCSS22,CortezICRA21,CortezTCST21,CortezAutomatica22} by exploiting structural properties of their dynamics. Other approaches have sought to extend Lyapunov backstepping \cite{Krstic} to CBFs for systems in strict-feedback form \cite{AndrewCDC22}.

Other works have sought to address the limitations outlined above by leveraging implicitly defined CBFs, often constructed by propagating forward the dynamics of the system in a receding-horizon fashion \cite{BreedenAutomatica23} under a ``backup" \cite{AmesIEEEA20,YuxiaoCDC21} or performance-based policy \cite{BreedenCDC22}. Such approaches have close connections with MPC, and, indeed, one may also leverage MPC techniques to construct CBFs in a receding horizon manner \cite{AmesCSM23,WabersichTAC22}. Although powerful, these techniques often require additional online computation that may prohibit their use for real-time control of high-dimensional systems.

To address these limitations, alternative approaches seek to shift the computational burden of constructing a CBF offline where one may leverage powerful optimization tools to build a CBF. For example, sum-of-squares programming has been used to construct CBFs for systems with polynomial dynamics \cite{ClarkCDC21,ClarkTAC22,DaiACC23,ZhaoLCSS23}. Other works have sought to bridge the gap between reachability analysis and CBFs \cite{HerbertCDC21,HerbertIROS22,ChoiTAC23}, and illustrate that a CBF for a general class of nonlinear systems can be constructed from the value function of a particular discounted optimal control problem. Although promising, these techniques are limited by the computation needed to solve sum-of-squares programs or compute value functions over a grid, both of which scale poorly with the state dimension. 

The computational challenges in constructing CBFs using offline optimization have motivated the use of learning-based techniques to learn CBFs from data. Such approaches model the CBF using a suitable class of function approximators, such as neural networks, and train such a model to satisfy the criteria of a CBF either directly  \cite{DawsonCoRL22,DawsonTRO23,SoArXiV23} or by using data from expert demonstrations \cite{RobeyCDC20,LarsCoRL20}. These learning-based approaches empirically perform well but also face the challenge of verifying if the trained model satisfies CBF conditions for safety, which may preclude their application to systems where safety must be rigorously certified.

\subsection{Control Barrier Functions via Reduced-Order Models}
Modern autonomous systems, such as flying, legged, and wheeled robots, are generally characterized by high-dimensional nonlinear dynamics. Although CBF-based controllers may, in principle, be applied to such systems, this first requires constructing a CBF for a complex high-dimensional nonlinear system -- a task that many of the aforementioned methods struggle with. Rather than directly constructing a CBF for a complicated system, an alternative approach is to construct a CBF for a much simpler system, and then attempt to relate the inputs that enforce safety of this simpler system back to the inputs of the original system. That is, one may use a \emph{reduced-order} representation of the original, full-order, dynamics for the purpose of control design, and then refine such a controller for the full-order system provided its dynamic behavior is sufficiently captured by the reduced-order model. 

Such control designs, despite leveraging simple models, have demonstrated success in different areas of robotics. In mobile robotics, single integrator \cite{ZhaoICCA17} and unicycle models \cite{LucaSpringer01} are often used as the basis for control designs of more complicated nonholonomic systems. In legged robotics, reduced-order models such as the spring-loaded inverted pendulum \cite{Raibert}, linear inverted pendulum \cite{KajitaICRA21}, and hybrid-linear inverted pendulum \cite{XiaobinTRO22} have demonstrated continued success in controlling walking robots with high-dimensional nonlinear dynamics.

Inspired by their success in robotics, there has been recent interest in using reduced-order models for safety-critical control design. In the context of CBFs, such ideas were introduced in \cite{DrewIROS21,DrewLCSS22} where CBFs designed for simple kinematic models were used to generate safe velocity commands to be tracked by more complicated robotic systems, such as drones \cite{DrewIROS21} and manipulators \cite{DrewLCSS22}. Such control designs were formalized in \cite{TamasRAL22} by illustrating that the combination of a CBF for a reduced-order model and a Lyapunov function certifying tracking of the reduced-order trajectory may be used to establish safety of the full-order system. Further extensions and applications of this approach have been reported in \cite{TamasACC23,DrewRAL22,Jeeseop-Multi-Agent}. Although not explicitly framed as safety-critical control based on reduced-order models, CBF backstepping \cite{AndrewCDC22} shares with these approaches the ability to construct CBFs for complicated systems from CBFs for simple models.

\subsection{Objective of this Paper}
The primary objective of this paper is to provide a tutorial presentation of CBF techniques based on reduced-order models. In doing so, we present a unified formulation of techniques in the literature that share a common foundation of constructing CBFs for complex systems from CBFs for much simpler systems. These ideas are illustrated through formal results, simple numerical examples, and high-level overviews of more complicated applications. The majority of the stated theoretical results have already been established, in one form or another, in the various works cited herein. For illustrative purposes, the proofs of selected results are provided in the Appendix. Other results are new but are also minor extensions or combinations of existing results. For completeness, the proofs of such results are also collected in the Appendix. All the numerical examples presented in this tutorial can be reproduced using open-source code available on Github\footnote{\url{https://github.com/maxhcohen/ReducedOrderModelCBFs.jl}}.

\subsection{Organization and Outline}

The remainder of this paper is organized as follows.

In Sec. \ref{sec:safety}, we provide a self-contained introduction to safety-critical control via CBFs. First, we review the characterization of safety via set invariance \cite{Blanchini} and barrier functions \cite{AmesTAC17} and then discuss how such ideas may be extended to design safety-critical controllers using CBFs. Next, we discuss how CBFs may be extended to disturbed systems using the framework of ISSf \cite{AmesLCSS19,AnilLCSS22,AnilTCST23}, leading to the synthesis of robust safety-critical controllers. Finally, we review the concept of a smooth safety filter \cite{Cohen-Smooth-Safety} -- a class of differentiable CBF-based controllers that will play an important role in synthesizing CBFs via ROMs. 

In Sec. \ref{sec:ROM}, we begin our exposition on the construction of CBFs via ROMs. Here, we first discuss some of the technical challenges in constructing CBFs for high-dimensional systems and then outline various classes of systems whose structure facilitates the synthesis of CBFs using ROMs. 

In Sec. \ref{sec:backstepping}, we present our first constructive technique for CBF synthesis, which exploits the idea of CBF backstepping as originally developed in \cite{AndrewCDC22}. We demonstrate how this approach applies to general classes of systems whose dynamics may be interpreted as a layered control architecture and compare this backstepping approach with existing high-order CBF approaches.

In Sec. \ref{sec:robotic-cbfs}, we demonstrate how CBF backstepping may be specialized to robotic systems whose dynamics also exhibit a particularly useful cascaded structure. When such a system is fully actuated, we illustrate how one may directly apply the backstepping approach presented in Sec. \ref{sec:backstepping} to generate CBFs. We then extend this approach, combining it with the notion of an \emph{energy-based} CBF \cite{DrewLCSS22}, which further exploits the structure of the robot dynamics to construct CBFs. Finally, using ideas inspired by those from \cite{SpongIROS94}, we show how CBFs may be constructed for certain classes of underactuated robotic systems.

In Sec. \ref{sec:tracking}, we illustrate how previous constructions can be understood as combining a CBF for a ROM with a Lyapunov function certifying tracking of the ROM by the full-order dynamics.
Such an approach relaxes many of the structural requirements imposed in the previous sections and replaces them with the, perhaps, less strict requirement of the existence of a tracking controller. Moreover, we demonstrate how this approach
leads to the paradigm of \emph{model-free} safety-critical control \cite{TamasRAL22} in which one need not directly rely on the full-order dynamics to construct safety-critical controllers. 

In Sec. \ref{sec:case-studies}, we revisit more complex application examples from the literature that leverage the constructive CBF techniques outlined in previous sections. These examples include safety-critical control of fixed-wing aircraft, flying, legged and wheeled robots, manipulators, and heavy-duty trucks---both in simulation and hardware experiments. 

In Sec. \ref{sec:conclusions}, we highlight the limitations of the paradigms presented in this tutorial and provide our perspective on open research directions.

\section{A Primer on Safety-Critical Control}\label{sec:safety}

\subsection{Notation}
We use $\mathbb{N}$, $\R$, $\R_{\geq0}$, $\R_{>0}$ to denote the set of natural numbers, real numbers, nonnegative real numbers, and positive real numbers, respectively. The notation $\R^n$ denotes the $n$-dimensional Euclidean vector space. Given a vector $\bx\in\R^n$ we write $\bx\T\in\R^{1\times n}$ to denote its transpose and $\bx\cdot\by=\bx\T\by$ to denote the inner product between vectors.
Given a continuously differentiable scalar function $h\,:\,\R^n\rightarrow\R$ we denote the gradient of $h$ as $\nabla h\,:\,\R^n\rightarrow\R^n$. We use $L_{\bf}h(\bx)\coloneqq \nabla h(\bx)\cdot \bf(\bx)$ to denote the \emph{Lie derivative} of a continuously differentiable scalar function $h\,:\,\R^n\rightarrow\R$ along a vector field $\bf\,:\,\R^n\rightarrow\R^n$. The same definition applies when taking the Lie derivative of $h$ along a matrix-valued function $\bg\,:\,\R^n\rightarrow\R^{n\times m}$ whose columns can be thought of as vector fields on $\R^n$. For a continuously differentiable function $\bk\,:\,\R^n\rightarrow\R^m$ we use $\pdv{\bk}{\bx}(\bx)\in\R^{m\times n}$ to denote the Jacobian matrix of $\bk$ evaluated at $\bx\in\R^n$. A continuous function $\alpha\,:\,\R\rightarrow\R$ is said to be an \emph{extended class} $\Kinf$ \emph{function}, denoted by ${\alpha\in\Kinf^{\rm e}}$, if ${\alpha(0)=0}$, $\alpha$ is strictly increasing, and $\lim_{s\rightarrow\pm\infty}\alpha(s)=\pm\infty$. A continuous function $\alpha\,:\,\R_{\geq0}\rightarrow\R_{\geq0}$ is said to be class $\Kinf$ function, denoted by $\alpha\in\Kinf$, if $\alpha(0)=0$, $\alpha$ is strictly increasing and $\lim_{s\rightarrow\infty}\alpha(s)=\infty$. We use $\relu(x)\coloneqq\max\{0,x\}$ to denote the $\relu$ activation function. For a manifold $\Q$, we use $\Tq$ to denote the tangent space to $\Q$ at a point $\bq\in\Q$ and $\TQ$ to denote the tangent bundle. We use $\|\bx\|$ to denote the Euclidean norm of a vector $\bx\in\R^n$ and $\|\bx\|_\C\coloneqq \inf_{\by\in \C}\|\bx - \by\|$ to denote the distance between a vector $\bx\in\R^n$ and a set $\C\subset\R^n$. Given a function $h\,:\,\R^n\rightarrow\R$ and set $\C\subset\R$ we denote the \emph{restriction} of $h$ to $\C$ by $h\vert_{\C}\,:\,\C\rightarrow\R$.
For a closed set $\C\subset\R^n$, we use $\partial\C$ to denote its boundary and $\mathrm{Int}(\C)$ to denote its interior. We use $\bzero$ to denote a vector or matrix of zeros of appropriate dimension and $\bI$ to denote an identity matrix of appropriate dimension, where all dimensions will be made clear from the context.

\subsection{Foundations of Safety-Critical Control}
In this subsection, we outline the foundations of safety-critical control based on the fundamental notion of set invariance. We begin by considering the dynamical system:
\begin{equation}\label{eq:dyn}
    \dot{\bx} = \bf(\bx),
\end{equation}
where $\bx\in\R^n$ is the system state and $\bf\,:\,\R^n\rightarrow\R^n$ is a locally Lipschitz vector field.
Then, for each initial condition $\bx_0\in\R^n$, the dynamics in \eqref{eq:dyn} generate a unique continuously differentiable trajectory $\bx\,:\,I(\bx_0)\rightarrow\R^n$ defined on some maximal interval of existence $I(\bx_0)\subseteq\R_{\geq0}$ satisfying:
\begin{equation}
    \begin{aligned}
        \dot{\bx}(t) = & \bf(\bx(t)) \\ 
        \bx(0) = & \bx_0,
    \end{aligned}
\end{equation}
for all ${t\in I(\bx_0)}$ \cite[Ch. 3]{Khalil}.

The main property of \eqref{eq:dyn} studied in this paper is \emph{safety}, which is formalized by requiring trajectories of \eqref{eq:dyn} to remain within a safe set $\C\subset\R^n$ at all times.
\begin{definition}[Safety \cite{AmesECC19}]\label{def:safety}
    A set ${\C\subset\R^n}$ is said to be \emph{forward invariant} for \eqref{eq:dyn} if for each initial condition ${\bx_0\in\C}$, the resulting trajectory ${\bx\,:\,I(\bx_0)\rightarrow\R^n}$ satisfies $\bx(t)\in \C$ for all $t\in I(\bx_0)$. System \eqref{eq:dyn} is said to be \emph{safe} on a set $\C\subset\R^n$ if $\C$ is forward invariant.
\end{definition}
Necessary and sufficient conditions for set invariance, and thus safety, can be characterized using the notion of tangent cones\footnote{For a general closed set $\C$ one may define various classes of tangent cones, all of which coincide when $\C$ is convex. Examples include the Bouligand tangent cone \cite{Bouligand1932} and the Clarke tangent cone. In this tutorial, our definition corresponds to the Bouligand tangent cone.} \cite{nagumo1942lage,Bony1969,Brezis1970,Redheffer1972}. Informally, the tangent cone $\mathcal{T}_{\C}(\bx)\subset\R^n$ to a closed set $\C\subset\R^n$ at a point $\bx\in\R^n$ is the set of all vectors $\bv\in\R^n$ emanating from $\bx$ such that if one were to move infinitesimally along $\bv$, then one would remain in $\C$. Hence, for $\bx\in\mathrm{Int}(\C)$ we have $\mathcal{T}_{\C}(\bx)=\R^n$, whereas for $\bx\notin\C$ we have $\mathcal{T}_{\C}(\bx)=\emptyset$, implying the tangent cone is nontrivial only on the boundary of $\C$. The above ideas can be formalized concisely using the following definition:
\begin{equation}\label{eq:tangent-cone}
    \mathcal{T}_{\C}(\bx) \coloneqq \bigg\{\bv\in\R^n\,:\,\liminf_{\delta\rightarrow 0^{+}}\frac{\|\bx + \delta\bv\|_{\C}}{\delta}=0 \bigg\}.
\end{equation}
The following result, known as \emph{Nagumo's Theorem}, leverages tangent cones to provide necessary and sufficient conditions for safety.
\begin{theorem}[Nagumo's Theorem~\cite{nagumo1942lage}]\label{theorem:nagumo}
    A closed set $\C\subset\R^n$ is forward invariant for \eqref{eq:dyn} if and only if for all $\bx\in\partial\C$:
    \begin{equation}
        \bf(\bx)\in\mathcal{T}_{\C}(\bx).
    \end{equation}
\end{theorem}
Intuitively, Nagumo's Theorem states that $\C$ is forward invariant if and only if the vector field characterizing \eqref{eq:dyn} points into or is tangent to $\C$ for each point on the boundary of $\C$. Modern proofs of Nagumo's Theorem can be found in \cite[Ch. 4]{Blanchini} and \cite[Ch. 4]{AbrahamMarsdenRatiu}. Unfortunately, obtaining a closed-form expression to \eqref{eq:tangent-cone} for general closed sets $\C$ is often not possible, making the general version of Nagumo's Theorem challenging to apply in practice. To obtain more practical conditions for safety, we must restrict the class of sets whose invariance we wish to certify. Throughout this paper, we focus on sets of the form:
\begin{equation}\label{eq:C}
\begin{aligned}
    \C = & \{\bx\in\R^n\,:\,h(\bx)\geq0\}, \\
    \partial\C = & \{\bx\in\R^n\,:\,h(\bx)=0\}, \\
    \mathrm{Int}(\C) = & \{\bx\in\R^n\,:\,h(\bx) > 0\}, 
\end{aligned}
\end{equation}
where $h\,:\,\R^n\rightarrow\R$ is continuously differentiable. Before illustrating how such sets yield convenient representations of tangent cones, we require the notion of a \emph{regular value}.
\begin{definition}[Regular value \cite{AbrahamMarsdenRatiu}]\label{definition:regular-value}
    A real number ${a\in\R}$ is said to be a \emph{regular value} of a continuously differentiable function $h\,:\,\R^n\rightarrow\R$ if
    $\nabla h(\bx)\neq\bzero$ whenever $h(\bx)=a$.
\end{definition}
When $\C$ is defined as in \eqref{eq:C} and zero is a regular value of $h$, the tangent cone is straightforward to compute. 
\begin{lemma}[\cite{AbrahamMarsdenRatiu}]\label{lemma:regular-value}
    Consider a set $\C\subset\R^n$ as in \eqref{eq:C} and suppose that zero is a regular value of $h$. Then:
    \begin{equation}\label{eq:Tc}
        \mathcal{T}_{\C}(\bx) = \{\bv\in\R^n\,:\,\nabla h(\bx)\cdot\bv \geq 0 \},\quad \forall \bx\in\partial\C.
    \end{equation}
\end{lemma}
This characterization of tangent cones leads to the following useful corollary of Nagumo's Theorem.
\begin{corollary}
    Let the conditions of Lemma \ref{lemma:regular-value} hold. Then, $\C$ is forward invariant for \eqref{eq:dyn} if and only if:
    \begin{equation}\label{eq:Lfh>0}
        h(\bx)=0 \implies \dot{h}(\bx) =  L_{\bf}h(\bx) \geq 0.
    \end{equation}
\end{corollary}
Note that when zero is not a regular value of $h$, the condition in \eqref{eq:Lfh>0} does not necessarily imply the forward invariance of $\C$ since, in such a situation, the tangent cone does not coincide with \eqref{eq:Tc}.

The preceding developments serve as the foundation for \emph{barrier functions} -- Lyapunov-like functions that can be used to verify the safety (rather than stability) of nonlinear systems.

\begin{definition}[Barrier function~\cite{AmesADHS15}]\label{def:barrier}
    A continuously differentiable function $h\,:\,\R^n\rightarrow\R$ defining a set $\C\subset\R^n$ as in \eqref{eq:C} is said to be a \emph{barrier function} for \eqref{eq:dyn} on $\C$ if zero is a regular value of $h$ and there exists $\alpha\in\Kinf^{\rm e}$ such that for all $\bx\in\R^n$:
    \begin{equation}\label{eq:barrier}
        \dot{h}(\bx) = L_{\bf}h(\bx) \geq - \alpha(h(\bx)).
    \end{equation}
\end{definition}

Note that since $\alpha(0)=0$, the condition in \eqref{eq:barrier} implies that in \eqref{eq:Lfh>0}, thereby providing a suitable generalization of invariance conditions beyond just the boundary of $\C$. Intuitively, the condition in \eqref{eq:barrier} requires the system to ``slow down" as it approaches the boundary of $\C$ and stop once it reaches the boundary. Although our definition of a barrier function requires zero to be a regular value of $h$, this is not strictly necessary. Indeed, the use of an extended class $\Kinf$ function in conjunction with requiring inequality \eqref{eq:barrier} to hold at points outside of $\C$ enables one to dispense with this regularity condition and establish forward invariance using the comparison lemma \cite{KondaLCSS21}, providing further generalizations of classical invariance tools. An additional benefit of requiring inequality \eqref{eq:barrier} to hold on a set larger than $\C$ -- in our case, all of $\R^n$ -- is that such a condition not only enforces invariance of $\C$, but also attractivity of $\C$. That is, $\C$ is \emph{asymptotically stable}\footnote{Note that forward invariance is a necessary condition for asymptotic stability of a set. Thus, barrier functions can also be seen as generalizing Lyapunov functions certifying stability of equilibrium points to Lyapunov functions certifying stability of sets.} for \eqref{eq:dyn} with $V(\bx)=\relu(-h(\bx))$ as a Lyapunov function. 
\begin{theorem}[\cite{AmesADHS15}]\label{theorem:barrier}
    If $h\,:\,\R^n\rightarrow\R$ is a barrier function for \eqref{eq:dyn} on a set ${\C\subset\R^n}$ as in \eqref{eq:C}, then $\C$ is forward invariant. Moreover, if
    $\C$ is compact or the vector field $\bf$ in \eqref{eq:dyn} is forward complete,
    then $\C$ is asymptotically stable.
\end{theorem}
In the above result, the requirement that \eqref{eq:barrier} holds on all of $\R^n$ is made only for ease of exposition -- Theorem \ref{theorem:barrier} and almost all other barrier-related results presented in this tutorial can be generalized to hold on a subset $\mathcal{D}\subseteq\R^n$ such that $\C\subset\mathcal{D}$. Finally, we note that the characterization of set invariance via barrier functions is tight in the sense that, under certain conditions, the existence of a barrier function is necessary and sufficient for forward invariance.

\begin{theorem}[\cite{AmesADHS15}]
    Let $h\,:\,\R^n\rightarrow\R$ be a continuously differentiable function defining a compact set $\C\subset\R^n$ as in \eqref{eq:C} and assume zero is a regular value of $h$. Then, $\C$ is forward invariant for \eqref{eq:dyn} if and only if $h\vert_{\C}\,:\,\C\rightarrow\R$ is a barrier function for \eqref{eq:dyn} on $\C$. 
\end{theorem}

The preceding generalizations of set invariance via barrier functions play an important role in synthesizing controllers enforcing safety,
discussed in the following section.

\subsection{Control Barrier Functions}
In the previous subsection, we laid the foundation for safety-critical control using the language of set invariance and illustrated how barrier functions provide a useful tool for verifying safety properties of dynamical systems. In this section, we focus our attention on control systems of the form:
\begin{equation}\label{eq:control-affine}
    \dot{\bx} = \bf(\bx) + \bg(\bx)\bu,
\end{equation}
where $\bf\,:\,\R^n\rightarrow\R^n$ is a locally Lipschitz vector field modeling the \emph{drift} of the system, $\bg\,:\,\R^n\rightarrow\R^{n\times m}$ is a locally Lipschitz mapping characterizing the \emph{control directions}, and $\bu\in\R^m$ is the control input. Defining a notion of safety for a control system, such as in \eqref{eq:control-affine}, rather than a closed-loop system, such as in \eqref{eq:dyn}, requires some modifications. Def. \ref{def:safety} cannot be directly applied to \eqref{eq:control-affine} since the trajectories of \eqref{eq:control-affine} cannot be determined, in general, until one specifies a control input $\bu$. The definition of safety for \eqref{eq:control-affine} is captured via the notion of \emph{controlled invariance}.

\begin{definition}[Controlled invariance \cite{Blanchini}]\label{def:controlled-invariance}
    A set $\C\subset\R^n$ is said to be \emph{feedback controlled invariant} for \eqref{eq:control-affine} if there exists a locally Lipschitz feedback controller $\bk\,:\,\R^n\rightarrow\R^m$ such that $\C$ is forward invariant for the closed-loop system:
    \begin{equation}\label{eq:closed-loop-control-affine}
        \dot{\bx} = \bf(\bx) + \bg(\bx)\bk(\bx) \eqqcolon \bf_{\mathrm{cl}}(\bx).
    \end{equation}
\end{definition}

Rather than verifying that an a priori designed controller renders $\C$ forward invariant using the barrier conditions outlined in the previous subsection, our objective in this subsection is to provide a general methodology to design controllers that enforce safety by construction. Towards this objective, the aforementioned barrier conditions suggest designing such a controller so as to satisfy:
\begin{equation}\label{eq:closed-loop-barrier}
    \underbrace{L_{\bf}h(\bx) + L_{\bg}h(\bx)\bk(\bx)}_{L_{\bf_\mathrm{cl}}h(\bx)} \geq - \alpha (h(\bx)),
\end{equation}
implying that such a controller enforces safety of the closed-loop system by Theorem \ref{theorem:barrier}. This observation
motivates the concept of a \emph{control barrier function} (CBF).

\begin{definition}[Control barrier function~\cite{AmesTAC17}]\label{def:cbf}
    A continuously differentiable function $h\,:\,\R^n\rightarrow\R$ defining a set $\C\subset\R^n$ as in \eqref{eq:C} is said to be a \emph{control barrier function} for \eqref{eq:control-affine} on $\C$ if there exists $\alpha\in\Kinf^{\rm e}$ such that for all $\bx\in\R^n$:
    \begin{equation}\label{eq:cbf}
       \sup_{\bu\in\R^m}\dot{h}(\bx,\bu)= \sup_{\bu\in\R^m}\big\{L_{\bf}h(\bx) + L_{\bg}h(\bx)\bu \big\} > - \alpha(h(\bx)).
    \end{equation}
\end{definition}
In contrast to Def. \ref{def:barrier}, we do not explicitly require zero to be a regular value of $h$ in the above definition since this property implicitly follows from the strict inequality in \eqref{eq:cbf}. Further motivation behind the use of this strict inequality is presented in Remark \ref{remark:strict}, and concerns the continuity of controllers synthesized from CBFs. The existence of a CBF implies that for each $\bx\in\R^n$ there exists an input $\bu\in\R^m$ enforcing the inequality:
\begin{equation*}
    L_{\bf}h(\bx) + L_{\bg}h(\bx)\bu > - \alpha(h(\bx)).
\end{equation*}
To use such inputs to enforce safety, we must be able to stitch them together into a locally Lipschitz feedback controller $\bk\,:\,\R^n\rightarrow\R^m$ satisfying \eqref{eq:closed-loop-barrier}. Fortunately, the existence of a CBF implies the existence of such a controller.
\begin{theorem}[\cite{AmesTAC17}]
    If $h\,:\,\R^n\rightarrow\R$ is a CBF for \eqref{eq:control-affine} on a set $\C\subset\R^n$ as in \eqref{eq:C}, then $\C$ is feedback controlled invariant.
    Furthermore, if a locally Lipschitz feedback controller ${\bk\,:\,\R^n\rightarrow\R^m}$ satisfies~\eqref{eq:closed-loop-barrier} for all $\bx \in \R^n$, then $\C$ is forward invariant for~\eqref{eq:closed-loop-control-affine}.
\end{theorem}

Although the above theorem guarantees the existence of a controller enforcing safety, it does not explicitly state how to construct one. The most popular approach to constructing CBF-based controllers is to incorporate \eqref{eq:closed-loop-barrier} as a constraint in an optimization problem parameterized by the system state. That is, the controller $\bx\mapsto \bk(\bx)$ is itself an optimization problem that returns, for each $\bx$, a control input $\bu=\bk(\bx)$ satisfying \eqref{eq:closed-loop-barrier}. This approach is motivated by the fact that such an inequality defines an affine constraint on the control input, implying $\bk(\bx)$ can often be cast as a quadratic program (QP) that, in many situations, 
admits a closed-form solution.

Perhaps the greatest utility of this QP-based perspective is the ability to use CBFs as a \emph{safety filter} for a desired control policy $\bk_{\mathrm{d}}\,:\,\R^n\rightarrow\R^m$ whose safety has not yet been established. Often, it is desirable to modify such a controller in a minimally invasive fashion while guaranteeing safety. This leads to the instantiation of safety-critical controllers via the following optimization problem:
\begin{equation}\label{eq:safety-filter-qp}
    \begin{aligned}
        \bk(\bx) = & \argmin_{\bu\in\R^m} && \tfrac{1}{2}\|\bu - \bk_{\mathrm{d}}(\bx)\|^2 \\
        & \mathrm{subject~to} && L_{\bf}h(\bx) + L_{\bg}h(\bx)\bu \geq - \alpha(h(\bx)),
    \end{aligned}
\end{equation}
which is a QP whose closed-form solution can be obtained
by defining:
\begin{equation}\label{eq:a-and-b}
    \begin{aligned}
        a(\bx) \coloneqq & L_{\bf}h(\bx) + L_{\bg}h(\bx)\bk_{\mathrm{d}}(\bx) + \alpha(h(\bx)) \\ 
        b(\bx) \coloneqq & \|L_{\bg}h(\bx)\|^2,
    \end{aligned}
\end{equation}
and applying the Karush-Kuhn Tucker conditions \cite{Boyd} to yield~\cite{AnilTCST23}:
\begin{equation}\label{eq:safety-filter-qp-closed-form}
\begin{aligned}
    \bk(\bx) = & \bk_{\mathrm{d}}(\bx) + \lambda(a(\bx),b(\bx))L_{\bg}h(\bx)\T \\
    \lambda(a,b) \coloneqq & \begin{cases}
        0 & b \leq 0  \\
        \relu(-a/b) & b > 0,
    \end{cases}
\end{aligned}
\end{equation}
where $\lambda$ is the Lagrange multiplier associated with the constraint in \eqref{eq:safety-filter-qp}. Note that, by \eqref{eq:safety-filter-qp-closed-form}, the controller in \eqref{eq:safety-filter-qp} allows the desired controller $\bk_{\mathrm{d}}$ to be applied so long as it satisfies the barrier condition \eqref{eq:closed-loop-barrier}, and provides a minimal correction to $\bk_{\mathrm{d}}$ when such a condition is not satisfied. Importantly, the closed-form expression to the QP \eqref{eq:safety-filter-qp} in \eqref{eq:safety-filter-qp-closed-form} obviates the need explicitly solve an optimization problem in the control loop, which enables the deployment of such controllers on hardware with limited computational resources.  Although this closed-form expression is only valid for a single CBF, whereas, in practice, one must often consider multiple CBFs, one often can combine multiple CBFs into one, allowing one to leverage the closed form solution even for arbitrarily complicated safety specifications \cite{Tamas-Smooth-CBFs}.

\begin{remark}[Strict inequality]\label{remark:strict}
    One may note that in \eqref{eq:barrier} and \eqref{eq:closed-loop-barrier} we have used a nonstrict inequality, whereas in the definition of a CBF \eqref{eq:cbf} we have opted for a \emph{strict} inequality. This difference is subtle but plays an important role in ensuring Lipschitz continuity of CBF-based controllers~\cite{jankovic2018robust}. In short, the strict inequality preserves Lipschitz continuity of CBF-based controllers at points where $L_{\bg}h(\bx)=\bzero$ (see \cite[Ch. 3.5.3]{Sepulchre} for a similar discussion in the context of control Lyapunov functions).  Such points arise often in practice. For example, any compact safe set\footnote{If $\C$ is compact and $h$ is continuously differentiable, then $h$ achieves a local maximum over $\C$. At such a local maximum the gradient of $h$ must vanish, implying $L_{\bg}h$ will also vanish.} will contain points such that $L_{\bg}h(\bx)=\bzero$. Note that, as a result, one may use a nonstrict inequality in \eqref{eq:cbf} if $L_{\bg}h(\bx)\neq\bzero$ for all $\bx\in\R^n$. Finally, we note that the strict inequality is a property of the dynamics and safe set irrespective of any particular controller -- its purpose is to restrict the class of functions that may serve as a CBF to those that can be used to synthesize a locally Lipschitz feedback controller. 
\end{remark}

Although constructing a controller given a CBF can be done systematically, constructing a CBF is often more challenging. To determine if a candidate CBF $h$ -- a continuously differentiable function defining \eqref{eq:C} -- is indeed a CBF,
one must verify that \eqref{eq:cbf} holds for each $\bx\in\R^n$.
To do so, one may compute the supremum in \eqref{eq:cbf}:
\begin{equation*}
    \sup_{\bu\in\R^m}\big\{L_{\bf}h(\bx) + L_{\bg}h(\bx)\bu\big\} = 
    \begin{cases}
        \infty & L_{\bg}h(\bx)\neq\bzero \\
        L_{\bf}h(\bx) &  L_{\bg}h(\bx)=\bzero
    \end{cases}
\end{equation*}
and verify that the above result is strictly greater than $-\alpha(h(\bx))$. This simplifies to verifying that:
\begin{equation*}
    L_{\bg}h(\bx)=\bzero \implies L_{\bf}h(\bx) > - \alpha(h(\bx)),
\end{equation*}
for all ${\bx\in\R^n}$. Intuitively, the CBF condition \eqref{eq:cbf} is a scalar inequality, which, when $L_{\bg}h(\bx)\neq\bzero$, is always possible to satisfy by simply taking $\bu$ as large or small as necessary. When $L_{\bg}h(\bx)=\bzero$, however, one must rely on the unforced dynamics of the system -- captured via $\bf$ -- to satisfy the CBF inequality. This discussion is formalized via the following lemma.
\begin{lemma}\label{lemma:Lgh=0}
    A continuously differentiable function $h\,:\,\R^n\rightarrow\R$ is a CBF for \eqref{eq:control-affine} on $\C$ if and only if zero is a regular value of $h$ and for all $\bx\in\R^n$:
    \begin{equation}\label{eq:Lgh=0}
        L_{\bg}h(\bx) = \bzero \implies L_{\bf}h(\bx) > - \alpha(h(\bx)).
    \end{equation}
\end{lemma}

\begin{remark}[Input constraints]
    Lemma \ref{lemma:Lgh=0} provides necessary and sufficient conditions for $h$ to be a CBF when the control input is \emph{unconstrained}, that is, when $\bu$ may take any value in $\R^m$. When additional inputs bounds are present in the sense that $\bu$ may only take values in a strict subset $\mathcal{U}\subset\R^m$, Lemma \ref{lemma:Lgh=0} provides necessary\footnote{If $h$ is not a CBF without input bounds, then it certainly will not be with input bounds.}, but not necessarily sufficient conditions that $h$ must satisfy to be a CBF. For ease of exposition, this tutorial will focus on the construction of CBFs \emph{without} additional input bounds. Many of the approaches discussed herein may be extended to include input bounds through the use of backup CBFs \cite{AmesIEEEA20,YuxiaoCDC21}, with more details on the unification of backup CBFs and ROMs discussed in \cite{TamasACC23}.
\end{remark}

For relatively simple systems, Lemma \ref{lemma:Lgh=0} provides a simple condition that one may check to certify that a continuously differentiable function $h$ defining a set $\C$ as in \eqref{eq:C} is indeed a CBF. 
The following example demonstrates such a procedure for a canonical example in the CBF literature: the inverted pendulum.

\begin{figure*}
    \centering
    \includegraphics{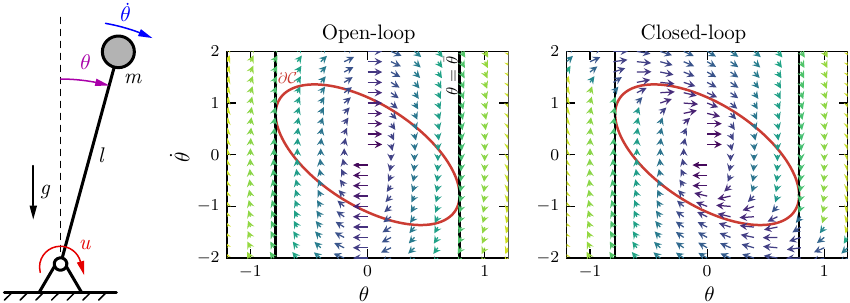}
    \caption{Vector field of the inverted pendulum in Example \ref{example:inverted-pendulum} without any controller (left) and with the safety filter from \eqref{eq:safety-filter-qp-closed-form} (right). In each plot, the red ellipse denotes the boundary of $\C$, the black vertical lines denote $|\theta|=\bar{\theta}=\frac{\pi}{4}$, and the arrows of varying color illustrate the system vector field. The varying colors of the arrows characterize the magnitude of each vector, with lighter colors corresponding to larger magnitudes.}
    \label{fig:inverted-pendulum}
\end{figure*}

\begin{example}[Inverted pendulum]\label{example:inverted-pendulum}
    We now consider the 
    example of an inverted pendulum with state $\bx=(\theta,\dot{\theta})$ and dynamics:
    \begin{equation*}
        \underbrace{
        \begin{bmatrix}
            \dot{\theta} \\ \ddot{\theta}
        \end{bmatrix}}_{\dot{\bx}}
        =
        \underbrace{
        \begin{bmatrix}
            \dot{\theta} \\ \frac{g}{l}\sin(\theta)
        \end{bmatrix}}_{\bf(\bx)}
        +
        \underbrace{
        \begin{bmatrix}
            0 \\ \frac{1}{ml^2}
        \end{bmatrix}}_{\bg(\bx)}\bu,
    \end{equation*}
    where $\theta\in\R$ denotes the angular position of the pendulum, $g$ the acceleration due to gravity, $l$ the length of the pendulum, and $m$ the mass of the pendulum.
    We establish a safety-critical controller for the inverted pendulum by following the corresponding example in~\cite{AnilLCSS22}.
    Our objective is to design a controller for the above system that keeps the pendulum upright in the sense that its angular position satisfies $|\theta|\leq\bar{\theta}$ for some $\bar{\theta}\in\R_{>0}$.
    
    To achieve this objective, we propose the CBF candidate:
    \begin{equation*}
        h(\bx) = \bar{\theta}^2 - \theta^2 - \frac{1}{2}(\dot{\theta} + \theta)^2,
    \end{equation*}
    which defines a candidate safe set $\C\subset\R^2$ as in \eqref{eq:C}. Note that if $(\theta,\dot{\theta})\in\C$, then $|\theta|\leq\bar{\theta}$ since:
    \begin{equation*}
        h(\bx)\geq0\implies \bar{\theta}^2 - \theta^2 \geq \frac{1}{2}(\dot{\theta} + \theta)^2 \geq 0 \implies \theta^2 \leq \bar{\theta}^2.
    \end{equation*}
    Hence, enforcing forward invariance of $\C$ ensures that $|\theta(t)|\leq \bar{\theta}$ for all $t$. To check if $h$ is a CBF we first compute:
    \begin{equation*}
        \nabla h(\bx) = \begin{bmatrix}
            -2\theta - (\dot{\theta} + \theta) \\ -(\dot{\theta} + \theta)
        \end{bmatrix},
    \end{equation*}
    and verify that zero is a regular value of $h$ by investigating the solution set of the linear system:
    \begin{equation*}
        \nabla h(\bx)=\bzero \iff
        \begin{bmatrix}
            3 & 1 \\ 1 & 1
        \end{bmatrix}
        \begin{bmatrix}
            \theta \\ \dot{\theta}
        \end{bmatrix}
        =
        \begin{bmatrix}
            0 \\ 0
        \end{bmatrix}.
    \end{equation*}
    The matrix in the above linear system is positive definite, thus the only solution is $(\theta,\dot{\theta})=\bzero$. Since the only point where the gradient of $h$ vanishes is at the origin, which does not lie on the boundary of $\C$, zero is a regular value of $h$. To use Lemma \ref{lemma:Lgh=0} and verify $h$ as a CBF, we must analyze the behavior of $\dot{h}$ when $L_{\bg}h(\bx)=0$. To this end, we note that:
    \begin{equation*}
        L_{\bg}h(\bx) = -\frac{\dot{\theta} + \theta}{ml^2} = 0 \implies \dot{\theta} + \theta =0.
    \end{equation*}
    Hence, when $L_{\bg}h(\bx)=0$, we also have:
    \begin{equation*}
        \begin{aligned}
            L_{\bf}h(\bx) = & \begin{bmatrix}
                -2\theta & 0
            \end{bmatrix}
            \begin{bmatrix}
                \dot{\theta} \\ \frac{g}{l}\sin(\theta)
            \end{bmatrix}
            = -2\theta\dot{\theta} = 2\theta^2,
        \end{aligned}
    \end{equation*}
    and $ h(\bx) = \bar{\theta}^2 - \theta^2$, implying that:
    \begin{equation*}
        L_{\bf}h(\bx) + \alpha(h(\bx)) = 2\theta^2 + \alpha(\bar{\theta}^2 - \theta^2).
    \end{equation*}
    By taking $\alpha(s)=\alpha_0 s$ as a linear extended class $\Kinf$ function, we see that:
    \begin{equation*}
        L_{\bf}h(\bx) + \alpha(h(\bx)) = (2 - \alpha_0)\theta^2 + \alpha_0\bar{\theta}^2 > 0,
    \end{equation*}
    for all $\bx\in\R^2$ for any $\alpha_0\in(0,2]$, implying that \eqref{eq:cbf} holds for all $\bx\in\R^2$ and, consequently, that $h$ is a CBF for the inverted pendulum. To accomplish the objective of keeping the pendulum upright, we synthesize a safety filter $\bk\,:\,\R^2\rightarrow\R$ using the QP in \eqref{eq:safety-filter-qp} with a nominal policy of $\bk_{\mathrm{d}}(\bx)=0$ and $\alpha_0=1$ whose closed-form solution is given by \eqref{eq:safety-filter-qp-closed-form}. The closed-loop vector field of the pendulum under the influence of the safety filter and the corresponding safe set is provided in Fig. \ref{fig:inverted-pendulum}.
\end{example}

The previous example illustrates the procedure required to construct a CBF for relatively simple systems. 
In Example \ref{example:inverted-pendulum}, our CBF was different than the safety constraint $\bar{\theta}^2 - \theta^2 \geq0$ we wished to satisfy and contained additional terms that depended on both the position and velocity of the pendulum. For relatively simple systems, such as the inverted pendulum, appending such terms to the original safety requirement to obtain a CBF can often be done through intuition or trial-and-error. For more complex high-dimensional systems, however, constructing such a ``handcrafted" CBF by carefully blending various states of the system into a single scalar function may be intractable. 

Motivated by these challenges, the primary objective of this paper is to outline a comprehensive methodology for systematically constructing CBFs for high-dimensional nonlinear systems based on reduced-order models. Ultimately, this methodology enables one to construct CBFs for complex systems from CBFs for much simpler systems, such as the inverted pendulum outlined above. Before presenting such constructions, we discuss in the following section how the results of the present section can be extended to handle uncertainties.

\subsection{Robust Safety-Critical Control}\label{sec:issf}
In the previous subsections, we discussed notions of safety for dynamical and control systems, implicitly assuming that the dynamics governing the system are fully known. In reality, however, any system will be affected by unmodeled dynamics and disturbances, which begs the question: how do safety properties degrade in the presence of uncertainties, and how may we design controllers so as to mitigate the effects of such uncertainties?
In this subsection, we discuss robust variants of CBFs via the notion of \emph{input-to-state safety} (ISSf) \cite{AmesLCSS19,AnilLCSS22,AnilTCST23}, which provides an answer to this question.

Our starting point is the uncertain control affine system:
\begin{equation}\label{eq:disturbed-control-affine}
    \dot{\bx} = \bf(\bx) + \bg(\bx)(\bu + \bd),
\end{equation}
where $\bd\in\R^m$ is a disturbance. As the disturbance enters the dynamics through the same channels as the control input, the disturbance is said to be \emph{matched}, implying that, if the disturbance were known, it could simply be canceled by the control input. Given a locally Lipschitz feedback controller $\bk\,:\,\R^n\rightarrow\R^m$ and a piecewise continuous disturbance signal $t\mapsto \bd(t)$, we obtain the closed-loop system:
\begin{equation}\label{eq:disturbed-closed-loop}
    \dot{\bx} = \bf(\bx) + \bg(\bx)(\bk(\bx) + \bd(t)),
\end{equation}
which, for each initial condition ${\bx_0\in\R^n}$, admits a piecewise continuously differentiable solution $\bx\,:\,I(\bx_0,\bd(\cdot))\rightarrow\R^n$ defined on some maximal interval of existence $I(\bx_0,\bd(\cdot))\subseteq\R_{\geq0}$.

In what follows, we assume bounded disturbance:
\begin{equation}\label{eq:disturbance-bound}
    \|\bd\|_{\infty} \coloneqq \sup_{t\geq0}\|\bd(t)\| \leq \delta,
\end{equation}
with some $\delta\geq0$.
Given this bound on $\bd$, we introduce a family of \emph{inflated} safe sets:
\begin{equation}\label{eq:C-delta}
    \C_{\delta} \coloneq \{\bx\in\R^n\,:\,h_{\delta}(\bx) \geq 0\},
\end{equation}
parameterized by $\delta$, where:
\begin{equation}\label{eq:h-delta}
    h_{\delta}(\bx) \coloneqq h(\bx) + \gamma(\delta),
\end{equation}
for a $\gamma\in\Kinf$ to be specified shortly. Our notion of safety for \eqref{eq:disturbed-closed-loop} is captured via the notion of ISSf.

\begin{definition}[Input-to-state safety~\cite{AmesLCSS19}]\label{def:ISSf}
    System~\eqref{eq:disturbed-closed-loop} is said to be \emph{input-to-state safe} (ISSf) on a set ${\C\subset\R^n}$ as in \eqref{eq:C} if there exists a ${\gamma\in\Kinf}$ such that for all $\delta\geq0$ and all ${t\mapsto \bd(t)}$ satisfying \eqref{eq:disturbance-bound}
    the set $\C_{\delta}\!\subset\!\R^n$ as in \eqref{eq:C-delta} is forward invariant for \eqref{eq:disturbed-closed-loop}. 
\end{definition}
The ISSf property implies a graceful degradation of safety in the presence of uncertainties -- potential safety violations are bounded by the magnitude of such uncertainties. 
Similar to previous subsections, controllers enforcing such a safety property may be constructed using CBFs.

\begin{definition}[ISSf control barrier function~\cite{AnilLCSS22}]\label{def:ISSf-CBF}
    A continuously differentiable function $h\,:\,\R^n\rightarrow\R$ defining a set $\C\subset\R^n$ as in \eqref{eq:C-delta} is said to be an \emph{input-to-state safe CBF} (ISSf-CBF) for \eqref{eq:disturbed-control-affine} on $\C$ if there exist ${\alpha\in\Kinf^{\rm e}}$ and ${\varepsilon\in\R_{>0}}$ such that for all ${\bx\in\R^n}$:
    \begin{equation}\label{eq:ISSf-CBF}
        \sup_{\bu\in\R^m}\big\{L_{\bf}h(\bx) + L_{\bg}h(\bx)\bu \big\} > - \alpha(h(\bx)) + \frac{1}{\varepsilon}\|L_{\bg}h(\bx)\|^2.
    \end{equation}
\end{definition}
The main difference between CBFs and ISSf-CBFs is the inclusion of $\tfrac{1}{\varepsilon}\|L_{\bg}h(\bx)\|^2$ in \eqref{eq:ISSf-CBF}, which imposes a stronger condition on the control input. This term serves to mitigate the impact of uncertainties via the tuning parameter $\varepsilon>0$ as shown in the following result.

\begin{theorem}[\cite{AnilLCSS22}]
    If $h\,:\,\R^n\rightarrow\R$ is an ISSf-CBF for \eqref{eq:disturbed-control-affine} on a set $\C\subset\R^n$ as in \eqref{eq:C}, then any locally Lipschitz controller $\bk\,:\,\R^n\rightarrow\R^m$ satisfying:
    \begin{equation}\label{eq:ISSf-CBF-closed-loop}
        L_{\bf}h(\bx) + L_{\bg}h(\bx)\bk(\bx) \geq - \alpha(h(\bx)) + \frac{1}{\varepsilon}\|L_{\bg}h(\bx)\|^2,
    \end{equation}
    renders the closed-loop system \eqref{eq:disturbed-closed-loop} ISSf on $\C$ with:
    \begin{equation}\label{eq:gamma-delta-ISSf-CBF}
        \gamma(\delta) = -\alpha^{-1}\left(-\frac{\varepsilon\delta^2}{4}\right).
    \end{equation}
\end{theorem}
According to \eqref{eq:gamma-delta-ISSf-CBF}, the inflated set $\C_{\delta}$ can be brought as close as desired to the original safe set $\C$ by decreasing $\varepsilon$, with $\C_{\delta}\rightarrow\C$ in the limit as $\varepsilon\rightarrow0$. Although, in principle, one can take $\varepsilon$ as close to zero as desired, doing so generally imposes a stronger condition on the control input, requiring larger control effort, which may not be achievable in practice.
Similar to CBFs, the ISSf-CBF condition \eqref{eq:ISSf-CBF-closed-loop} can be interpreted as an affine constraint that the control input must satisfy, leading to the construction of ISSf enforcing controllers via QPs as in \eqref{eq:safety-filter-qp}. Note that when the uncertainties $\bd$ are matched, as in \eqref{eq:disturbed-control-affine}, and $L_{\bg}h(\bx)=\bzero$, neither the control input nor uncertainties may impact the system, implying the criterion for constructing CBFs in Lemma \ref{lemma:Lgh=0} also applies to ISSf-CBFs.

\subsection{Smooth Safety Filters}\label{sec:smooth}

In what follows, many of our results will require smooth (differentiable as many times as necessary) CBF controllers. This may seem restrictive since the vast majority of CBF controllers -- including the ones discussed in this tutorial thus far -- are computed as the solution to an optimization problem and are inherently nonsmooth. However, when the problem data itself is smooth (i.e., the dynamics $\bf,\bg$, CBF $h$, and extended class $\Kinf$ function $\alpha$), it is always possible to construct a smooth CBF controller.

\begin{lemma}[\cite{Cohen-Smooth-Safety}]\label{lemma:smooth-safety}
    Consider system \eqref{eq:control-affine} with
    ${\bf:\R^n\!\rightarrow\!\R^n}$, $\bg\,:\,\R^n\rightarrow\R^{n\times m}$ smooth functions
    and let $h\,:\,\R^n\rightarrow\R$ be a smooth CBF for \eqref{eq:control-affine} on a set $\C\subset\R^n$ as in \eqref{eq:C} with a smooth $\alpha\in\Kinf^{\rm e}$. Then, there exists a smooth feedback controller ${\bk\,:\,\R^n\rightarrow\R^m}$ such that
    \eqref{eq:closed-loop-barrier} holds
    for all ${\bx\in\R^n}$.
\end{lemma}

\begin{figure*}
    \centering
    \includegraphics{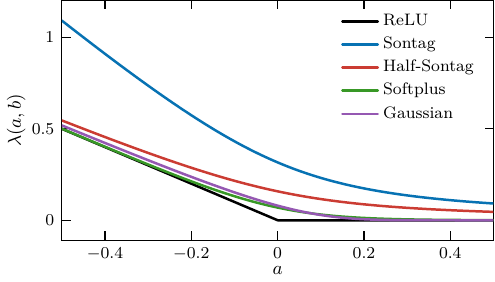}
    \hfill
    \includegraphics{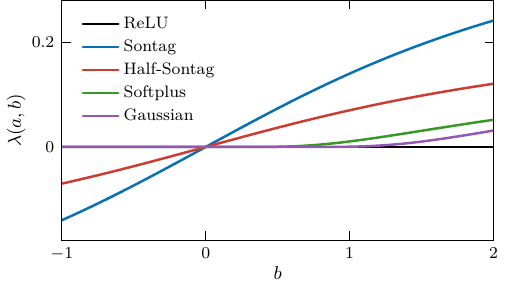}
    \caption{Smooth universal formulas for safety-critical control compared to the $\relu$ function associated with quadratic programs. The left plot illustrates the variation of $\lambda(a,b)$ with respect to $a$ for a fixed $b>0$ while the right plot illustrates the variation of $\lambda(a,b)$ with respect to $b$ for a fixed $a>0$ for each of the formulas in \eqref{eq:smooth-formulas}.}
    \label{fig:smooth-formulas}
\end{figure*}

The class of smooth controllers considered in this tutorial inherit the same structure as the closed-form QP controller \eqref{eq:safety-filter-qp-closed-form}:
\begin{equation}
    \bk(\bx) = \bk_{\mathrm{d}}(\bx) + \lambda(a(\bx),b(\bx))L_{\bg}h(\bx)\T,
\label{eq:safety-filter-smooth}
\end{equation}
where $\bk_{\mathrm{d}}\,:\,\R^n\rightarrow\R^m$ is a nominal controller and $a$ and $b$ are as in \eqref{eq:a-and-b}. Any smooth controller of the form \eqref{eq:safety-filter-smooth} satisfying the CBF inequality \eqref{eq:closed-loop-barrier} is said to be a \emph{smooth safety filter}. The fact that the QP controller in \eqref{eq:safety-filter-qp} is nonsmooth stems from the presence of the $\relu$ activation function in the Lagrange multiplier $\lambda$ in \eqref{eq:safety-filter-qp-closed-form}, which has the interpretation of ``activating" the safety filter when the nominal controller fails to guarantee satisfaction of the CBF constraint in \eqref{eq:closed-loop-barrier}. This non-smoothness can be removed by modifying the Lagrange multiplier $\lambda$ using various ``smooth universal formulas" such as \cite{Cohen-Smooth-Safety}: 
\begin{equation}\label{eq:smooth-formulas}
    \begin{aligned}
        \lambda(a,b) = & 
        \begin{cases}
            0 & b = 0 \\
            \frac{-a + \sqrt{a^2 + \sigma b^2}}{b} & b \neq 0
        \end{cases} && \;\; \text{(Sontag)} \\ 
        \lambda(a,b) = &  
        \begin{cases}
            0 & b = 0 \\
            \frac{-a + \sqrt{a^2 + \sigma b^2}}{2b} & b \neq 0
        \end{cases} && \;\; \text{(Half-Sontag)} \\ 
        \lambda(a,b) = &  
        \begin{cases}
            0 & b \leq 0 \\ 
            \sigma\log(1 + e^{-\frac{a}{\sigma b}}) & b > 0
        \end{cases} && \;\; \text{(Softplus)} \\ 
        \lambda(a,b) = &  
        \begin{cases}
            0 & b \leq 0 \\ 
            \sigma \frac{\mathrm{pdf}_{\mathcal{N}(0,1)}\big(\frac{a}{\sigma b}\big)}{\mathrm{cdf}_{\mathcal{N}(0,1)}\big(\frac{a}{\sigma b}\big)} & b > 0
        \end{cases} && \;\; \text{(Gaussian)}, \\ 
    \end{aligned}
\end{equation}
where $\sigma>0$ and $\mathrm{pdf}_{\mathcal{N}(0,1)}(\cdot)$ and $\mathrm{cdf}_{\mathcal{N}(0,1)}(\cdot)$ denote the probability density function and cumulative distribution function of a zero-mean Gaussian distribution with unit variance \cite{PioCDC19}. Each of these functions can be shown to be smooth on the set\footnote{In \cite{Cohen-Smooth-Safety} this set was originally taken as a subset of $\R\times\R_{\geq0}$ since, in the context of CBFs, $b\coloneqq \|L_{\bg}h(\bx)\|^2\geq0$ for all $\bx\in\R^n$, but can be extended to a subset of $\R^2$ to discuss smoothness of \eqref{eq:smooth-formulas} independent of their relation to CBFs.}:
\begin{equation*}
        \mathcal{S} = \{(a,b)\in\R^2\,:\,a>0\;\text{or}\;b > 0\},
\end{equation*}
and may be interpreted as a smooth over-approximation of the original Lagrange multiplier from \eqref{eq:safety-filter-qp-closed-form} as illustrated in
Fig. \ref{fig:smooth-formulas}. The safety properties of these smooth universal formulas -- including how closely they may approximate the QP-based controller \eqref{eq:safety-filter-qp} -- can be established using the techniques introduced in \cite{Cohen-Smooth-Safety}. 

\begin{remark}
    In the context of control Lyapunov functions (CLFs), it is often stated that Sontag's formula \cite{SontagSCL89} is smooth everywhere except possibly the origin, where one can generally only guarantee continuity under the small control property \cite[Ch. 3.5.3]{Sepulchre}. However, this phenomenon is unique to CLFs and does not arise in the context of CBFs provided one is willing to use a strict inequality in \eqref{eq:cbf}. Indeed, as discussed in Remark \ref{remark:strict}, to guarantee even continuity of CBF or CLF based controllers, one must generally use a strict inequality in the definition of a CBF/CLF, otherwise, the controller may not be continuous when $b=0$. This follows from the observation that $\lambda(a,0)=0$ and the limit of $\lambda(a,b)$ as $b$ approaches zero is zero under the condition that $b=0\implies a> 0$, where $\lambda$ is any of the formulas from \eqref{eq:safety-filter-qp-closed-form} and \eqref{eq:smooth-formulas}. In contrast, if one only requires $b=0\implies a\geq 0$ this limit may not exist. Now, when using a CLF, the strict inequality does not hold at the origin since CLFs are positive definite, and thus one requires an additional property to guarantee continuity, which comes in the form of the small control property. However, in the context of CBFs, under the presumption that zero is a regular value of $h$, which implicitly holds when defining a CBF as in \eqref{eq:cbf}, the strict inequality holds for all $\bx\in\mathbb{R}^n$, which ensures continuity of the QP-based controller at all points and smoothness of the other formulas at all points.
\end{remark}

As each of the formulas in \eqref{eq:smooth-formulas} is an over-approximation of the Lagrange multiplier from \eqref{eq:safety-filter-qp-closed-form}, the resulting smooth safety filter in \eqref{eq:safety-filter-smooth} enforces \emph{strict} satisfaction of the CBF inequality \eqref{eq:closed-loop-barrier}, which will become important when constructing CBFs from reduced-order models. Our discussion on smooth safety filters is formalized in the following result.

\begin{theorem}[\cite{Cohen-Smooth-Safety}]\label{theorem:smooth-safety}
    Let the conditions of Lemma \ref{lemma:smooth-safety} hold. Then, for each $\lambda\,:\,\R^2\rightarrow\R$ in \eqref{eq:smooth-formulas}, the controller $\bk\,:\,\R^n\rightarrow\R^m$ in \eqref{eq:safety-filter-smooth} is smooth and satisfies:
    \begin{equation}
        L_{\bf}h(\bx) + L_{\bg}h(\bx)\bk(\bx) > - \alpha(h(\bx)),
    \end{equation}
    for all $\bx\in\R^n$ and therefore renders the set $\C\subset\R^n$ from \eqref{eq:C} forward invariant for the closed-loop system.
\end{theorem}

\section{Reduced-Order Models and Layered Control Architectures}\label{sec:ROM}

In this section, we begin our formal presentation of synthesizing CBFs via reduced-order models (ROMs). First, we motivate our eventual constructions by discussing the challenges associated with synthesizing CBFs for high-dimensional systems. We then introduce various classes of control systems that may be interpreted as \emph{layered} control architectures. These include, for example, robotic systems, where the dynamics of higher layers act as ROMs for the lower layer dynamics, the states of which are, in turn, viewed as control inputs to the aforementioned ROM. 

\subsection{Challenges in Constructing CBFs}
Our main focus in this tutorial is on high-dimensional nonlinear control systems whose dynamics may be viewed as a \emph{layered architecture} in which states of lower layers are viewed as control inputs for higher layers. This perspective is motivated by the fact that constructing CBFs for high-dimensional systems may be challenging -- such CBFs must generally take into account the behavior of the full-order dynamics to ensure safety. As demonstrated throughout this tutorial, these challenges can often be overcome by exploiting the layered structure present in many relevant systems to recursively construct a CBF for a complex system from a CBF for a much simpler one.

Many of the challenges associated with constructing CBFs are often related to the \emph{relative degree} of a function $h\,:\,\R^n\rightarrow\R$ defining a candidate safe set as in \eqref{eq:C}. 

\begin{definition}[Relative degree]
    A smooth function ${h\,:\,\R^n\rightarrow\R}$ is said to have \emph{relative degree} ${r\in\mathbb{N}}$ for \eqref{eq:control-affine} on a set $\mathcal{D}\subseteq\R^n$ if:
    \begin{enumerate}
        \item $L_{\bg}L_{\bf}^{r-i}h(\bx)=\bzero$ for all $\bx\in\mathcal{D}$ and $i\in\{2,\dots,r\}$;
        \item $L_{\bg}L_{\bf}^{r-1}h(\bx) \neq \bzero$ for some $\bx\in\mathcal{D}$,
    \end{enumerate}
    where higher-order Lie derivatives are defined as:
    \begin{equation*}
    \begin{aligned}
        L_{\bf}^0h(\bx) & \coloneqq h(\bx),\;
        && \quad L_{\bf}^ih(\bx) \coloneqq \pdv{L_{\bf}^{i-1}h}{\bx}\bf(\bx), \\
        L_{\bg}L_{\bf}h(\bx) & \coloneqq \pdv{L_{\bf}h}{\bx}\bg(\bx),\;
        && L_{\bg}L_{\bf}^ih(\bx) \coloneqq \pdv{L_{\bf}^ih}{\bx}\bg(\bx).
    \end{aligned}
    \end{equation*}
    If the second condition holds for all $\bx\in\mathcal{D}$, then $h$ is said to have \emph{uniform} relative degree $r\in\mathbb{N}$ for \eqref{eq:control-affine} on $\mathcal{D}$.
\end{definition}

When $h$ has uniform relative degree 1 for \eqref{eq:control-affine} on $\R^n$, i.e., if $L_{\bg}h(\bx)\neq\bzero$ for all $\bx\in\R^n$, then $h$ is a CBF for \eqref{eq:control-affine} (with $\bu\in\R^m$) since it is always possible to pick $\bu\in\R^m$ as large or small as necessary to satisfy \eqref{eq:cbf}. When $h$ has relative degree 1, but not uniform relative degree 1, $h$ is a CBF for \eqref{eq:control-affine} provided $L_{\bf}h(\bx)>-\alpha(h(\bx))$ whenever $L_{\bg}h(\bx)=\bzero$. When $h$ has relative degree larger than 1, then ${L_{\bg}h(\bx)=\bzero}$ for all $\bx\in\R^n$ and $h$ is unlikely to be a CBF for \eqref{eq:control-affine} unless the unforced dynamics of the system are already safe in the sense that $L_{\bf}h(\bx)>-\alpha(h(\bx))$ for all $\bx\in\R^n$. Thus, the ability to construct a CBF for a given system is tightly coupled to the construction of a relative degree one function whose zero superlevel set contains the set of states deemed to be safe. 

\begin{example}[Double integrator]\label{example:double-integrator}
    We illustrate many of the ideas presented in this tutorial using the simplest possible example of a higher-dimensional system -- the double integrator with state $\bx=(\bq,\bxi)\in\R^N$ and dynamics:
    \begin{equation}\label{eq:double-integrator}
        \underbrace{
        \begin{bmatrix}
            \dot{\bq} \\ \dot{\bxi}
        \end{bmatrix}}_{\dot{\bx}}
        =
        \underbrace{
        \begin{bmatrix}
            \bxi \\ \bzero
        \end{bmatrix}}_{\bf(\bx)}
        +
        \underbrace{
        \begin{bmatrix}
            \bzero \\ \bI
        \end{bmatrix}}_{\bg(\bx)}
        \bu.
    \end{equation}
    Here, $\bq\in\R^{n}$ represents the position/configuration of the system and $\bxi\in\R^{p}$ captures the velocity. Often, one desires to design a feedback controller for \eqref{eq:double-integrator} so that the resulting configuration trajectory $t\mapsto\bq(t)$ satisfies $\bq(t)\in\C_0$ for all $t\geq0$, where $\C_{0}\subset\R^{n}$ is a configuration constraint set that may, for example, capture the obstacle-free space in a collision avoidance problem. We assume this set may be characterized as the zero superlevel set of a continuously differentiable function $h_0\,:\,\R^{n}\rightarrow\R$ as:
    \begin{equation*}
        \C_0 = \{\bq\in\R^{n}\,:\,h_0(\bq)\geq0\}.
    \end{equation*}
    Given the objective of keeping the configuration in the above set, and the ability of CBFs to render such sets forward invariant, one may be tempted to simply take $h(\bx)=h_0(\bq)$ and $\C=\C_0\times\R^p$ as a CBF candidate and corresponding safe set for \eqref{eq:double-integrator}. Yet, this function may not serve as a CBF for \eqref{eq:double-integrator}, in general, since it has a relative degree larger than one:
    \begin{equation*}
        L_{\bg}h(\bx) = 
        \underbrace{
        \begin{bmatrix}
            \nabla h_0(\bq)\T & \bzero    
        \end{bmatrix}}_{\nabla h(\bx)\T}
        \underbrace{
        \begin{bmatrix}
            \bzero \\ \bI
        \end{bmatrix}}_{\bg(\bx)}
        =
        \bzero.
    \end{equation*}
    To remedy this, one must choose $h$ to additionally depend on $\bxi$, which could be done in a similar fashion to Example \ref{example:inverted-pendulum} so that $h$ has relative degree one and defines a set $\C$ such that rendering $\C$ forward invariant is sufficient to ensure satisfaction of the original configuration constraint in $\C_0$.
\end{example}

The previous example, although extremely simple, underscores one of the primary challenges\footnote{The other primary challenge is verifying \eqref{eq:cbf} when $\bu\in\mathcal{U}\subset\R^m$.} in constructing CBFs: a CBF, in general, must depend on all of the states of the system. For the double integrator in Example \ref{example:double-integrator}, it is often possible to construct a relative degree one function containing all of the system states to serve as CBF whose corresponding safe set contains the configuration constraint set of interest, as was done in Example \ref{example:inverted-pendulum} for the inverted pendulum. For more complex systems, however, capturing all of the states necessary to ensure safety in a single scalar function may be intractable. In the remainder of this tutorial, we outline various methodologies to systematically build CBFs for complex systems using ROMs -- lower dimensional representations of the original system that capture its high-level dynamics, but that are simple enough to construct CBFs for. Naturally, such methodologies require more structure than is present in the general control affine system \eqref{eq:control-affine} considered thus far. As hinted at earlier, these constructions are applicable to systems admitting a layered architecture in which the dynamics of higher layers act as ROMs for the lower-layer dynamics, the states of which are viewed as control inputs to the higher-layer dynamics. In the remainder of this section, we outline relevant classes of dynamics that satisfy such structural assumptions.

\subsection{Multi-layer Cascaded Dynamics}
The first layered control architecture we consider is the two-layer cascaded control system:
\begin{equation}\label{eq:two-layer-dyn}
    \begin{aligned}
        \dot{\bq} = & \bf_0(\bq) + \bg_0(\bq)\bxi \\
        \dot{\bxi} = & \bf_{1}(\bq,\bxi) + \bg_{1}(\bq,\bxi)\bu.
    \end{aligned}
\end{equation}
where $\bq\in\R^n$ represents the state of the top layer, $\bxi\in\R^p$ represents the states of the bottom layer, $\bu\in\R^m$ is the control input, and $\bf_0\,:\,\R^n\rightarrow\R^n$, $\bg_0\,:\,\R^n\rightarrow\R^{n\times p}$, $\bf_{1}\,:\,\R^n\times\R^p\rightarrow\R^p$, $\bg_{1}\,:\,\R^n\times\R^p\rightarrow\R^{p\times m}$ are locally Lipschitz mappings capturing the dynamics of the multi-layered system. For many physical systems of interest, $\bq$ may represent the system's position/configuration and $\bxi$ is the system's velocity, implying the top-layer dynamics:
\begin{equation}\label{eq:ROM}
    \dot{\bq} = \bf_0(\bq) + \bg_0(\bq)\bxi
\end{equation}
capture the kinematics of the system. Note that by defining $\bx\coloneqq (\bq,\bxi)\in\R^{n}\times\R^p=\R^{N}$, we may write \eqref{eq:two-layer-dyn} in standard control affine form:
\begin{equation}\label{eq:two-layer-dyn-control-affine}
    \underbrace{
    \begin{bmatrix}
        \dot{\bq} \\ \dot{\bxi}
    \end{bmatrix}}_{\dot{\bx}}
    =
    \underbrace{
    \begin{bmatrix}
         \bf_0(\bq) + \bg_0(\bq)\bxi \\  \bf_{1}(\bq,\bxi),
    \end{bmatrix}}_{\bf(\bx)}
    +
    \underbrace{
    \begin{bmatrix}
        \bzero \\ \bg_{1}(\bq,\bxi)
    \end{bmatrix}}_{\bg(\bx)}
    \bu,
\end{equation}
cf.~\eqref{eq:double-integrator}.
Here, we view \eqref{eq:ROM} as a ROM, with state $\bq$ and control input $\bxi$, for the multi-layered system \eqref{eq:two-layer-dyn} with the ultimate objective of building a CBF for the corresponding control affine system \eqref{eq:two-layer-dyn-control-affine} from a CBF for the ROM \eqref{eq:ROM}.

For ease of exposition, most of our discussion will focus on cascaded dynamics with two-layers as in \eqref{eq:two-layer-dyn}; however, the approaches we discuss are also applicable to more general multi-layer systems:
\begin{equation}\label{eq:multi-layer-r}
    \begin{aligned}
        \dot{\bq} = & \bf_0(\bq) + \bg_0(\bq)\bxi_1 \\ 
        \dot{\bxi}_1 = & \bf_1(\bq,\bxi_1) + \bg_1(\bq,\bxi_1)\bxi_2 \\ 
        \dot{\bxi}_2 = & \bf_2(\bq,\bxi_1,\bxi_2) + \bg_2(\bq,\bxi_1,\bxi_2)\bxi_3 \\
        \vdots & \\
        \dot{\bxi}_r = & \bf_{r}(\bq,\bxi_1,\bxi_2,\dots,\bxi_r) + \bg_r(\bq,\bxi_1,\bxi_2,\dots,\bxi_r)\bu,
    \end{aligned}
\end{equation}
with an arbitrary number of layers $r\in\mathbb{N}$. In traditional control-theoretic literature, such systems are said to be in \emph{strict feedback form} and can also be put into general control affine form \eqref{eq:control-affine} with state $\bx=(\bq,\bxi_1,\dots,\bxi_{r})$ as:
\begin{equation*}
    \!\underbrace{
    \begin{bmatrix}
        \dot{\bq} \\ \dot{\bxi}_1 \\ \vdots \\ \dot{\bxi}_r
    \end{bmatrix}}_{\dot{\bx}}\!
    =
    \!\underbrace{
    \begin{bmatrix}
        \bf_0(\bq) + \bg_0(\bq)\bxi_1 \\ \bf_1(\bq,\bxi_1) + \bg_1(\bq,\bxi_1)\bxi_2 \\ \vdots \\ \bf_{r}(\bq,\bxi_1,\bxi_2,\dots,\bxi_r)
    \end{bmatrix}}_{\bf(\bx)}\!
    +
    \!\underbrace{
    \begin{bmatrix}
        \bzero \\ \bzero \\ \vdots \\ \bg_r(\bq,\bxi_1,\bxi_2,\dots,\bxi_r)
    \end{bmatrix}}_{\bg(\bx)}\!
    \bu.
\end{equation*}

\subsection{Robotic Systems}
A particularly relevant class of systems whose dynamics exhibit a layered structure is mechanical systems, which can be used to model the majority of robotic systems. To introduce the dynamics of such systems, let $\bq\in\mathcal{Q}$ denote the generalized configuration of a mechanical system with $n$ degrees of freedom, where $\mathcal{Q}\subseteq\R^n$ is the configuration manifold. The dynamics of such systems are modeled using the Euler-Lagrange equations:
\begin{equation}\label{eq:robot-dyn}
    \bD(\bq)\ddot{\bq} + \bC(\bq,\dot{\bq})\dot{\bq} + \bG(\bq) = \bB\bu,
\end{equation}
where $\dot{\bq}\in T_{\bq}\mathcal{Q}$ is the generalized velocity, $\bD(\bq)\in\R^{n\times n}$ is the positive definite inertia matrix, $\bC(\bq,\dot{\bq})\in\R^{n\times n}$ is the Coriolis matrix, $\bG(\bq)\in\R^n$ represents gravitational and other potential effects, and $\bB\in\R^{n\times m}$ is the actuation matrix. By defining $\bx=(\bq,\dot{\bq})\in\TQ\subseteq\R^{2n}$, the above dynamics may be cast in control affine form \eqref{eq:control-affine} as:
\begin{equation}\label{eq:robot-dyn-control-affine}
    \underbrace{
    \begin{bmatrix}
        \dot{\bq} \\ \ddot{\bq}
    \end{bmatrix}}_{\dot{\bx}}
    = \underbrace{
    \begin{bmatrix}
        \dot{\bq} \\ -\bD(\bq)^{-1}\Big(\bC(\bq,\dot{\bq})\dot{\bq} +  \bG(\bq) \Big)
    \end{bmatrix}}_{\bf(\bx)}
    +
    \underbrace{
    \begin{bmatrix}
        \bzero \\ \bD(\bq)^{-1}\bB
    \end{bmatrix}}_{\bg(\bx)}
    \bu.
\end{equation}
When $m=n$ and $\bB$ is invertible, the robotic system \eqref{eq:robot-dyn} is said to be \emph{fully actuated}, otherwise it is said to be \emph{underactuated}. The dynamics in \eqref{eq:robot-dyn} are also a special case of the two-layer cascaded system in \eqref{eq:two-layer-dyn}, which can be recovered by defining:
\begin{equation*}
    \begin{aligned}
        \bf_0(\bq) = & \bzero, \quad
        \bf_1(\bq,\dot{\bq}) = -\bD(\bq)^{-1}\Big(\bC(\bq,\dot{\bq})\dot{\bq} +  \bG(\bq) \Big), \\
        \bg_0(\bq) = & \bI, \quad
        \bg_1(\bq,\dot{\bq}) = \bD(\bq)^{-1}\bB,
    \end{aligned}
\end{equation*}
which implies that the ROM for the full-order robotic system \eqref{eq:robot-dyn} takes the form of a single integrator:
\begin{equation}\label{eq:robotic-ROM}
    \dot{\bq} = \bxi,
\end{equation}
where the generalized velocity is viewed as a control input.

Although the structure of \eqref{eq:robot-dyn} dictates that its ROM is a single integrator, one may also employ more general ROMs. In particular, one may consider more general ROMs for \eqref{eq:robot-dyn} of the form:
\begin{equation}\label{eq:ROM-unicycle}
    \dot{\bq} = \bf_0(\bq) + \bg_0(\bq)\bxi,
\end{equation}
with control input $\bxi\in\R^p$, where $\bf_0\,:\,\R^n\rightarrow\R^n$ and $\bg_0\,:\,\R^n\rightarrow\R^{n\times p}$ capture the dynamics of the ROM. For example, \eqref{eq:ROM-unicycle} may be used to represent unicycle-like dynamics:
\begin{equation*}
    \underbrace{
    \begin{bmatrix}
        \dot{x} \\ \dot{y} \\ \dot{\theta}
    \end{bmatrix}}_{\dot{\bq}}
    =
    \underbrace{
    \begin{bmatrix}
        \cos(\theta) & 0 \\ \sin(\theta) & 0 \\ 0 & 1
    \end{bmatrix}}_{\bg_0(\bq)}
    \underbrace{
    \begin{bmatrix}
        v \\ \omega
    \end{bmatrix}}_{\bxi},
\end{equation*}
where $(x,y)\in\R^2$ denotes planar position, $\theta\in[0,2\pi)$ heading, $v\in\R$ forward velocity, and $\omega\in\R$ angular velocity. For ease of exposition, our presentation regarding robotic systems will focus on the single integrator ROM, and we will indicate how various results can be modified to account for more general ROMs, such as those described by \eqref{eq:ROM-unicycle}.

\section{Safe Backstepping}\label{sec:backstepping}
Backstepping is a recursive control design tool that has demonstrated success in constructing control Lyapunov functions (CLFs) \cite{freeman1992backstepping, Krstic} for nonlinear systems that possess a layered structure \eqref{eq:two-layer-dyn}. 
The main idea behind backstepping is to treat the states of lower layers as ``virtual" control inputs to the top layer, and then design a virtual controller for the top layer that would accomplish the given control objective.
However, as this controller is only ``virtual," in the sense that it cannot be directly applied to the top layer, one must ``backstep" through the dynamics to reach the actual control input. This backstepping process often requires differentiating through the virtual controllers designed at intermediate layers until the original input is reached. 
Once this input is reached, the control objective reduces to enforcing convergence of the bottom layer states to the aforementioned virtual controller, which, ultimately, leads to the satisfaction of the original control objective. 
As this procedure implies the existence of a controller satisfying the control objective for the overall system,
this enables
the construction of a certificate function, such as a CLF, that certifies the ability of the system to complete the given control objective. 
Thus, backstepping may be interpreted as a procedure to generate a certificate function for a potentially complex high-dimensional system from a certificate function for a much simpler lower-dimensional system.

In principle, there is nothing preventing one from applying a similar methodology to safety-critical control, rather than stabilization. Yet, backstepping has only recently been explored in the context of CBFs \cite{AndrewCDC22} despite the fact that CBFs, in their modern form, have existed for almost a decade \cite{AmesCDC14,AmesADHS15}. The reason, perhaps, for this delayed adoption of backstepping in the context of CBFs may be due to the emphasis in the CBF literature on optimization-based controllers, which are generally nonsmooth. Other reasons may be the development of viable alternatives, such as extended CBFs \cite{SreenathACC16,WeiCDC19,WeiTAC22}, that construct CBF-like functions for high-dimensional systems. In the remainder of this section, we demonstrate how recent results on smooth CBF-based controllers \cite{PioCDC19,Cohen-Smooth-Safety}, such as those outlined in Sec. \ref{sec:smooth}, provide a pathway towards the development of CBF backstepping and illustrate the advantages of such an approach over existing methods that construct CBFs for high-order systems.

\subsection{Backstepping with Control Barrier Functions}

Now we revisit backstepping in the context of safety-critical control with CBFs~\cite{AndrewCDC22}.
As a first step, we consider the top layer in \eqref{eq:ROM} as a ROM, where $\bxi$ -- the state of the bottom layer -- is viewed as a ``virtual" control input to the top layer. We wish to design this input to render:
\begin{equation}\label{eq:C0}
    \C_0 \coloneqq \{\bq\in\R^n\,:\,h_0(\bq)\geq0\},
\end{equation}
for some continuously differentiable $h_0\,:\,\R^n\rightarrow\R$, forward invariant for the top layer. To this end, we assume that $h_0$ is a CBF for this ROM in the sense that:
\begin{equation*}
    \sup_{\bxi\in\R^p}\big\{L_{\bf_0}h_0(\bq) + L_{\bg_0}h_0(\bq)\bxi \big\} > -\alpha(h_0(\bq)),
\end{equation*}
for all $\bq\in\R^n$ for some $\alpha\in\Kinf^e$. Provided $\bf_0$, $\bg_0$, $h_0$, and $\alpha$ are smooth, Theorem \ref{theorem:smooth-safety} then implies the existence of a smooth controller $\bk_0\,:\,\R^n\rightarrow\R^p$ satisfying:
\begin{equation}\label{eq:backstepping-strict-inequality}
    L_{\bf_0}h_0(\bq) + L_{\bg_0}h_0(\bq)\bk_0(\bq) > - \alpha(h_0(\bq)),
\end{equation}
for all $\bq\in\R^n$. This controller may be designed, for example, using the formulas in \eqref{eq:safety-filter-smooth} and \eqref{eq:smooth-formulas}. The interpretation of \eqref{eq:backstepping-strict-inequality} is that setting $\bxi=\bk_0(\bq)$ would ensure the forward invariance of $\C_0$ for the top-level dynamics if we could directly control $\bxi$.

As we cannot directly control $\bxi$, however, we must backstep through $\bk_0$ to determine the inputs $\bu$ that drive $\bxi$ to $\bk_0(\bq)$. Hence, the problem of constructing a CBF for the full-order system is reduced to that of tracking the output of the ROM. For the full-order dynamics in \eqref{eq:two-layer-dyn}, we leverage $\bk_0$ to propose the CBF candidate:
\begin{equation}\label{eq:h-backstepping}
    h(\bq,\bxi) \coloneqq h_0(\bq) - \frac{1}{2\mu}\|\bxi - \bk_0(\bq)\|^2,
\end{equation}
with parameter $\mu\in\R_{>0}$, which is used to define the safe set for the full-order system:
\begin{equation}\label{eq:C-backstepping}
    \C = \{(\bq,\bxi)\in\R^{n+p}\,:\, h(\bq,\bxi) \geq0\}.
\end{equation}
Importantly, note that
$(\bq,\bxi) \in \C$ implies $\bq \in \C_0$
since $h_0(\bq)\geq h(\bq,\bxi)$ for all $(\bq,\bxi)\in\R^{n+p}$.
Therefore, rendering $\C$ forward invariant for the full-order dynamics ensures that $\bq(t)\in\C_{0}$ for all $t\in I(\bq_0,\bxi_0)$.

To determine if the candidate CBF in \eqref{eq:h-backstepping} is indeed a CBF for the full-order dynamics in \eqref{eq:two-layer-dyn} with state $\bx=(\bq,\bxi)$, we recall from Lemma \ref{lemma:Lgh=0} that one need only to consider the system behavior when $L_{\bg}h(\bx)=\bzero$. To this end, we compute:
\begin{equation*}
    \nabla h(\bx) = 
    \begin{bmatrix}
        \nabla h_0(\bq) + \tfrac{1}{\mu}\pdv{\bk_0}{\bq}(\bq)\T(\bxi-\bk_0(\bq)) \\ 
        -\frac{1}{\mu}(\bxi-\bk_0(\bq)),
    \end{bmatrix}
\end{equation*}
and
\begin{equation*}
    \begin{aligned}
        L_{\bg}h(\bx) =  -\frac{1}{\mu}(\bxi-\bk_0(\bq))\T \bg_{1}(\bq,\bxi),
    \end{aligned}
\end{equation*}
noting that, if $\bg_{1}(\bq,\bxi)$ is pseudo-invertible for all $(\bq,\bxi)\in\R^{n+p}$, then:
\begin{equation*}
    L_{\bg}h(\bx)=\bzero \implies \bxi-\bk_0(\bq)=\bzero \implies h(\bx)= h_0(\bq).
\end{equation*}
Thus, when $L_{\bg}h(\bx)=\bzero$, we have:
\begin{equation*}
\begin{aligned}
    L_{\bf}h(\bx) = & L_{\bf_0}h_0(\bq) + L_{\bg_0}h_0(\bq)\bxi \\ 
    = & L_{\bf_0}h_0(\bq) + L_{\bg_0}h_0(\bq)\bk_0(\bq) \\ 
    > & -\alpha(h_0(\bq)) \\ 
    = & -\alpha(h(\bx)),
\end{aligned}
\end{equation*}
which implies that $h$ is a CBF for the full-order dynamics by Lemma \ref{lemma:Lgh=0}. This is formalized via the following theorem, which captures the main result with regard to CBF backstepping.

\begin{theorem}[\cite{AndrewCDC22}]\label{theorem:cbf-backstepping}
    Consider the two-layer dynamics in \eqref{eq:two-layer-dyn}, the constraint set $\C_0\subset\R^n$ in \eqref{eq:C0}, and suppose there exists a continuously differentiable controller $\bk_0\,:\,\R^n\rightarrow\R^p$ and $\alpha\in\Kinf^{\rm e}$ satisfying \eqref{eq:backstepping-strict-inequality}. If $\bg_{1}(\bq,\bxi)$ is pseudo-invertible for all $(\bq,\bxi)\in\R^{n+p}$, then $h\,:\,\R^n\times\R^p\rightarrow\R$ as defined in \eqref{eq:h-backstepping} is a CBF for the corresponding control affine system \eqref{eq:two-layer-dyn-control-affine} on the set $\C\subset\R^n\times\R^p$ as in \eqref{eq:C-backstepping}.
\end{theorem}

The preceding theorem facilitates the construction of CBFs for high-dimensional nonlinear systems that exhibit a layered structure as in \eqref{eq:two-layer-dyn}. Although these constructions have been presented for the special case of a two-layered system, similar to Lyapunov backstepping \cite{Krstic}, this approach may be recursively used to construct a CBF for a system with an arbitrary number $r\in\mathbb{N}$ of layers~\cite{AndrewCDC22} as defined in \eqref{eq:multi-layer-r}. The following examples illustrate the steps needed to construct a CBF using backstepping on the double integrator from Example \ref{example:double-integrator}.

\begin{example}[Double integrator]\label{example:double-int-1d}
    Consider a one-dimensional double integrator 
    with dynamics of the form \eqref{eq:two-layer-dyn},
    where ${\bq=x\in\R}$ represents the position and ${\bxi=v\in\R}$ represents velocity, while $\bx=(x,v)$ is the full-order state.
    Let the objective of designing a feedback controller be to keep the system's position $x$ in the interval $[-1,1]\subset\R$. This objective can be formalized by requiring the system's position to remain in the set:
    \begin{equation*}
        \C_0 = \{x\in\R\,:\,h_0(x) = 1- x^2 \geq 0\}.
    \end{equation*}
    Recall from Example \ref{example:double-integrator}, however, that this function may not serve as a CBF for the full-order system \eqref{eq:two-layer-dyn-control-affine} since, with $h(\bx)=h_0(x)$, we have $L_{\bg}h(\bx)=0$.
    
    To remedy this, we take a backstepping-based approach, where we view the top-layer dynamics:
    \begin{equation*}
        \dot{x} = \underbrace{0}_{f_0(x)} + \underbrace{1}_{g_0(x)}\times v
    \end{equation*}
    as a ROM with control input $v$. To check if $h_0$ is a CBF for the ROM, we compute:
    \begin{equation*}
        L_{g_0}h_0(x) = -2x,
    \end{equation*}
    so that when $L_{g_0}h_0(x)=0$, we have $x=0$ and:
    \begin{equation*}
        L_{f_0}h_0(x) + \alpha(h_0(x)) = \alpha(1 - x^2) = \alpha(1) > 0.
    \end{equation*}
    Hence, by Lemma \ref{lemma:Lgh=0}, $h_0$ is a CBF for the ROM for any ${\alpha\in\Kinf^{\rm e}}$, which for simplicity, we take as ${\alpha(s)=s}$. As $h_0$ is a CBF for the ROM, then, by Theorem \ref{theorem:smooth-safety}, there exists a smooth controller $k_0\,:\,\R\rightarrow\R$ satisfying \eqref{eq:backstepping-strict-inequality}.
    Furthermore, since $g_1(x,v)=1$ is invertible, the function:
    \begin{equation*}
        h(\bx)=h(x,v) = h_0(x) - \frac{1}{2\mu}\big(v - k_0(x)\big)^2,
    \end{equation*}
    is a CBF for the full-order dynamics on the set:
    \begin{equation}
        \C = \{(x,v)\in\R^2\,:\,h(x,v)\geq0\},
    \end{equation}
    by Theorem \ref{theorem:cbf-backstepping}.
    
    This safe set is illustrated for different values of $\mu$ in Fig.~\ref{fig:double-int-safe-set}, where the smooth controller $k_0$ is defined as in \eqref{eq:safety-filter-smooth} with $\lambda$ chosen as the Softplus universal formula \eqref{eq:smooth-formulas} with $\sigma=0.1$ and $k_{\rm d}(x)=0$. 
    Note that as $\mu$ is increased, the safe set $\C$ approaches the original constraint set $\C_0$ at the cost of including larger velocities, which may require compensation with larger control efforts. 

    \begin{figure}
    \centering
    \includegraphics{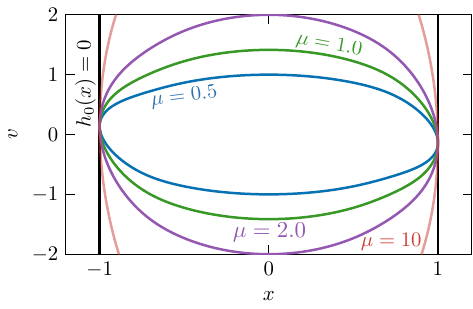}
    \caption{Safe set constructed for the one-dimensional double integrator via backstepping. Here, the colored curves represent the zero level set of $h$ as defined in \eqref{eq:h-backstepping} for various $\mu$, where $k_0$ is constructed using the Softplus universal formula from \eqref{eq:smooth-formulas} with $\sigma=0.1$. Note that as $\mu$ is increased the resulting safe set approaches the original constraint set $\C_0$ from \eqref{eq:C0}.}
    \label{fig:double-int-safe-set}
\end{figure}
\end{example}

\begin{figure}
    \centering
    \includegraphics{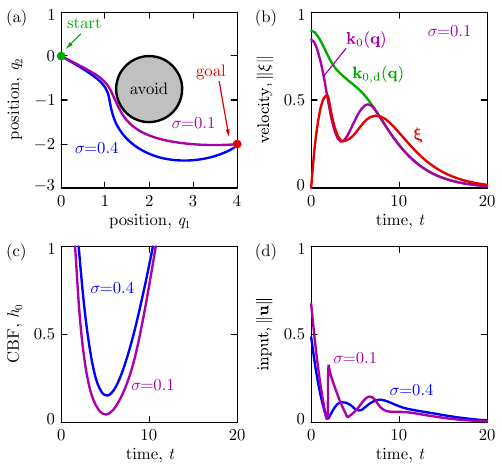}
    \caption{Results of the double integrator obstacle avoidance scenario from Example \ref{example:double-int-obstacle-avoidance}. (a) The trajectories of the double integrator's position, (b) its velocities, (c) the values of the safety constraint $h_0$ along the system's trajectory, and (d) the norm of the control input over time. This figure has been adapted from~\cite{AndrewCDC22}.}
    \label{fig:double_int_backstepping}
\end{figure}

\begin{example}[Obstacle avoidance \cite{AndrewCDC22}]\label{example:double-int-obstacle-avoidance}
    We now continue Example \ref{example:double-integrator} but present the details of constructing a CBF for an obstacle avoidance problem, which is used as an opportunity to illustrate the effect of the smooth safety filter on the corresponding CBF.
    This example was previously presented in the context of safe backstepping in~\cite{AndrewCDC22}.
    As demonstrated in Example~\ref{example:double-integrator}, any function that depends only on position
    is not a CBF for the double integrator. 
    Yet, by viewing a
    single integrator $\dot{\bq}=\bxi$ as a reduced-order representation of the full-order double integrator dynamics, we may still design a controller that uses a CBF constructed from the function characterizing the distance to the obstacle:
    \begin{equation*}
        h_0(\bq) = \frac{1}{2}\big(\|\bq - \bq_{\mathrm{o}}\|^2 - R_{\mathrm{o}}^2\big),
    \end{equation*}
    where $\bq_{\mathrm{o}}\in\R^2$ is the obstacle's center and $R_{\mathrm{o}}\in\R_{>0}$ is its radius, which is a valid CBF for the single integrator.
    
    To construct a CBF for the double integrator from its reduced-order single integrator model, we leverage the safe backstepping approach outlined in this section. First, we construct a smooth safety filter $\bk_0\,:\,\R^2\rightarrow\R^2$ for the single integrator via \eqref{eq:safety-filter-smooth}, where $\lambda$ is chosen as the Gaussian smooth universal formula \eqref{eq:smooth-formulas} and $\alpha(s)=s$, which filters out unsafe controls from the desired reduced-order controller $\bk_{0,\mathrm{d}}(\bq)\coloneqq K_{\mathrm{p}}(\bq_{\mathrm{g}} - \bq)$, where $\bq_{\mathrm{g}}\in\R^2$ is a goal location and $K_{\mathrm{p}}\in\R_{>0}$ is a gain. This smooth safety filter is then used to construct a CBF for the double integrator using \eqref{eq:h-backstepping} with $\mu=1$.
    Finally, the CBF is used to synthesize a QP-based safety filter $\bk\,:\,\R^{4}\rightarrow\R^2$ for the full-order system using \eqref{eq:safety-filter-qp}.
    
    The results of this procedure are displayed in Fig.~\ref{fig:double_int_backstepping} that is repeated from~\cite{AndrewCDC22}.
    Simulations are shown for various choices of $\sigma$ in the smooth universal formula \eqref{eq:smooth-formulas}. Note that as $\sigma$ approaches zero, the behavior of the smooth safety filter approaches that of a QP controller, where $\lambda$ depends on the $\relu$ activation function, leading to less smooth control signals.
\end{example}

\subsection{Comparison to Extended Control Barrier Functions}\label{sec:HOCBF}

Control barrier backstepping may be interpreted as a systematic methodology to construct a CBF for a high-dimensional system from a high relative degree safety constraint $h_0(\bq)\geq0$ that depends only on the states of the top layer, the end result of which is a \emph{relative degree one} CBF $h(\bq,\bxi_1,\dots,\bxi_r)$ for a higher dimensional control system. The construction of this CBF requires only a CBF for the top layer of \eqref{eq:multi-layer-r} and a few controllability assumptions, namely that each $\bg_i$ for $i\in\{1,\dots,r\}$ is pseudo-invertible. 

This approach is similar in spirit to other high-order CBF techniques that build a relative degree one CBF-like function from a high relative degree safety constraint $h_0(\bq)\geq0$ defining a set $\C_0\subset\R^n$ as in \eqref{eq:C0}, but have important technical differences as we discuss next. Such approaches are typically predicated on constructing an \emph{extended} CBF (also referred to as an \emph{exponential} \cite{SreenathACC16} or \emph{high order} \cite{WeiCDC19,WeiTAC22} CBF) by computing the derivative of $h_0$ along the system vector fields until the control input appears, reminiscent of classical input-output linearization. For example, when considering the two-layer cascaded system \eqref{eq:two-layer-dyn}, $h_0$ has relative degree two, thus one computes:
\begin{equation}
    h(\bx) = L_{\bf_0}h_0(\bq) + L_{\bg_0}h_0(\bq)\bxi + \alpha_0 h_0(\bq),
\label{eq:extended_CBF}
\end{equation}
where $\alpha_0\in\R_{>0}$ and $\bx=(\bq,\bxi)$, as an extended CBF candidate, which now has relative degree one and defines a set $\C\subset\R^{n}\times\R^{p}$ as its zero superlevel set.

Note, however, that unlike the backstepping-based approach, $\hat{\C}_0 = \C_0 \times \R^p$ is not a subset of $\C$ and one must instead consider the intersection $\hat{\C}_0\cap\C\subset\R^{n}\times\R^p$ as the candidate safe set of interest. To guarantee safety, this extended CBF must then satisfy:
\begin{equation}\label{eq:extended-CBF-inequality}
    \sup_{\bu\in\R^m}\big\{L_{\bf}h(\bx) + L_{\bg}h(\bx)\bu \big\} > - \alpha(h(\bx)),
\end{equation}
for all $\bx\in\hat{\C}_0\cap\C$ for some $\alpha\in\Kinf^e$, which can be used to develop feedback controllers enforcing forward invariance of $\hat{\C}_0\cap\C$. Similar to CBFs, the satisfaction of \eqref{eq:extended-CBF-inequality} can also be verified by checking that $L_{\bf}h(\bx) > -\alpha(h(\bx))$ whenever $L_{\bg}h(\bx)=\bzero$. Unfortunately, as illustrated in the following example \cite{TanTAC22,Cohen}, an extended CBF satisfying \eqref{eq:extended-CBF-inequality} may not exist even for relatively simple safety constraints. 

\begin{figure}
    \centering
    \includegraphics{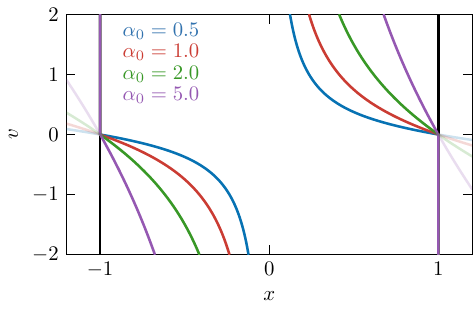}
    \caption{Safe set constructed for the one-dimensional double integrator using the extended CBF approach. Here, the colored curves represent the boundary of $\hat{\C}_0\cap\C$ for different choices of $\alpha_0$, the black lines denote the boundary of $\hat{\C}_0$, and the transparent curves of corresponding color denote the boundary of $\C$ for different choices of $\alpha_0$.}
    \label{fig:hocbf-example}
\end{figure}

\begin{example}[\cite{Cohen}]\label{example:hocbf-example}
    We now consider the same system and safety constraint $h_0$ and corresponding constraint set $\C_0$ as in Example \ref{example:double-int-1d}, but attempt to construct a safe set using an extended CBF rather than using backstepping. Since $h_0$ has relative degree larger than one based on Example \ref{example:double-integrator}, we calculate the extended CBF candidate in~\eqref{eq:extended_CBF}:
    \begin{equation*}
            h(\bx)
            = -2xv + \alpha_0 - \alpha_0x^2,
    \end{equation*}
    which defines a set $\C$ as its zero superlevel set, and a candidate safe set as $\hat{\C}_0\cap\C$ with ${\hat{\C}_0 = \C_0 \times \R}$. This candidate safe set for different choices of $\alpha_0$ is illustrated in Fig. \ref{fig:hocbf-example}. Similar to Example \ref{example:double-int-1d}, one may force $\hat{\C}_0\cap\C$ closer to $\hat{\C}_0$ by increasing $\alpha_0$.
    
    To check if $h$ satisfies the criteria in \eqref{eq:extended-CBF-inequality} for all ${\bx\in\hat{\C}_0\cap\C}$, we must ensure that $L_{\bf}h(\bx) + \alpha(h(\bx))>0$ whenever $L_{\bg}h(\bx)=0$. To this end, we compute:
    \begin{equation*}
        L_{\bg}h(\bx)=
        \underbrace{
        \begin{bmatrix}
            -2v - 2\alpha_0 x & -2x
        \end{bmatrix}}_{\nabla h(\bx)\T}
        \underbrace{
        \begin{bmatrix}
            0 \\ 1
        \end{bmatrix}}_{\bg(\bx)}
        = - 2x,
    \end{equation*}
    noting that $L_{\bg}h(\bx)=0$ implies $x=0$. Hence, when $L_{\bg}h(\bx)=0$, we also have:
    \begin{equation*}
        \begin{aligned}
            L_{\bf}h(\bx) \!+\! \alpha(h(\bx)) 
            = - 2v^2 \!+\! \alpha(\alpha_0),
        \end{aligned}
    \end{equation*}
    implying \eqref{eq:extended-CBF-inequality} only holds at points such that:
    \begin{equation*}
        v^2 < \frac{\alpha(\alpha_0)}{2}.
    \end{equation*}
    That is, when $x=0$, \eqref{eq:extended-CBF-inequality} only holds provided the magnitude of the velocity is bounded above by a function of $\alpha_0$ and $\alpha$. In practice, one may tune $\alpha_0$ and $\alpha$ so that \eqref{eq:extended-CBF-inequality} is only violated for arbitrarily large velocities, yet, such points will still be contained in $\hat{\C}_0\cap\C$ (see Fig. \ref{fig:hocbf-example}), implying \eqref{eq:extended-CBF-inequality} does not hold for all $\bx\in\hat{\C}_0\cap\C$ and, consequently, that $h$ is not an extended CBF.
\end{example}

The previous example demonstrates that one must take care when using the extended CBF methodology, as seemingly benign safety constraints may generate a function that cannot serve as an extended CBF no matter the choice of extended class $\Kinf$ functions. The consequence of this is that controllers synthesized from such invalid extended CBFs may not be well-defined even in the case when the control input is unconstrained. In contrast, the backstepping methodology outlined above produces, by construction, a relative degree one function that is guaranteed to be a CBF for the full-order system. The price to pay for this correct-by-construction approach is that it requires the full-order dynamics to exhibit a particular cascaded structure. In the following subsection, we extend this approach to a more general class of cascaded systems.

\subsection{Mixed Relative Degree Backstepping}
Another advantage of CBF backstepping over existing high order CBF approaches is the ability to handle layered systems with a \emph{mixed relative degree} -- that is, systems where inputs may enter not only at the bottom layer as in \eqref{eq:multi-layer-r}, but also at intermediate layers. Such mixed relative degree systems with two layers take the form: 
\begin{equation}\label{eq:two-layer-dyn-mixed}
    \begin{aligned}
        \dot{\bq} = & \bf_0(\bq) + \bg_{0}^{\bxi}(\bq)\bxi + \bg_{0}^{\bu}(\bq)\bu_0 \\
        \dot{\bxi} = & \bf_1(\bq,\bxi) + \bg_{1}^{\bu}(\bq,\bxi)\bu_1,
    \end{aligned}
\end{equation}
where $\bx=(\bq,\bxi)\in\R^n\times\R^p$ is the system state, $\bu=(\bu_0,\bu_1)\in\R^{m_0}\times\R^{m_1}$ is the control input, and $\bf_0\,:\,\R^n\rightarrow\R^n$, $\bg_0^{\bxi}\,:\,\R^n\rightarrow\R^{n\times p}$, $\bg_0^{\bu}\,:\,\R^n\rightarrow\R^{n\times m_0}$, $\bf_1\,:\,\R^n\times\R^p\rightarrow\R^p$, $\bg_1^{\bu}\,:\,\R^n\times\R^p\rightarrow\R^{p\times m_1}$ characterize the dynamics. Similar to the previous layered architecture, this system admits a control affine representation \eqref{eq:control-affine} as:
\begin{equation}\label{eq:two-layer-dyn-mixed-control-affine}
\underbrace{
    \begin{bmatrix}
        \dot{\bq} \\ \dot{\bxi}
    \end{bmatrix}}_{\dot{\bx}}
    =
    \underbrace{
    \begin{bmatrix}
        \bf_0(\bq) + \bg_{0}^{\bxi}(\bq)\bxi \\  \bf_1(\bq,\bxi)
    \end{bmatrix}}_{\bf({\bx})}
    +
    \underbrace{
    \begin{bmatrix}
        \bg_0^\bu(\bq) & \bzero \\ \bzero & \bg_1^\bu(\bq,\bxi)
    \end{bmatrix}}_{\bg(\bx)}
    \underbrace{
    \begin{bmatrix}
        \bu_0 \\ \bu_1
    \end{bmatrix}}_{\bu}.
\end{equation}

For this system, we consider a function $h_0\,:\,\R^n\rightarrow\R$ on the top layer states defining a constraint set $\C_0\subset\R^n$ as in \eqref{eq:C0}. The mixed relative degree characterization of \eqref{eq:two-layer-dyn-mixed} follows from the fact that the safety constraint $h_0$ may have different relative degrees with respect to different components of the control vector $\bu=(\bu_0,\bu_1)$. To address this challenge, we suppose the existence of smooth feedback controllers $\bk_{0}^{\bxi}\,:\,\R^n\rightarrow\R^p$, $\bk_{0}^{\bu}\,:\,\R^n\rightarrow\R^m$ and $\alpha\in\Kinf^{\rm e}$ satisfying:
\begin{equation}\label{eq:backstepping-strict-inequality-mixed}
        L_{\bf_0}h_0(\bq) \!+\! L_{\bg_{0}^{\bxi}}h_0(\bq)\bk_0^{\bxi}(\bq) \!+\! L_{\bg_{0}^{\bu}}h_0(\bq)\bk_0^{\bu}(\bq) > - \alpha(h_0(\bq)),
\end{equation}
for all $\bq\in\R^n$. With the above condition, we propose the CBF candidate~\cite{AndrewCDC22}:
\begin{equation}\label{eq:h-backstepping-mixed}
    h(\bq,\bxi) = h_0(\bq) - \frac{1}{2\mu}\|\bxi - \bk_{0}^{\bxi}(\bq)\|^2,
\end{equation}
which is used to define a candidate safe set $\C$ as in \eqref{eq:C-backstepping}. Once again, note that $(\bq,\bxi)\in\C$ implies $\bq\in\C_0$ since $h_0(\bq) \geq h(\bq,\bxi)$ for all $(\bq,\bxi)\in\R^{n}\times\R^{p}$. With these conditions, we may state the following result formalizing the construction of CBFs for mixed relative degree systems.

\begin{theorem}[\cite{AndrewCDC22}]\label{theorem:cbf-backstepping-mixed}
    Consider the dynamics in \eqref{eq:two-layer-dyn-mixed}, the set ${\C_0\subset\R^n}$ in \eqref{eq:C0}, and suppose there exist smooth feedback controllers ${\bk_{0}^{\bxi}\,:\,\R^n\rightarrow\R^p}$, ${\bk_{0}^{\bu}\,:\,\R^n\rightarrow\R^m}$ and $\alpha\in\Kinf^{\rm e}$ satisfying \eqref{eq:backstepping-strict-inequality-mixed}. If $g_1^{\bu}(\bq,\bxi)$ is pseudo-invertible for all $(\bq,\bxi)\in\R^{n+p}$, then $h\,:\,\R^n\times\R^p\rightarrow\R$ as defined in \eqref{eq:h-backstepping-mixed} is a CBF for the corresponding control affine system \eqref{eq:two-layer-dyn-mixed-control-affine} on the set $\C\subset\R^n\times\R^p$ as in \eqref{eq:C-backstepping}.
\end{theorem}
The proof of this theorem largely follows the same procedure as that of Theorem \ref{theorem:cbf-backstepping} and is provided in the Appendix for completeness. Similar to \eqref{eq:multi-layer-r}, one may recursively apply Theorem \ref{theorem:cbf-backstepping-mixed} to construct CBFs for mixed relative degree systems with an arbitrary number of layers: 
\begin{equation}\label{eq:multi-layer-mixed-r}
    \begin{aligned}
        \dot{\bq} = & \bf_0(\bq) + \bg_{0}^{\bxi}(\bq)\bxi_1 + \bg_{0}^{\bu}(\bq)\bu_0 \\
        \dot{\bxi}_1 = & \bf_1(\bq,\bxi_1) + \bg_{1}^{\bxi}(\bq,\bxi_1)\bxi_2 + \bg_{1}^{\bu}(\bq,\bxi_1)\bu_1 \\
        \vdots & \\
        \dot{\bxi}_r = & \bf_r(\bq,\bxi_1,\dots,\bxi_r) + \bg_{r}^{\bu}(\bq,\bxi_1,\dots,\bxi_r)\bu_r.
    \end{aligned}
\end{equation}

\begin{figure}
    \centering
    \includegraphics{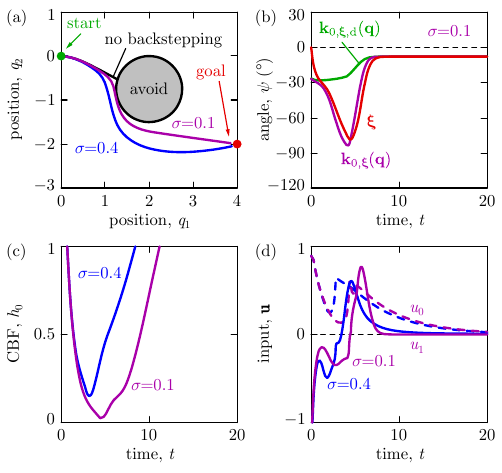}
    \caption{Simulation results for the unicycle from Example \ref{example:unicycle}. Each plot has a similar interpretation to those in Fig. \ref{fig:double_int_backstepping}. This figure has been adapted from~\cite{AndrewCDC22}.}
    \label{fig:unicycle_backstepping}
\end{figure}

\begin{example}[Unicycle \cite{AndrewCDC22}]\label{example:unicycle}
    A classic example of a mixed-relative degree system is the unicycle:
    \begin{equation*}
    \begin{aligned}
        \dot{x} = & v\cos(\psi) \\ 
        \dot{y} = & v\sin(\psi) \\ 
        \dot{\psi} = & \omega,
    \end{aligned}
    \end{equation*}
    where $(x,y)\in\R^2$ denote planar position, $\psi\in\R$ the heading angle, $v\in\R$ the linear velocity, and $\omega\in\R$ the angular velocity. Here, the state is $\bx\coloneqq(x,y,\psi)$ while the control input is $\bu\coloneqq(v,\omega)=(u_0,u_1)$. As written, the above dynamics are not in the form of \eqref{eq:two-layer-dyn-mixed}, but can be transformed into such a system with a few modifications. First, we define:
    \begin{equation*}
        \bq \coloneqq \begin{bmatrix}
            x \\ y
        \end{bmatrix}
        =
        \begin{bmatrix}
            q_1 \\ q_2
        \end{bmatrix},
        \quad \bxi \coloneqq \begin{bmatrix}
            \cos(\psi) \\ \sin(\psi)
        \end{bmatrix}
        =
        \begin{bmatrix}
            \xi_1 \\ \xi_2
        \end{bmatrix},
    \end{equation*}
    which implies that:
    \begin{equation*}
    \begin{aligned}
        \dot{\bq} & = \bxi u_0\eqqcolon\bv,\\
        \dot{\bxi} & = \begin{bmatrix}
            -\xi_2 \\ \xi_1
        \end{bmatrix}u_1.
    \end{aligned}
    \end{equation*}
    where $\bv$ denotes the planar velocity vector.
    Note that, as opposed to~\eqref{eq:two-layer-dyn-mixed}, the first equation is not affine w.r.t.~$(\bxi,u_0)$ but is affine in $\bv$.
    Thus we conduct backstepping by viewing the single integrator with input $\bv$ as a reduced-order model for the unicycle, and by converting $\bv$ to $(\bxi,u_0)$.
    
    Our control objective for this system is the same as that in Example \ref{example:double-int-obstacle-avoidance}: we wish to design a controller that enforces convergence of the position to a goal location while avoiding an obstacle. This obstacle avoidance task can be captured using the same safety constraint $h_0$ as in Example \ref{example:double-int-obstacle-avoidance}. We then synthesize a smooth safety filter $\bk_0\,:\,\R^2\rightarrow\R^2$ for the single integrator using the same approach as in Example \ref{example:double-int-obstacle-avoidance}, which outputs safe velocity commands $\bv = \bk_0(\bq)$. To use such commands in backstepping,
    we decompose $\bv = \bk_0(\bq)$ into $\bxi = \bk_0^{\bxi}(\bq)$ and $u_0 = k_0^{\bu}(\bq)$ as:
    \begin{equation*}
        \bk_0(\bq) = \underbrace{\frac{\bk_0(\bq)}{\|\bk_0(\bq)\|}}_{\bk_0^{\bxi}(\bq)} \underbrace{\|\bk_0(\bq)\|}_{k_0^{\bu}(\bq)},
    \end{equation*}
    which is valid so long as $\bk_0(\bq)\neq\bzero$.
    Then the desired value $\bk_0^{\bxi}(\bq)$ of $\bxi$ is used to construct a CBF for the full-order system as in \eqref{eq:h-backstepping-mixed}. This CBF is subsequently used to synthesize a safety filter $\bk\,:\,\R^3\rightarrow\R^2$ for the unicycle equipped with the desired controller:
    \begin{equation*}
        \bk_{\mathrm{d}}(\bx) = \begin{bmatrix}
            K_{\mathrm{p}}\|\bq - \bq_{\mathrm{g}}\| \\ -K_{\psi} \big( \sin(\psi) - \sin(\psi_0(\bq)) \big)
        \end{bmatrix},
    \end{equation*}
    where ${K_{\mathrm{p}},K_{\psi}\in\R_{>0}}$ are gains and ${\psi_0\,:\,\R^2\rightarrow\R}$, defined by
    ${\bk_0^{\bxi}(\bq) = \begin{bmatrix} \cos(\psi_0(\bq)) & \sin(\psi_0(\bq)) \end{bmatrix}\T}$, computes the desired heading angle.
    The results of applying such a controller $\bu=\bk(\bx)$ to the unicycle are provided in Fig. \ref{fig:unicycle_backstepping}, where all extended class $\Kinf$ functions involved are chosen as the identity function.
\end{example}

\section{Constructive Safety for Robotic Systems}\label{sec:robotic-cbfs}
We now turn our attention to a special case of the cascaded control systems considered in the previous section -- robotic systems with dynamics in \eqref{eq:robot-dyn}. These dynamics comply with the structure outlined in Sec. \ref{sec:backstepping}, implying the developed backstepping results may be applied to \eqref{eq:robot-dyn} by converting such systems into the form of \eqref{eq:two-layer-dyn} as detailed in Sec. \ref{sec:ROM}. However, given the relevance of CBFs in the context of robotics, and the fact that \eqref{eq:robot-dyn} possess certain structural properties that further facilitate the construction of CBFs, we outline in this section how the previous developments may be specialized to robotic systems. 

As in the previous section, we wish to design a feedback controller for the full-order system that keeps the system inside a subset of the configuration space:
\begin{equation}\label{eq:C0-robotic}
    \C_{0} \coloneqq \{ \bq\in\mathcal{Q}\,:\,h_0(\bq)\geq0\},
\end{equation}
where $h_0\,:\,\mathcal{Q}\rightarrow\R$ is a continuously differentiable configuration constraint. Although we wish to keep the configuration in $\C_0$, such an objective may not be possible without taking into account the full-order dynamics \eqref{eq:robot-dyn}. That is, similar to Example \ref{example:double-integrator}, $\C_0$ is unlikely to be a controlled invariant set for \eqref{eq:robot-dyn} since for $h(\bx)=h_0(\bq)$ we would have:
\begin{equation*}
    L_{\bg}h(\bx)=\begin{bmatrix}
        \nabla h_0(\bq)\T & \bzero
    \end{bmatrix}
    \begin{bmatrix}
        \bzero \\ \bD(\bq)^{-1}\bB
    \end{bmatrix}
    =
    \bzero,
\end{equation*}
for all ${\bx\in\TQ}$. In what follows, we outline various approaches to construct CBFs for the full-order dynamics \eqref{eq:robot-dyn} from the configuration constraint \eqref{eq:C0-robotic} under different assumptions regarding the system's actuation capability.

\subsection{Safe Backstepping for Robotic Systems}\label{sec:robotic-backstepping}
To remedy that $h_0$ is not a CBF, we first follow the backstepping-based approach outlined in the previous section, where we suppose the existence of a continuously differentiable controller $\bk_0\,:\,\mathcal{Q}\rightarrow\R^n$ satisfying:
\begin{equation}\label{eq:k0-robotic}
    \nabla h_0(\bq) \cdot \bk_0(\bq) > - \alpha(h_0(\bq)),
\end{equation}
for all $\bq\in\mathcal{Q}$. Similar to Sec. \ref{sec:backstepping}, we think of \eqref{eq:robotic-ROM}
as a reduced-order model for the full-order system \eqref{eq:robot-dyn} with input $\bxi\in\R^n$ and $\bk_0$ representing a controller we would apply to the reduced-order dynamics if we could simply set $\dot{\bq}=\bk_0(\bq)$. Thus, $\bk_0$ may be interpreted as a \emph{desired velocity} that we wish the full-order system to track. This controller is used to construct the \emph{energy-based CBF candidate}:
\begin{equation}\label{eq:ECBF}
    h(\bq,\dot{\bq}) = h_0(\bq) - \frac{1}{\mu}V(\bq,\dot{\bq}),
\end{equation}
where $\mu\in\R_{>0}$ and:
\begin{equation}\label{eq:V-ECBF}
    V(\bq,\dot{\bq}) \coloneqq \frac{1}{2}(\dot{\bq} - \bk_0(\bq))\T\bD(\bq)(\dot{\bq} - \bk_0(\bq)),
\end{equation}
whose form is inspired by that of the system's kinetic energy.
This energy-based CBF candidate defines:
\begin{equation}\label{eq:C-ECBF}
    \C \coloneqq \{(\bq,\dot{\bq})\in\TQ\,:\,h(\bq,\dot{\bq})\geq0\},
\end{equation}
as a candidate safe set, which ensures that $\bq\in\C_0$ whenever $(\bq,\dot{\bq})\in\C$ since $h_0(\bq)\geq h(\bq,\dot{\bq})$ for all $(\bq,\dot{\bq})\in\TQ$. Verifying this CBF candidate requires checking the behavior of $\dot{h}$ when $L_{\bg}h(\bx)=\bzero$, where $\bg\,:\,\TQ\rightarrow\R^{n\times m}$ is defined as in \eqref{eq:robot-dyn-control-affine} and $\bx=(\bq,\dot{\bq})$. To this end, we compute:
\begin{equation*}
    \pdv{h}{\dot{\bq}}(\bq,\dot{\bq}) = -\frac{1}{\mu}(\dot{\bq} - \bk_0(\bq))\T\bD(\bq),
\end{equation*}
noting that:
\begin{equation*}
    \begin{aligned}
        L_{\bg}h(\bx) = & 
        \underbrace{
        \begin{bmatrix}
            \pdv{h}{\bq}(\bq,\dot{\bq}) & \pdv{h}{\dot{\bq}}(\bq,\dot{\bq})
        \end{bmatrix}}_{\nabla h(\bx)\T}
        \underbrace{
        \begin{bmatrix}
            \bzero \\ \bD(\bq)^{-1}\bB
        \end{bmatrix}}_{\bg(\bx)} \\
        = & -\frac{1}{\mu}(\dot{\bq} - \bk_0(\bq))\T\bB.
    \end{aligned}
\end{equation*}
Thus, when \eqref{eq:robot-dyn} is fully actuated, we have:
\begin{equation*}
    L_{\bg}h(\bx) = \bzero \implies \dot{\bq} - \bk_0(\bq) = \bzero \implies h(\bq,\dot{\bq})= h_0(\bq),
\end{equation*}
so that, when $L_{\bg}h(\bx) = \bzero$, we have:
\begin{equation*}
    \begin{aligned}
        L_{\bf}h(\bx) = \nabla h_0(\bq)\cdot \dot{\bq} =  \nabla h_0(\bq)\cdot \bk_0(\bq) > &  -\alpha(h_0(\bq)) \\ = & -\alpha(h(\bq,\dot{\bq})),
    \end{aligned}
\end{equation*}
which implies that $h$ is a CBF for the corresponding control affine dynamics \eqref{eq:robot-dyn-control-affine} by Lemma \ref{lemma:Lgh=0}. The preceding discussion is formalized in the following lemma.

\begin{lemma}\label{lemma:cbf-robotic}
     Consider system \eqref{eq:robot-dyn}, a configuration constraint $h_0\,:\,\Q\rightarrow\R$ defining a set $\C_0\subset\Q$ as in \eqref{eq:C0-robotic}, and suppose there exists a continuously differentiable function $\bk_0\,:\,\Q\rightarrow\R$ satisfying \eqref{eq:k0-robotic}. If \eqref{eq:robot-dyn} is fully actuated, then $h\,:\,\TQ\rightarrow\R$ as in \eqref{eq:ECBF} is a CBF for the corresponding control affine system \eqref{eq:robot-dyn-control-affine} on $\C\subset\TQ$ as in \eqref{eq:C-ECBF}.
\end{lemma}

\begin{remark}
    The preceding result can also be applied to reduced-order models other than the single integrator in \eqref{eq:robotic-ROM}, such as the general control affine ROM in \eqref{eq:ROM-unicycle}. 
    To construct a CBF for \eqref{eq:robot-dyn} from this reduced-order model, however, one must modify \eqref{eq:k0-robotic} to:
    \begin{equation*}
        \nabla h_0(\bq)\cdot (\underbrace{\bf_0(\bq) + \bg_0(\bq)\bk_0(\bq)}_{\eqqcolon\bf_{0,\mathrm{cl}}(\bq)}) > - \alpha(h_0(\bq)),
    \end{equation*}
    and \eqref{eq:V-ECBF} to:
    \begin{equation*}
        V(\bq,\dot{\bq}) = \frac{1}{2}(\dot{\bq} - \bf_{0,\mathrm{cl}}(\bq))\T\bD(\bq)(\dot{\bq} - \bf_{0,\mathrm{cl}}(\bq)).
    \end{equation*}
\end{remark}

\begin{figure}
    \centering
    \includegraphics{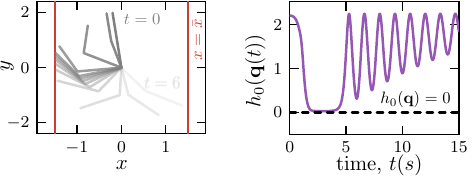}
    \caption{Simulation results corresponding to the double pendulum from Example \ref{example:double_pendulum}. The left plot illustrates the evolution of the pendulum in Cartesian space, where the red lines denote the boundary of the configuration constraint set, while the right plot illustrates the value of the configuration constraint along the system's trajectory.}
    \label{fig:double-pendulum}
\end{figure}

\begin{example}[Double pendulum]\label{example:double_pendulum}
    To illustrate the systematic construction of CBFs for robotic systems, we apply the results of this subsection to a fully actuated double pendulum with configuration $\bq=(\theta_1,\theta_2)$ denoting the angular position of the first $\theta_1$ and second $\theta_2$ link.
    Our objective is to design a feedback controller that keeps the $x$-component of Cartesian position $(x,y)$ of the pendulum's tip within a certain range $|x|\leq\bar{x}$. To this end, we first define $\bp\,:\,\Q\rightarrow\R^2$ associating to each configuration $\bq\in\Q$ the Cartesian position of the pendulum's tip as:
    \begin{equation*}
        \bp(\bq) = l_1\begin{bmatrix}
            \sin(\theta_1) \\ -\cos(\theta_1)
        \end{bmatrix}
        +l_2\begin{bmatrix}
            \sin(\theta_1 + \theta_2) \\ -\cos(\theta_1 + \theta_2)
        \end{bmatrix}=
        \begin{bmatrix}
            x\\ y
        \end{bmatrix}.
    \end{equation*}
    Denoting by $p_x(\bq)=x$, we propose:
    \begin{equation*}
        h_0(\bq) = \bar{x}^2 - p_{x}(\bq)^2,
    \end{equation*}
    as a configuration constraint defining the configuration constraint set $\C_0\subset\Q$ as in \eqref{eq:C0-robotic}, which we use as a CBF to define a smooth safety filter $\bk_0\,:\,\Q\rightarrow\R^2$ as in \eqref{eq:safety-filter-smooth} for the single integrator reduced-order model \eqref{eq:robotic-ROM} using the Softplus universal formula \eqref{eq:smooth-formulas} with $\sigma=0.1$ and $\alpha(s)=s$. 
    This system is fully actuated, hence:
    \begin{equation*}
        h(\bq,\dot{\bq}) = h_0(\bq) - \frac{1}{2\mu}(\dot{\bq} - \bk_0(\bq))\T\bD(\bq)(\dot{\bq} - \bk_0(\bq)),
    \end{equation*}
    is a CBF for the full-order dynamics \eqref{eq:robot-dyn-control-affine} by Lemma \ref{lemma:cbf-robotic}. This CBF is then used to construct a QP-based safety filter \eqref{eq:safety-filter-qp} for the corresponding control affine system \eqref{eq:robot-dyn-control-affine} and nominal controller $\bk_{\mathrm{d}}(\bq,\dot{\bq})=-\dot{\bq}$ that adds damping to the system. To demonstrate the effectiveness of this CBF, we simulate the system from an upright position with the objective of bringing the pendulum to a downward position while keeping the pendulum within the safe set, the results of which are provided in Fig. \ref{fig:double-pendulum}. Note that the pendulum initially falls towards the boundary of the safe set, stops itself before crossing the boundary, and then allows the tip of the pendulum to slide along the boundary of the safe set until reaching a downward position.
\end{example}

\subsection{Energy-based Control Barrier Functions}
At this point, one could directly use $h$ from \eqref{eq:ECBF} as a CBF for the control affine representation of the robot dynamics \eqref{eq:robot-dyn-control-affine}; however, such an approach presents certain limitations. In particular, such an approach requires computing the vector fields $\bf$ and $\bg$ in \eqref{eq:robot-dyn-control-affine}, requiring inversions of the inertia matrix $\bD$, which may be costly for high-dimensional robotic systems. In what follows, we demonstrate how one may directly leverage \eqref{eq:robot-dyn} without first converting such dynamics into control affine form to compute controllers enforcing safety. Such constructions are facilitated by the formal notion of an energy-based CBF.

\begin{definition}\label{def:ECBF}
    The continuously differentiable function $h\,:\,\TQ\rightarrow\R$ defined as in \eqref{eq:ECBF} that defines a set $\C\subset\TQ$ as in \eqref{eq:C-ECBF} is said to be an \emph{energy-based control barrier function} for \eqref{eq:robot-dyn} on $\C$ if there exists $\alpha\in\Kinf^e$ such that for all $(\bq,\dot{\bq})\in\TQ$
    \begin{equation*}
        \begin{aligned}
            \sup_{\bu\in\R^m}\bigg\{\frac{1}{\mu}(\dot{\bq} - \bk_0(\bq))\T\bigg[\bD(\bq)\pdv{\bk_0}{\bq}(\bq)\dot{\bq} + \bC(\bq,\dot{\bq})\bk_0(\bq)\\ + \bG(\bq) - \bB\bu \bigg] +\nabla h_0(\bq)\cdot\dot{\bq}\bigg\} > - \alpha(h(\bq,\dot{\bq})).
        \end{aligned}
    \end{equation*}
\end{definition}
By defining:
\begin{equation}\label{eq:a-and-b-robotic}
    \begin{aligned}
        a(\bq,\dot{\bq}) \coloneqq & \nabla h_0(\bq)\cdot\dot{\bq} + \frac{1}{\mu}(\dot{\bq} - \bk_0(\bq))\T \bigg[\bD(\bq)\pdv{\bk_0}{\bq}(\bq)\dot{\bq} \\ & + \bC(\bq,\dot{\bq})\bk_0(\bq) + \bG(\bq)\bigg] + \alpha(h(\bq,\dot{\bq})), \\ 
        b(\bq,\dot{\bq}) \coloneqq & \frac{1}{\mu^2}\|(\dot{\bq} - \bk_0(\bq))\T\bB\|^2,
    \end{aligned}
\end{equation}
the validity of an energy-based CBF candidate may be assessed using the same approach as for standard CBFs. Namely, $h$ is an energy-based CBF provided that:
\begin{equation*}
    b(\bq,\dot{\bq}) = 0 \implies a(\bq,\dot{\bq}) > 0.
\end{equation*}
When $\bk_0\,:\,\Q\rightarrow\R^n$ and $\alpha\in\Kinf^e$ satisfy \eqref{eq:k0-robotic}, and \eqref{eq:robot-dyn} is fully actuated, the above condition holds since:
\begin{equation*}
    b(\bq,\dot{\bq}) = 0 \implies (\dot{\bq} - \bk_0(\bq))\T\bB = \bzero \implies \dot{\bq} = \bk_0(\bq),
\end{equation*}
so that when $b(\bq,\dot{\bq})=0$, we have:
\begin{equation*}
    \begin{aligned}
        a(\bq,\dot{\bq}) = & \nabla h_0(\bq)\cdot \dot{\bq} + \alpha(h(\bq,\dot{\bq})) \\ 
        = & \nabla h_0(\bq)\cdot\bk_0(\bq) + \alpha(h_0(\bq)) > 0,
    \end{aligned}
\end{equation*}
where the second equality follows from $\dot{\bq}=\bk_0(\bq)$ and the inequality from \eqref{eq:k0-robotic}. With the above calculations, we have the following result regarding the construction of energy-based CBFs.

\begin{lemma}\label{lemma:ECBF}
    Let the assumptions of Lemma \ref{lemma:cbf-robotic} hold. Then, $h\,:\,\TQ\rightarrow\R$ as defined in \eqref{eq:ECBF} is an energy-based CBF for \eqref{eq:robot-dyn} on the set $\C\subset\TQ$ as defined in \eqref{eq:C-ECBF}.
\end{lemma}

Although the above result formalizes the construction of energy-based CBFs, we have yet to show that they may be used to synthesize controllers enforcing safety. The following theorem shows that this is indeed the case.

\begin{theorem}\label{theorem:ECBF}
    If $h\,:\,\TQ\rightarrow\R$ is an energy-based CBF for \eqref{eq:robot-dyn} on a set $\C\subset\TQ$ as in \eqref{eq:C}, the any locally Lipschitz controller $\bk\,:\,\TQ\rightarrow\R^m$ satisfying:
    \begin{equation}\label{eq:ECBF-inequality}
        \begin{aligned}
            \frac{1}{\mu}(\dot{\bq} - \bk_0(\bq))\T\bigg[\bD(\bq)\pdv{\bk_0}{\bq}(\bq)\dot{\bq} + \bC(\bq,\dot{\bq})\bk_0(\bq)\\ + \bG(\bq) - \bB\bk(\bq,\dot{\bq}) \bigg] + \nabla h_0(\bq)\cdot\dot{\bq} \geq - \alpha(h(\bq,\dot{\bq})),
        \end{aligned}
    \end{equation}
    for all $(\bq,\dot{\bq})\in\TQ$ renders $\C$ forward invariant for the closed-loop system \eqref{eq:robot-dyn} with $\bu=\bk(\bq,\dot{\bq})$.
\end{theorem}

The proof of this result, presented in the Appendix, exploits the following property of robotic systems in \eqref{eq:robot-dyn}.
\begin{property}\label{property:skew-symmetric}
    The inertia and Coriolis matrices in \eqref{eq:robot-dyn} satisfy the skew-symmetric property:
    \begin{equation}
        \bv\T(\dot{\bD}(\bq,\dot{\bq}) - 2\bC(\bq,\dot{\bq}))\bv =0,
    \end{equation}
    for all $(\bq,\dot{\bq})\in\TQ$ and any $\bv\in\R^n$.
\end{property}

Once an energy-based CBF has been constructed, a controller satisfying \eqref{eq:ECBF-inequality} may be synthesized by incorporating \eqref{eq:ECBF-inequality} as a constraint into an optimization problem to instantiate the safety filter:
\begin{equation}\label{eq:safety-filter-qp-robotic}
    \begin{aligned}
        \min_{\bu\in\R^m}\quad  &  \frac{1}{2}\|\bu - \bk_{\mathrm{d}}(\bq,\dot{\bq})\|^2 \\ 
        \text{s.t.} \quad &  \frac{1}{\mu}(\dot{\bq} - \bk_0(\bq))\T\bigg[\bD(\bq)\pdv{\bk_0}{\bq}(\bq)\dot{\bq} + \bC(\bq,\dot{\bq})\bk_0(\bq)\\ & + \bG(\bq) - \bB\bu \bigg] +\nabla h_0(\bq)\cdot\dot{\bq} \geq - \alpha(h(\bq,\dot{\bq}))
    \end{aligned}
\end{equation}
where $\bk_{\mathrm{d}}\,:\,\TQ\rightarrow\R^m$ is a desired control policy, whose closed-form solution is given similarly to \eqref{eq:safety-filter-qp-closed-form} by:
\begin{equation*}
    \bk(\bq,\dot{\bq})= \bk_{\mathrm{d}}(\bq,\dot{\bq}) - \frac{1}{\mu}\lambda\left(a(\bq,\dot{\bq}),b(\bq,\dot{\bq})\right)\bB\T(\dot{\bq} - \bk_0(\bq)),
\end{equation*}
where $a\,:\,\TQ\rightarrow\R$ and $b\,:\,\TQ\rightarrow\R$ are defined as in \eqref{eq:a-and-b-robotic}, and $\lambda\,:\,\R^2\rightarrow\R$ is defined with the $\relu$ activation function as in \eqref{eq:safety-filter-qp-closed-form}.
This controller no longer contains the inverse of the inertia matrix $\bD$.
Another advantage of directly leveraging the robot dynamics in \eqref{eq:robot-dyn} is that this approach enables the use of safety-enforcing controllers other than the QP-based controller in \eqref{eq:safety-filter-qp-robotic}. For example, when $\alpha\in\Kinf^e$ is Lipschitz continuous with Lipschitz constant $\ell\in\R_{>0}$ and \eqref{eq:robot-dyn} is fully actuated, one can verify that:
\begin{equation}
\begin{aligned}
    \bk(\bq,\dot{\bq}) = & \bB^{-1}\bigg[\bD(\bq)\pdv{\bk_0}{\bq}(\bq)\dot{\bq} + \bC(\bq,\dot{\bq})\bk_0(\bq) + \bG(\bq) \\ & + \mu\nabla h_0(\bq) - \frac{\gamma}{2}\bD(\bq)(\dot{\bq} - \bk_0(\bq)) \bigg],
\end{aligned}
\end{equation}
satisfies \eqref{eq:ECBF-inequality} for any $\gamma\geq\ell$.

\begin{remark}
    The energy-based CBFs outlined in this section are a generalization of those originally introduced in \cite{DrewLCSS22}. In particular, earlier notions of such CBFs are recovered by taking $\bk_0(\bq)=\bzero$ in \eqref{eq:ECBF} to obtain:
    \begin{equation}\label{eq:ECBF-original}
        h(\bq,\dot{\bq}) = h_0(\bq) - \frac{1}{2\mu}\dot{\bq}\T\bD(\bq)\dot{\bq}.
    \end{equation}
    A limitation of the above CBF candidate becomes evident when verifying if \eqref{eq:ECBF-original} is indeed a CBF via Lemma \ref{lemma:Lgh=0}. When \eqref{eq:robot-dyn} is fully actuated, we have:
    \begin{equation*}
        L_{\bg}h(\bx) = -\frac{1}{\mu}\dot{\bq}\T \bB=\bzero \implies \dot{\bq}=\bzero,
    \end{equation*}
    implying that when $L_{\bg}h(\bx)=\bzero$, we also have:
    \begin{equation*}
        \begin{aligned}
            L_{\bf}h(\bx) + \alpha(h(\bx)) = & \nabla h_0(\bq) \cdot \dot{\bq} + \alpha(h_0(\bq)) = \alpha(h_0(\bq)),
        \end{aligned}
    \end{equation*}
    which is only \emph{strictly} greater than zero on the interior of the safe set and is thus not a CBF on any set\footnote{Recall that although Def. \ref{def:cbf} requires \eqref{eq:cbf} to hold for all $\bx\in\R^n$, one may also require \eqref{eq:cbf} to only hold on a set $\mathcal{D}$ containing $\C$.} $\mathcal{D}\supseteq\C$. Although, in principle, one may relax the strict inequality in Def. \ref{def:cbf} to a nonstrict one so that \eqref{eq:ECBF-original} may serve as a CBF on $\C$, the lack of the strict satisfaction of \eqref{eq:cbf} may lead to controllers that are discontinuous when $\dot{\bq}=\bzero$. 
\end{remark}

\subsection{Underactuated Robotic Systems}\label{sec:underactuated}
The previous results in this section formalize the construction of CBFs for fully actuated robotic systems and illustrate that when the control input is unconstrained, it is always possible to construct a CBF for the full-order dynamics \eqref{eq:robot-dyn} by simply building a CBF for a reduced-order model. These results are not surprising given that fully actuated systems are feedback equivalent to double integrators -- a class of systems for which CBFs can be readily constructed as detailed in Sec. \ref{sec:backstepping}. The construction of CBFs becomes more challenging when \eqref{eq:robot-dyn} is underactuated; however, under certain assumptions, similar approaches to those outlined thus far may still be employed with the help of ideas introduced in \cite{SpongIROS94} (see also \cite[Ch. 3]{Tedrake}). To introduce these ideas, we rewrite \eqref{eq:robot-dyn} as:
\begin{equation}\label{eq:robot-dyn-alt}
    \bD(\bq)\ddot{\bq} + \bH(\bq,\dot{\bq}) = \bB\bu,
\end{equation}
where $\bD$ and $\bB$ are as in \eqref{eq:robot-dyn} and $\bH(\bq,\dot{\bq})\coloneqq \bC(\bq,\dot{\bq}) \dot{\bq} + \bG(\bq)$ collects the Coriolis and gravitational terms from \eqref{eq:robot-dyn}. We now suppose that \eqref{eq:robot-dyn-alt} is underactuated (i.e., $m<n$) and that the configuration can be partitioned into actuated $\bq_{1}\in\Q_1\subset\R^{n_1}$ and passive $\bq_2\in\Q_2\subset\R^{n_2}$ components in the sense that $\ddot{\bq}_{1}$ may be directly influenced by the control input while $\ddot{\bq}_{2}$ may only be indirectly influenced through the evolution of $\bq_{1}$. Under this assumption, we may represent the dynamics as:
\begin{equation}\label{eq:robot-dyn-partitioned}
    \underbrace{
    \begin{bmatrix}
        \bD_{11}(\bq) & \bD_{12}(\bq) \\ 
        \bD_{21}(\bq) & \bD_{22}(\bq)
    \end{bmatrix}}_{\bD(\bq)}
    \underbrace{
    \begin{bmatrix}
        \ddot{\bq}_{1} \\ \ddot{\bq}_{2}
    \end{bmatrix}}_{\ddot{\bq}}
    +
    \underbrace{
    \begin{bmatrix}
        \bH_1(\bq,\dot{\bq}) \\ \bH_{2}(\bq,\dot{\bq})
    \end{bmatrix}}_{\bH(\bq,\dot{\bq})}
    =
    \underbrace{
    \begin{bmatrix}
        \bB_{1} \\ \bzero
    \end{bmatrix}}_{\bB}
    \bu,
\end{equation}
where $\bD_{11}(\bq)\in\R^{n_{1}\times n_{1}}$ and $\bD_{22}(\bq)\in\R^{n_{2}\times n_{2}}$ are uniformly positive definite since $\bD$ is as well. We now suppose that our configuration constraint set $\C_0\subset\Q$ can be characterized as the zero superlevel set of a continuously differentiable function $h_0\,:\,\Q\rightarrow\R$ as in \eqref{eq:C0-robotic} that depends only on either the actuated or passive components of the configuration. For example, if our component of interest is $\bq_1$ -- the actuated component -- we assume that:
\begin{equation}\label{eq:C0-robotic-actuated}
    \C_0 = \{\bq\in\Q\,:\,h_{0,1}(\bq_1)\geq0\},
\end{equation}
whereas if our component of interest is $\bq_2$ -- the passive component -- we assume that:
\begin{equation}\label{eq:C0-robotic-passive}
    \C_0 = \{\bq\in\Q\,:\,h_{0,2}(\bq_2)\geq0\},
\end{equation}
where $h_{0,i}\,:\,\Q_i\rightarrow\R$, $i\in\{1,2\}$ is continuously differentiable.
Our objective is now to use the decomposition in \eqref{eq:robot-dyn-partitioned} to derive a new set of equations that depends only on the acceleration of one of the components of the configuration, depending on the configuration constraint.

We begin with the simpler situation in which our configuration constraint depends on the actuated components of the configuration. Our objective is to derive an equivalent representation of \eqref{eq:robot-dyn-alt} that depends only on $\ddot{\bq}_1$. To this end, we note that since $\bD_{22}(\bq)$ is invertible, we may use the second equation in \eqref{eq:robot-dyn-partitioned} to solve for $\ddot{\bq}_2$ as:
\begin{equation}
    \ddot{\bq}_2 = -\bD_{22}(\bq)^{-1}\left[\bD_{21}(\bq)\ddot{\bq}_1 + \bH_{2}(\bq,\dot{\bq})\right].
\end{equation}
This expression may now be substituted back into the first equation to obtain:
\begin{equation}\label{eq:robot-dyn-q1}
\begin{aligned}
    \bar{\bD}_{1}(\bq)\ddot{\bq}_1 + \bar{\bH}_{1}(\bq, \dot{\bq}) = \bB_{1}\bu,
\end{aligned}
\end{equation}
which depends only on $\ddot{\bq}_1$, where 
\begin{equation*}
    \begin{aligned}
        \bar{\bD}_{1}(\bq) \coloneqq & \bD_{11}(\bq) - \bD_{12}(\bq)\bD_{22}(\bq)^{-1}\bD_{21}(\bq), \\ 
        \bar{\bH}_{1}(\bq, \dot{\bq}) \coloneqq & \bH_{1}(\bq,\dot{\bq}) - \bD_{12}(\bq)\bD_{22}(\bq)^{-1}\bH_2(\bq,\dot{\bq}).
    \end{aligned}
\end{equation*}
Note that $\bar{\bD}_1$ is simply the Schur complement of $\bD$ and is symmetric and positive definite since $\bD$ is as well \cite{SpongIROS94}. Given the dynamics in \eqref{eq:robot-dyn-q1}, we propose the CBF candidate:
\begin{equation}
\begin{aligned}\label{eq:h1-robotic}
     h(\bq,\dot{\bq}) = & h_{0,1}(\bq_1) \\ &  - \frac{1}{2\mu}(\dot{\bq}_1 - \bk_{0,1}(\bq_1))\T\bar{\bD}_1(\bq)(\dot{\bq}_1 - \bk_{0,1}(\bq_1)),
\end{aligned}
\end{equation}
where $\mu\in\R_{>0}$ and $\bk_{0,1}\,:\,\Q_1\rightarrow\R^{n_1}$ is a continuously differentiable controller satisfying:
\begin{equation}\label{eq:k01-robotic}
    \nabla h_{0,1}(\bq_1)\cdot \bk_{0,1}(\bq_1) > - \alpha(h_{0,1}(\bq_1)),
\end{equation}
for all $\bq_1\in\Q_1$ for some $\alpha\in\Kinf^e$. This CBF candidate may be used to define a candidate safe set $\C\subset\TQ$ for the robotic system as in \eqref{eq:C-ECBF}. The following theorem illustrates that this function is a CBF for the control affine representation of this underactuated robotic system. 

\begin{theorem}\label{theorem:CBF-underactuated-1}
    Consider system \eqref{eq:robot-dyn-partitioned} and a configuration constraint set $\C_0\subset\Q$ as in \eqref{eq:C0-robotic-actuated}. Provided $\bB_1\in\R^{n_1\times m}$ is pseudo-invertible and $\bk_{0,1}\,:\,\Q_1\rightarrow\R^{n_1}$ satisfies \eqref{eq:k01-robotic}, then the function $h\,:\,\TQ\rightarrow\R$ as defined in \eqref{eq:h1-robotic} is a CBF for the corresponding control affine system \eqref{eq:robot-dyn-control-affine}.
\end{theorem}
A proof of this theorem is provided in the Appendix and follows a similar argument to the results of Sec. \ref{sec:robotic-backstepping}. Note that, under the assumption that $\bB_1$ is pseudo-invertible, system \eqref{eq:robot-dyn-q1} effectively acts as a fully actuated system since one may directly command any desired $\ddot{\bq}_1$ to achieve the control objective, and is reminiscent of the collocated feedback linearization method outlined in \cite{SpongIROS94}.

The fact that we may construct a CBF for the actuated subsystem in \eqref{eq:robot-dyn-partitioned} under similar assumptions to those in the previous section should not be too surprising. A more interesting situation, however, arises when our configuration constraint is a function of the passive components of the configuration as in \eqref{eq:C0-robotic-passive}. Under the following condition, a similar approach to that just introduced may be used to construct a CBF from a configuration constraint on the passive components of the configuration.

\begin{definition}[\cite{SpongIROS94}]
    System \eqref{eq:robot-dyn-partitioned} is said to \emph{strongly inertially coupled} on a set $\mathcal{D}\subset\Q$ if $\bD_{21}(\bq)$ is pseudo-invertible for all $\bq\in\mathcal{D}$.
\end{definition}
Provided the above condition is satisfied, we may rewrite the first equation in \eqref{eq:robot-dyn-partitioned} in terms of $\ddot{\bq}_2$ by first solving the second equation in \eqref{eq:robot-dyn-partitioned} for $\ddot{\bq}_1$ to obtain:
\begin{equation*}
    \ddot{\bq}_1 = -\bD_{21}(\bq)^{\dag}\big[\bD_{22}(\bq)\ddot{\bq}_2 + \bH_2(\bq,\dot{\bq}) \big],
\end{equation*}
where $\bD_{21}(\bq)^\dag$ denotes the pseudo-inverse of $\bD_{21}(\bq)$. The above expression can then be substituted into the first equation in \eqref{eq:robot-dyn-partitioned} to obtain:
\begin{equation}\label{eq:robot-dyn-q2}
    \bar{\bD}_2(\bq)\ddot{\bq}_{2} + \bar{\bH}_2(\bq,\dot{\bq}) = \bB_1\bu,
\end{equation}
where
\begin{equation*}
    \begin{aligned}
        \bar{\bD}_2(\bq) \coloneqq & \bD_{12}(\bq) - \bD_{11}(\bq)\bD_{21}(\bq)^\dag\bD_{22}(\bq) \\ 
        \bar{\bH}_2(\bq,\dot{\bq}) \coloneqq & \bH_{1}(\bq,\dot{\bq}) - \bD_{11}(\bq)\bD_{21}(\bq)^\dag\bH_{2}(\bq,\dot{\bq}),
    \end{aligned}
\end{equation*}
which now depends only on $\ddot{\bq}_2$, and is a valid representation of \eqref{eq:robot-dyn-partitioned} on the set where \eqref{eq:robot-dyn-partitioned} is strongly inertially coupled. As discussed in \cite{SpongIROS94}, $\bar{\bD}_2$ also has full rank on the set where the strong inertial coupling condition holds. Given the dynamics in \eqref{eq:robot-dyn-q2}, we propose the CBF candidate:
\begin{equation}
\begin{aligned}\label{eq:h2-robotic}
     h(\bq,\dot{\bq}) = & h_{0,2}(\bq_2) - \frac{1}{2\mu}\left\Vert\bar{\bD}_2(\bq)(\dot{\bq}_2 - \bk_{0,2}(\bq_2)) \right\Vert^2 \\ 
\end{aligned}
\end{equation}
where $\mu\in\R_{>0}$ and $\bk_{0,2}\,:\,\Q_2\rightarrow\R^{n_2}$ is a continuously differentiable controller satisfying:
\begin{equation}\label{eq:k02-robotic}
    \nabla h_{0,2}(\bq_2)\cdot \bk_{0,2}(\bq_2) > - \alpha(h_{0,2}(\bq_2)),
\end{equation}
for all $\bq_2\in\Q_2$ for some $\alpha\in\Kinf^e$. As in the previous case, this CBF candidate may be used to define a candidate safe set $\C\subset\TQ$ for the robotic system as in \eqref{eq:C-ECBF}. Now, under the additional assumption that \eqref{eq:robot-dyn-partitioned} is strongly inertially coupled on $\C_0$, Theorem \ref{theorem:CBF-underactuated-1} may be extended to construct a CBF from a configuration constraint that depends on the passive components of the configuration.

\begin{theorem}\label{theorem:CBF-underactuated-2}
    Consider system \eqref{eq:robot-dyn-partitioned} and a configuration constraint set $\C_0\subset\Q$ as in \eqref{eq:C0-robotic-passive}. Provided $\bB_1\in\R^{n_1\times m}$ is pseudo-invertible, $\bk_{0,2}\,:\,\Q_2\rightarrow\R^{n_2}$ satisfies \eqref{eq:k01-robotic}, and \eqref{eq:robot-dyn-partitioned} is strongly inertially coupled on $\C_0$, then the function $h\,:\,\TQ\rightarrow\R$ as defined in \eqref{eq:h2-robotic} is a CBF for the corresponding control affine system \eqref{eq:robot-dyn-control-affine}.
\end{theorem}
The above theorem, whose proof follows the same steps as those in the proof of Theorem \ref{theorem:CBF-underactuated-1}, is, effectively, an extension of the non-collocated feedback linearization method from \cite{SpongIROS94} to safety-critical control. The following example illustrates how one may apply these results to a classic underactuated robotic system.

\begin{figure}
    \centering
    \includegraphics{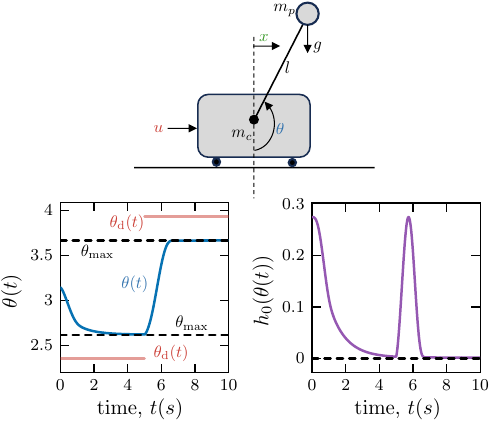}
    \caption{Results of the cartpole simulation from Example \ref{example:cartpole}. Here, the left plot displays the evolution of the pole's position and the right plot illustrates the evolution of the configuration constraint along the trajectory of the system, both of which demonstrate the resulting safe behavior.}
    \label{fig:cartpole}
\end{figure}

\begin{example}[Cartpole]\label{example:cartpole}
    We now demonstrate the design of CBFs for underactuated robotic systems using an example borrowed from \cite{DrewLCSS22}, which involves designing a safety-critical controller for the cartpole system as illustrated in Fig. \ref{fig:cartpole}. The configuration of this system is given by $\bq=(x,\theta)$, where $x\in\R$ is the position of the cart and $\theta\in[0,2\pi)$ the angular position of the pole, and the input corresponds to a force applied to the cart. The dynamics are of the form \eqref{eq:robot-dyn} with:
    \begin{equation*}
        \begin{aligned}
            & \bD(\bq) = \begin{bmatrix}
                m_{\rm c} + m_{\rm p} & m_{\rm p}l\cos(\theta) \\ m_{\rm p}l\cos(\theta) & m_{\rm p}l^2
            \end{bmatrix}, \quad
            \bB =
            \begin{bmatrix}
                1 \\ 0
            \end{bmatrix},  \\
            & \bC(\bq,\dot{\bq}) = \begin{bmatrix}
                0 & -m_{\rm p}l\dot{\theta}\sin(\theta) \\ 
                0 & 0
            \end{bmatrix}, \quad
            \bG(\bq) = \begin{bmatrix}
                0 \\ m_{\rm p}gl\sin(\theta)
            \end{bmatrix} 
        \end{aligned}
    \end{equation*}
    where $m_{\rm c}\in\R_{>0}$ denotes the mass of the cart, $m_{\rm p}\in\R_{>0}$ denotes the mass of the pole, $l\in\R_{>0}$ denotes the length of the pole, and $g\in\R_{>0}$ is the acceleration due to gravity. These dynamics may also be represented as in \eqref{eq:robot-dyn-partitioned} with $x$ and $\theta$ corresponding to the actuated and passive components of the configuration, respectively, implying one may directly influence $\ddot{x}$ via control inputs, whereas $\ddot{\theta}$ may only be indirectly influenced by actuating the cart. Our control objective is to constrain the angular position of the pole to lie within $\theta\in[\tfrac{5\pi}{6},\tfrac{7\pi}{6}]$, which may be expressed as the safety constraint:
    \begin{equation*}
        h_0(\theta) = \left(\frac{\pi}{6}\right)^2 - (\theta - \pi)^2,
    \end{equation*}
    where $\theta=\pi$ corresponds to the pole being upright, which defines a configuration constraint set $\C_0\subset\Q$ as in \eqref{eq:C0-robotic-passive}. As our safety constraint depends only on $\theta$, we attempt to rewrite the cartpole dynamics as in \eqref{eq:robot-dyn-q2}. To do so, we must ensure that the cartpole dynamics are strongly inertially coupled, at least on $\C_0$, which follows from the fact that $D_{21}(\bq) = m_{\rm p}l\cos(\theta)$ is only zero for $\theta=\pm\pi/2$  and is not contained in $\C_0$. Hence, we represent the cartpole dynamics as in \eqref{eq:robot-dyn-q2} for all $\bq\in\C_0$ with:
    \begin{equation*}
        \begin{aligned}
            \bar{D}_2(\bq) = & m_{\rm p}l\cos(\theta) - \frac{(m_{\rm c} + m_{\rm p})m_{\rm p}l^2}{m_{\rm p}l\cos(\theta)} \\
            \bar{H}_2(\bq,\dot{\bq}) = & -m_{\rm p}l\dot{\theta}^2\sin(\theta) - \frac{(m_{\rm c} + m_{\rm p})(m_{\rm p}gl\sin(\theta))}{m_{\rm p}l\cos(\theta)},
        \end{aligned}
    \end{equation*}
    which are valid so long as $\cos(\theta)\neq0$. With this representation of the dynamics, we form our CBF candidate as in \eqref{eq:h2-robotic}, where $k_{0,2}\,:\,[0,2\pi)\rightarrow\R$ is constructed using the Softplus universal formula from Sec. \ref{sec:smooth}. Since the dynamics are strongly inertially coupled on $\C_0$ and $B_1=1$ is invertible, the function $h$ from \eqref{eq:h2-robotic} is a CBF for the control-affine representation of this system \eqref{eq:robot-dyn-control-affine}. This CBF is used to construct a QP-based safety filter $\bk$ as in \eqref{eq:safety-filter-qp} for the nominal controller:
    \begin{equation*}
        k_{\mathrm{d}}(\bq, \dot{\bq}) = -K_{\theta}(\theta - \theta_{\mathrm{d}}(t)) - K_{\dot{\theta}}\dot{\theta},
    \end{equation*}
    where $K_{\theta},K_{\dot{\theta}}\in\R_{>0}$ are gains, which attempts to track a desired trajectory $\theta_{\mathrm{d}}\,:\,\R_{\geq0}\rightarrow\R$ for the pole's angular position. The results of applying this safety filter to the cartpole are provided in Fig. \ref{fig:cartpole}. Note that the desired pole position lies outside of $\C_0$ so that the performance objective is directly in conflict with the safety objective. Despite this, and the fact that one cannot directly actuate the angular position of the pole, safety is guaranteed through the careful construction of a CBF. 
\end{example}

\section{Stable Tracking of Safe Reduced Order Models}\label{sec:tracking}

In the previous sections, we outlined various methodologies to construct CBFs for high-dimensional systems with cascaded dynamics. Although these approaches enable the systematic construction of CBFs for relevant classes of systems, they are heavily model-dependent in the sense that one must leverage the full-order dynamics of the system to compute controllers enforcing safety. In practice, such models may be imperfect or may be computationally intensive to compute, limiting their use in controllers that must run in real time. Moreover, in many situations, one may not even have direct access to the control input for the full-order system, and may only be able to pass reference commands to black-box modules within the existing autonomy stack that compute such control inputs. 

In this section, we present a suite of techniques to address these aforementioned challenges. Such techniques are, in a certain sense, a generalization of the ideas introduced thus far and enable the application of these ideas to more complex systems, but also lead to a fundamentally different approach to safety-critical control. Our developments here are facilitated by the realization that the paradigm of safety-critical control based on ROMs can be understood as certifying the ability of the full-order system to track a suitably designed ROM. Earlier, we implicitly combined a CBF for a ROM with a Lyapunov-like function
to produce a CBF for the overall system. In this section, we make such an idea more explicit. 

The benefit of making this unification of barrier and Lyapunov functions explicit lies in the ability to decouple the design of the safety-critical control architecture from the full-order model. This decoupling leads to a notion of \emph{model-free} safety-critical control in the sense that the safety-critical component of the control architecture may be designed and implemented independent of the full-order dynamics. Safety of the full-order dynamics can then be guaranteed so long as such dynamics track commands generated by the ROM. The synthesis of such tracking controllers may require knowledge of the full-order dynamics; however, tracking controllers for many relevant classes of systems, such as those in robotics, are well established and may be readily applied within this model-free safety-critical control paradigm to enforce safety.

\subsection{Lyapunov-certified Tracking}
To illustrate the ideas introduced earlier in a more general context, consider again the two-layered system from \eqref{eq:two-layer-dyn}, 
which may also be written in standard control affine form \eqref{eq:control-affine} with state $\bx=(\bq,\bxi)$ as noted in \eqref{eq:two-layer-dyn-control-affine}. As we did earlier, we consider the top-level dynamics:
\begin{equation*}
    \dot{\bq} = \bf_0(\bq) + \bg_0(\bq)\bxi,
\end{equation*}
as a reduced-order representation of the full-order system for which we wish to design a smooth controller $\bk_0\,:\,\R^n\rightarrow\R^p$ that would enforce safety of the ROM if its dynamics were directly controllable. Rather than leveraging $\bk_0$ to backstep through these dynamics to compute a safe controller, here we consider the existence of a \emph{tracking controller} $\bk\,:\,\R^n\times\R^p\rightarrow\R^m$ that is able drive the state $\bxi$ to $\bk_0(\bq)$.
Accordingly, we assume that there exists a Lyapunov function $V\,:\,\R^n\times\R^p\rightarrow\R_{\geq0}$ for the full-order dynamics:
\begin{equation*}
    \begin{aligned}
        \dot{\bq} = & \bf_0(\bq) + \bg_0(\bq)\bxi \\
        \dot{\bxi} = & \bf_{1}(\bq,\bxi) + \bg_{1}(\bq,\bxi)\bk(\bq,\bxi),
    \end{aligned}
\end{equation*}
satisfying:
\begin{subequations}\label{eq:tracking-Lyap}
    \begin{equation}\label{eq:tracking-Lyap-1}
        \gamma_1\|\bxi - \bk_0(\bq)\|^2 \leq V(\bq,\bxi) \leq \gamma_2\|\bxi - \bk_0(\bq)\|^2
    \end{equation}
    \begin{equation}\label{eq:tracking-Lyap-2}
        \dot{V}(\bq,\bxi) = L_{\bf}V(\bx) + L_{\bg}V(\bx)\bk(\bx)  \leq - \gamma V(\bq,\bxi),
    \end{equation}
\end{subequations}
for positive constants $\gamma_1,\gamma_2,\gamma>0$. This Lyapunov function certifies the ability of the full-order dynamics to track commands generated by the reduced dynamics, represented as the outputs of the reduced-order controller $\bk_0\,:\,\R^n\rightarrow\R^p$.

To see how this tracking controller and corresponding Lyapunov function may be used to establish safety of the overall system, we write the top layer dynamics from \eqref{eq:two-layer-dyn} as:
\begin{equation}\label{eq:top-layer-dyn-dist}
    \begin{aligned}
        \dot{\bq} = \bf_0(\bq) + \bg_0(\bq)(\bk_0(\bq) + \bd),
    \end{aligned}
\end{equation}
where:
\begin{equation}
    \bd \coloneqq \bxi - \bk_0(\bq),
\end{equation}
is the tracking error for the full order system, which is treated as a disturbance that must be rejected by the top layer to ensure safety. To account for this disturbance, we now require $\bk_0$ to satisfy:
\begin{equation}\label{eq:top-layer-issf}
    L_{\bf_0}h_0(\bq) + L_{\bg_0}h_0(\bq)\bk_0(\bq) \geq - \alpha h_0(\bq) + \frac{1}{\varepsilon}\|L_{\bg_0}h_0(\bq)\|^2,
\end{equation}
where $h_0\,:\,\R^n\rightarrow\R$ defines the set $\C_0\subset\R^n$ as in \eqref{eq:C0} and $\alpha,\varepsilon>0$. That is, rather than requiring $h_0$ to be a CBF for the top layer dynamics, we now require $h_0$ to be an \emph{ISSf-CBF} (see Sec. \ref{sec:issf}) for the top layer. 
Following a similar procedure as before, we now define:
\begin{equation}\label{eq:h-Lyap}
    h(\bq,\bxi)  = h_0(\bq) - \frac{1}{\mu \gamma_1} V(\bq,\bxi),
\end{equation}
as a candidate barrier function for the closed-loop system, which defines the candidate safe set:
\begin{equation}\label{eq:C-Lyap-tracking}
    \C = \{(\bq,\bxi)\in\R^{n}\times\R^p\,:\,h(\bq,\bxi) \geq 0\},
\end{equation}
as its zero superlevel set. As $V$ is positive definite, we have $h(\bq,\bxi)\geq0\implies h_0(\bq)\geq0$ so that enforcing forward invariance of $\C$ in \eqref{eq:C-Lyap-tracking} is sufficient to ensure that $h_0(\bq(t))\geq0$. The following theorem provides conditions under which $h$ is a barrier function for the closed-loop system.

\begin{theorem}\label{theorem:cbf-tracking}
    Consider the dynamics in \eqref{eq:two-layer-dyn}, the constraint set $\C_0\subset\R^n$ in \eqref{eq:C0}, and suppose there exists a continuously differentiable controller $\bk_0\,:\,\R^n\rightarrow\R^p$ and positive constants $\alpha,\varepsilon>0$ satisfying \eqref{eq:top-layer-issf}. Furthermore, suppose there exists a tracking controller $\bk\,:\,\R^n\times\R^p\rightarrow\R^m$ and Lyapunov function $V\,:\,\R^n\times\R^p\rightarrow\R_{\geq0}$ satisfying \eqref{eq:tracking-Lyap} for positive constants $\gamma_1,\gamma_2,\gamma>0$. Provided:
    \begin{equation}\label{eq:Lyap-safety-conditions}
        \gamma \geq  \alpha + \frac{\varepsilon\mu}{4},
    \end{equation}
    then $\C\subset\R^n\times\R^p$ as defined in \eqref{eq:C-Lyap-tracking} is forward invariant for the closed-loop control affine system \eqref{eq:two-layer-dyn-control-affine} with $\bu=\bk(\bq,\bxi)$.
\end{theorem}

The previous theorem, whose proof is provided in the Appendix, states that, with good enough tracking performance, safety may be enforced on the full-order dynamics by simply tracking the outputs of a safe ROM. The condition in \eqref{eq:Lyap-safety-conditions} requires that the rate of convergence of the tracking error -- captured via $\gamma$ -- must be larger than the rate at which the ROM may approach the boundary of the constraint set -- captured via $\alpha$. For a fixed tracking controller, one may satisfy \eqref{eq:Lyap-safety-conditions} by designing an appropriate ROM by decreasing $\alpha$, which limits how quickly the ROM may approach the boundary of the constraint set, and decreasing $\varepsilon$, which corresponds to robustifying the ROM to larger tracking errors. Hence, for a fixed tracking controller satisfying \eqref{eq:tracking-Lyap}, one may always ensure safety at the cost of using a more conservative ROM. 

As argued earlier, the benefit of the preceding result is that the safety-critical portion of the control architecture only relies on the reduced-order dynamics.
As opposed, the results from earlier sections established the existence of CBF for the full-order system, the dynamics of which one must ultimately leverage to synthesize a controller enforcing safety. Here, one may instead leverage an existing tracking controller that may already be integrated into the system's autonomy stack to track commands produced by the reduced-order controller and guarantee safety.

These safety guarantees, of course, are conditioned on the ability of such a tracking controller to perfectly track reference commands. In practice, however, perfect tracking -- the satisfaction of \eqref{eq:tracking-Lyap-2} -- is often not achievable and instead, our tracking controller $\bk$ may only achieve:
\begin{equation}\label{eq:tracking-Lyap-ISS}
    \dot{V}(\bq,\bxi) \leq - \gamma V(\bq,\bxi) + \delta,
\end{equation}
for positive constants $\gamma,\delta>0$. That is, the tracking controller enforces \emph{input-to-state-stability} (ISS) of the tracking error dynamics rather than exponential stability as in \eqref{eq:tracking-Lyap-2}. The inability of the full-order dynamics to perfectly track the reduced-order model leads us to consider the modified barrier candidate:
\begin{equation}\label{eq:h-Lyap-ISS}
    h(\bq,\bxi) = h_0(\bq) -\frac{1}{\mu \gamma_1}\Big(V(\bq,\bxi) - \frac{\delta}{\alpha} \Big),
\end{equation}
which defines a candidate safe set $\C$ as in \eqref{eq:C-Lyap-tracking}. Compared to \eqref{eq:h-Lyap}, the above barrier candidate inflates the original safe set proportional to $\delta$ to account for imperfect tracking.
The following result illustrates that under similar conditions to those in Theorem \ref{theorem:cbf-tracking}, this tracking controller enforces ISSf of the overall system with respect to the ISSf barrier function \eqref{eq:h-Lyap-ISS}.

\begin{theorem}\label{theorem:cbf-tracking-iss}
    Consider the dynamics in \eqref{eq:two-layer-dyn}, the constraint set $\C_0\subset\R^n$ in \eqref{eq:C0}, and suppose there exists a continuously differentiable controller $\bk_0\,:\,\R^n\rightarrow\R^p$ and positive constants $\alpha,\varepsilon>0$ satisfying \eqref{eq:top-layer-issf}. Furthermore, suppose there exists a tracking controller $\bk\,:\,\R^n\times\R^p\rightarrow\R^m$ and Lyapunov function $V\,:\,\R^n\times\R^p\rightarrow\R_{\geq0}$ satisfying \eqref{eq:tracking-Lyap-1} and \eqref{eq:tracking-Lyap-ISS} for positive constants $\gamma_1,\gamma_2,\gamma,\delta>0$. Provided \eqref{eq:Lyap-safety-conditions} holds, then
    then $\C\subset\R^n\times\R^p$ as defined in \eqref{eq:C-Lyap-tracking}, with $h\,:\,\R^n\times\R^p\rightarrow\R$ from \eqref{eq:h-Lyap-ISS}, is forward invariant for the closed-loop control affine system \eqref{eq:two-layer-dyn-control-affine} with $\bu=\bk(\bq,\bxi)$.
\end{theorem}

The proof of this result follows the same steps as those employed in the proof of Theorem \ref{theorem:cbf-tracking}. As this result establishes forward invariance of an inflated safe set, rather than the original safe set defined by \eqref{eq:h-Lyap}, it effectively establishes ISSf of the full-order dynamics. Note that for both Theorems \ref{theorem:cbf-tracking} and \ref{theorem:cbf-tracking-iss} the parameters of the ROM and tracking controller must satisfy the same condition \eqref{eq:Lyap-safety-conditions}; however, the safe sets for each of these results -- characterized as the zero superlevel sets of \eqref{eq:h-Lyap} and \eqref{eq:h-Lyap-ISS}, respectively -- are different. Compared to \eqref{eq:h-Lyap}, the safe set defined by \eqref{eq:h-Lyap-ISS} is inflated by an additional margin proportional to $\delta/\alpha$. One can bring the resulting inflated safe set closer to the original safe set by increasing $\alpha$, resulting in a more ``aggressive" ROM; however, to guarantee ISSf, the increase in $\alpha$ must be compensated for with larger $\gamma$, which requires the tracking controller to enforce faster convergence of the system to commanded references.
Furthermore, by increasing robustness through decreasing $\varepsilon$, one may take larger values of $\mu$ in~\eqref{eq:Lyap-safety-conditions}, making the corresponding forward invariant set given by~\eqref{eq:h-Lyap-ISS} closer to the original constraint set given by $h_0$.
Before proceeding, we illustrate how one may apply these results with the help of the following example.

\begin{figure}[t]
    \centering
    \includegraphics{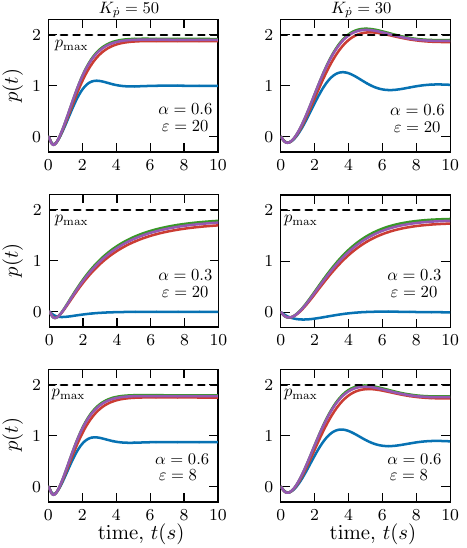}
    \caption{Model-free safety-critical control of the planar Segway from Example \ref{example:segway_smooth}. The plots display the evolution of the Segway's position generated by the controller in \eqref{eq:segway_controller_2D} with $K_{\dot{p}}=50$ (left) and  $K_{\dot{p}}=30$ (right) for different choices of $\alpha$ and $\varepsilon$}. The curves of different colors represent the trajectories under different smooth safety filters for the ROM, where the colors have the same interpretation as in Fig. \ref{fig:smooth-formulas}.
    \label{fig:segway-smooth}
\end{figure}

\begin{example}[Planar Segway]\label{example:segway_smooth}
    We demonstrate the model-free safety-critical control paradigm by using the example of a Segway control problem from~\cite{TamasRAL22}.
    Consider the planar Segway model in Fig.~\ref{fig:segway_2D}(a) with configuration
    ${\bq = (p,\,\varphi) \in \mathcal{Q} = \mathbb{R} \times [0,2\pi)}$ including the position $p$ and pitch angle $\varphi$ of the Segway.
    We seek to drive the Segway with a desired speed $\dot{p}_{\rm d}$ until reaching a wall at position $p_{\rm max}$ where the Segway must stop automatically such that ${p \leq p_{\rm max}}$.
    The dynamics of the Segway are given by~\eqref{eq:robot-dyn} with
    ${u \in \R}$ being the voltage on the Segway's motors and:
    \begin{align*}
    \bD(\bq) & \!=\!
    \begin{bmatrix}
    m_{0} & m L \cos \varphi \\
    m L \cos \varphi & J_{0}
    \end{bmatrix}\!, \;\;
    \bG(\bq) \!=\!
    \begin{bmatrix}
    0 \\
    - m g L \sin \varphi
    \end{bmatrix}\!, \\
    \bC(\bq,\dot{\bq}) & \!=\!
    \begin{bmatrix}
    b_{\rm t}/R & - b_{\rm t} - m L \dot{\varphi} \sin \varphi \\
    - b_{\rm t} & b_{\rm t} R
    \end{bmatrix}\!, \;\;
    \bB \!=\!
    \begin{bmatrix}
    K_{\rm m}/R \\ -K_{\rm m}
    \end{bmatrix}\!,
    \end{align*}
    where $R$ and $L$ are geometric dimensions, $m$, $m_0$, $J_0$ are mass and inertia parameters, $g$ is acceleration from gravity, while $b_{\rm t}$ and $K_{\rm m}$ are motor parameters, all given in~\cite{TamasRAL22}. Note that although these dynamics are in the form of \eqref{eq:robot-dyn}, they are underactuated, which complicates the backstepping-like methods developed in previous sections.

    To address this challenge, we proceed to leverage the model-free safety-critical control approach developed in this section, where
    we use the single integrator ${\dot{\bq} = \bxi}$ as a ROM to provide safety against collision with the wall, with desired controller ${\bk_{0,{\rm d}}(\bq) = \begin{bmatrix} \dot{p}_{\rm d} & 0 \end{bmatrix}\T}$ and CBF:
    \begin{equation*}
    h_0(\bq) = p_{\rm max} - p,
    \end{equation*}
    that satisfies ${L_{\bg_0}h_0(\bq) \neq \bzero}$. This CBF is then used to construct a smooth safety filter $\bk_0\,:\,\Q\rightarrow\R^2$ as in Sec. \ref{sec:smooth} for the ROM. The output of this smooth safety filter represents a safe velocity for the Segway: the robot may travel with the desired speed $\dot{p}_{\rm d}$ until getting close to the wall, where it must reduce its speed according to its distance from the wall.
    The safe velocity can be tracked by an on-board controller designed for the full system~\eqref{eq:robot-dyn} that also stabilizes the Segway upright:
    \begin{equation}\label{eq:segway_controller_2D}
        \bk(\bq,\dot{\bq}) = K_{\dot{p}} (\dot{p} - k_0(\bq)) + K_{\varphi} \varphi + K_{\dot{\varphi}} \dot{\varphi}.
    \end{equation}
    with gains $K_{\dot{p}}$, $K_{\varphi}$, $K_{\dot{\varphi}}$, where $k_0(\bq)$ is the first component of $\bk_0(\bq)$ and represents a safe forward velocity. This controller satisfies the conditions of Theorem \ref{theorem:cbf-tracking-iss} using:
    \begin{equation*}
        V(\bq,\dot{\bq}) = \frac{1}{2}(\dot{\bq} - \bk_0(\bq))\T \bD(\bq)(\dot{\bq} - \bk_0(\bq)),
    \end{equation*}
    as an ISS Lyapunov function, wherein the constants $\gamma$ and $\delta$ from \eqref{eq:tracking-Lyap-ISS} may be determined using a similar analysis to that performed in \cite{TamasRAL22}.
    
    The results of applying this controller to the Segway for different choices of gains in \eqref{eq:segway_controller_2D} and different choices of $\alpha$ and $\varepsilon$ used in synthesizing the smooth safety filter $\bk_0$ are provided in Fig. \ref{fig:segway-smooth}. In particular, the left and right columns in Fig. \ref{fig:segway-smooth} illustrate the behavior of the system for $K_{\dot{p}}=50$ and  $K_{\dot{p}}=30$, respectively, for different choices of $\alpha$ and $\varepsilon$. Here, safety is maintained for larger $K_{\dot{p}}$, resulting in larger $\gamma$ in \eqref{eq:tracking-Lyap-ISS}, whereas safety is violated for small values of $K_{\dot{p}}$. Intuitively, larger values of $K_{\dot{p}}$ allow the full-order dynamics to respond faster to commands generated by the ROM and maintain safety (cf.~\eqref{eq:Lyap-safety-conditions}). This highlights the fact that, although the controller \eqref{eq:segway_controller_2D} ultimately applied to this system does not directly leverage the full-order Segway dynamics, tuning this tracking controller to enforce safety may require exploiting model knowledge. In practice, however, it may not be possible to modify an existing tracking controller to satisfy \eqref{eq:Lyap-safety-conditions} as it may represent a ``black-box" module already be integrated into the system's autonomy stack. In such a situation, one can only tune the behavior of the reduced-order model via $\alpha$ and $\varepsilon$, to satisfy the conditions required by \eqref{eq:Lyap-safety-conditions}. The effect of changing $\alpha$ for the two tracking controllers is illustrated in the middle row of Fig. \ref{fig:segway-smooth}, where the tracking controller that originally did not enforce safety ($K_{\dot{p}}=30$) maintains safety with a lower value of $\alpha$. Intuitively, decreasing $\alpha$ causes the reduced-order model to approach the boundary of the constraint set more slowly, requiring less aggressive tracking by the full-order dynamics to ensure safety. Alternatively, one may tune the reduced-order model by decreasing $\varepsilon$ (bottom row of Fig. \ref{fig:segway-smooth}), which effectively adds an additional robustness margin to the reduced-order model, causing it to stop short of the original constraint boundary.
\end{example}

\subsection{Safely Tracking Nonsmooth ROMs}
Thus far, the safety-critical control via ROM paradigm has relied on the use of \emph{smooth} ROMs, implying that one must leverage the smooth safety filters from Sec. \ref{sec:smooth} to design a safe ROM controller $\bk_0\,:\,\R^n\rightarrow\R^p$. Although these smooth safety filters can be tuned to approximate the QP-based safety filter from \eqref{eq:safety-filter-qp} arbitrarily closely, in practice, such controllers tend to be more conservative than their QP counterparts. Our restriction to smooth controllers at the ROM level was necessary in our backstepping approach since such controllers were explicitly used to define a CBF for the full-order system, which must be continuously differentiable\footnote{Note that nonsmooth versions of CBFs do exist \cite{GlotfelterLCSS17,UsevitchCDC20} and have been used to address multiple safety constraints \cite{GlotfelterTAC20}.}. Smoothness also played an important role in the previous subsection wherein we explicitly combined a ROM CBF and a smooth Lyapunov function to build a CBF for the full-order system; however, as shown in this subsection, the existence of a smooth Lyapunov function is not necessary to establish such results.

We now relax this smoothness requirement, which facilitates the use of QP-based controllers for the ROM, by assuming that the tracking error $\bd$ is bounded as:
\begin{equation}\label{eq:tracking-error-bound}
    \|\bd\|^2\leq Me^{-\gamma t} + \delta,
\end{equation}
for nonnegative constants $M,\gamma,\delta\geq0$. This bound reflects the ability of the full-order system to exponentially track the reduced-order model up to a bound $\delta$ and is analogous to the ISS condition in \eqref{eq:tracking-Lyap-ISS}, albeit without the explicit use of a Lyapunov function. One may set various constants in \eqref{eq:tracking-error-bound} equal to zero to reflect the tracking capabilities of the full-order system: $\delta=0$ reflects perfect tracking and $M=0$ reflects bounded, but not convergent tracking.  Rather than building a barrier function for the full-order system from a Lyapunov function, we directly utilize \eqref{eq:tracking-error-bound} to propose the time-varying barrier candidate:
\begin{equation}\label{eq:h-time-varying}
    h(\bq,\bxi,t) = h_0(\bq) - \frac{M}{\mu}e^{-\gamma t}  + \frac{\varepsilon \delta}{4\alpha},
\end{equation}
for a positive constant $\mu>0$, which defines the time-varying safe set:
\begin{equation}\label{eq:C-time-varying}
    \C(t) \coloneqq \{(\bq,\bxi)\in\R^n\times\R^p\,:\,h(\bq,\bxi,t)\geq0\},
\end{equation}
associating to each time $t$ a set $\C(t)\subset\R^{n}\times\R^{p}$ of safe states.
The following theorem shows that, under similar conditions to the preceding results, $h$ as in \eqref{eq:h-time-varying} is an ISSf barrier function for the closed-loop system.
\begin{theorem}\label{theorem:cbf-tracking-nonsmooth}
    Consider the dynamics in \eqref{eq:two-layer-dyn}, the constraint set $\C_0\subset\R^n$ in \eqref{eq:C0}, and suppose there exists a controller $\bk_0\,:\,\R^n\rightarrow\R^p$ and positive constants $\alpha,\varepsilon>0$ satisfying \eqref{eq:top-layer-issf}. Furthermore, suppose there exists a tracking controller $\bk\,:\,\R^n\times\R^p\rightarrow\R^m$ enforcing the tracking error bound in \eqref{eq:tracking-error-bound} for constants $M,\gamma,\delta\geq0$. Provided that \eqref{eq:Lyap-safety-conditions} holds
    then $\C(t)\subset\R^n\times\R^p$ as defined in \eqref{eq:C-time-varying} is forward invariant for the corresponding closed-loop control affine system \eqref{eq:two-layer-dyn-control-affine} with $\bu=\bk(\bq,\bxi)$.
\end{theorem}
For completeness, the proof of this theorem is provided in the Appendix. The following example shows how the preceding results allow for leveraging a QP-based controller for the ROM from Example \ref{example:segway_smooth}.

\begin{figure}
    \centering
    \includegraphics{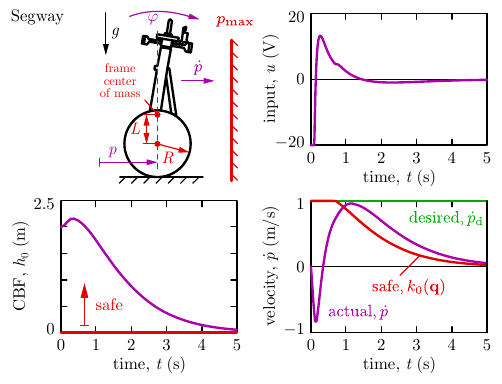}
    \caption{Model-free safety-critical control of a Segway in simulation, with results from~\cite{TamasRAL22}.
    A planar Segway model is controlled to stop in front of a wall, by the help of a CBF-based safe velocity command and a velocity-tracking controller.}
    \label{fig:segway_2D}
\end{figure}

\begin{example}[Planar Segway]\label{example:segway_nonsmooth}
    We now return to Example \ref{example:segway_smooth}, where we seek to use a QP-based controller \eqref{eq:safety-filter-qp} for the ROM rather than a smooth safety filter. The QP solution~\eqref{eq:safety-filter-qp-closed-form} leads to the following safety-critical controller for the ROM:
    \begin{equation*}
    \bk_0(\bq) = \begin{bmatrix}
        k_0(\bq) \\ 0
    \end{bmatrix}, \quad 
    k_0(\bq) = \min \{ \dot{p}_{\rm d}, \alpha(p_{\rm max} - p) - \tfrac{1}{\varepsilon} \},
    \end{equation*}
    with ${\alpha>0}$. Although this controller is nonsmooth, we may leverage the same exact tracking controller \eqref{eq:segway_controller_2D} as in the previous example, and leverage Theorem \ref{theorem:cbf-tracking-nonsmooth} to establish safety of the full-order dynamics.

    Figure~\ref{fig:segway_2D} shows the corresponding simulation results from~\cite{TamasRAL22}.
    The Segway's motion is safe, as established by Theorem~\ref{theorem:cbf-tracking-nonsmooth}.
    Once again, the safe velocity expression does not use the full model~\eqref{eq:robot-dyn}, but only exploits the underlying multi-layer structure with a corresponding trivial ROM that has no parameters.
    This ultimately leads to a {\em model-free} method with a simple explicit ``min'' formula to provide safety for a robotic system.
    Meanwhile, the tracking controller does not involve the expressions in the model~\eqref{eq:robot-dyn} either, however, as discussed in Example \ref{example:segway_smooth}, appropriate selection of the gains $K_{\dot{p}}$, $K_{\varphi}$, $K_{\dot{\varphi}}$ may require model information. Furthermore, when directly tuning the gains of the tracking controller is not feasible, one may directly modify the parameters of the reduced-order model to ensure safety as demonstrated in Example \ref{example:segway_smooth}.
\end{example}

\section{Case Studies}\label{sec:case-studies}
Thus far, we have introduced a variety of different CBF techniques based on the idea of leveraging ROMs to extend a CBF for a simple system to one for a complex system. In each of our illustrations of these techniques, we have chosen relatively simple examples that are just rich enough to capture the main ideas introduced herein. Yet, the motivation for introducing such ideas in the first place was to provide a viable pathway to safety-critical control of complex, high-dimensional autonomous systems.

The safety-critical controllers established above through the use of CBF theory have been implemented on a wide variety of such systems, and, in this section, we revisit more complex application examples from the literature that use these methods.
These examples include safety-critical control of fixed-wing aircraft, flying, legged and wheeled robots, manipulators, and heavy-duty trucks---both in simulation and hardware experiments.

\subsection{Run-time Assurance on Fixed-wing Aircraft}

\begin{figure}[!t]
    \centering
    \includegraphics{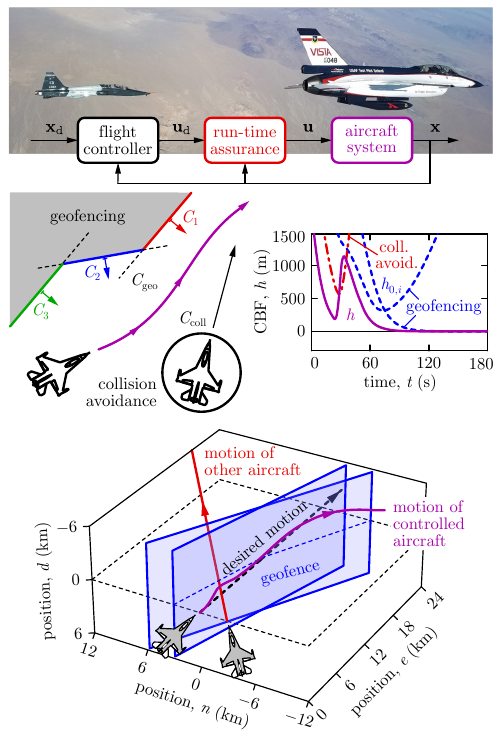}
    \caption{Run-time assurance on fixed-wing aircraft to guarantee safety with respect to collision avoidance and geofencing.
    The results -- repeated from~\cite{TamasTCST23} -- demonstrate that safety-critical flight controllers, which use backstepping-based CBFs and leverage the multi-layer structure of the underlying dynamics, are able to generate maneuvers to prevent collision with other aircraft and entry into restricted airspace.}
    \label{fig:airplane}
\end{figure}

We demonstrate the application of safe backstepping with CBFs by revisiting the work in~\cite{TamasTCST23}, wherein a fixed-wing aircraft was controlled in a safety-critical fashion with the objective of preventing collision with other aircraft or entry into a restricted airspace bounded by a ``geofence''.
The overall control pipeline is illustrated in Fig.~\ref{fig:airplane}.
The aircraft uses a desired flight controller, that tracks a trajectory with stable flight, and a run-time assurance (RTA) system, that overrides this desired flight controller whenever necessary for collision avoidance and geofencing.
The RTA is formulated as a safety filter using CBFs constructed by backstepping.

The controller synthesis is based on a kinematic model, that is used to design acceleration and angular velocity commands for the aircraft in a provably safe fashion.
This model has a multi-layer cascaded structure similar to~\eqref{eq:multi-layer-mixed-r}:
\begin{equation*}
\begin{aligned}
    \dot{\br} & = \bv(\bzeta), \\
    \dot{\bzeta} & = \bf_{\bzeta}(\bzeta,\phi,A_{\rm T},Q), \\
    \dot{\phi} & = f_{\phi}(\bzeta,\phi,Q,P),
\end{aligned}
\end{equation*}
with state ${\bx \!=\! (\br,\bzeta,\phi) \!\in\! \R^7}$ and input ${\bu \!=\! (A_{\rm T},P,Q) \!\in\! \R^3}$; see detailed description in~\cite{TamasTCST23}.
According to this model, the position ${\br \in \R^3}$ evolves according to the expression of the velocity $\bv$, given by the state ${\bzeta \in \R^3}$ that includes speed, pitch angle and yaw angle.
The evolution of $\bzeta$ depends on the roll angle ${\phi \in \R}$,
the longitudinal acceleration ${A_{\rm T} \in \R}$ and the angular velocity ${Q \in \R}$ about the right axis of the aircraft (related to pitching up or down), where $A_{\rm T}$ and $Q$ are viewed as control inputs.
Finally, the evolution of the last state $\phi$ involves the angular velocity $P$ about the front axis (related to rolling), which is considered to be the third control input.
Overall, the dynamics have a 3-layer cascaded structure, where inputs enter at the second and third layers.
Importantly, the right-hand side functions $\bf_{\bzeta}$ and $f_{\phi}$ are affine in
the control inputs $A_{\rm T}$, $P$, $Q$ and in certain expressions of the states.

This structure can be exploited to synthesize a CBF via backstepping for use in collision avoidance and geofencing.
For collision avoidance, consider the distance:
\begin{equation*}
    h_{0,i}(\br) = \| \br - \br_{i} \| - \rho_{i},
\end{equation*}
between the controlled aircraft and multiple other aircraft with index $i$,
whose position is ${\br_i \in \R^3}$, while ${\rho_i>0}$ are collision radii.
For geofencing, the distance between the aircraft and a planar geofence boundary with position $\br_i$ and normal vector $\bn_i$ can be utilized:
\begin{equation*}
    h_{0,i}(\br) = \bn_{i}^\top (\br - \br_{i}) - \rho_{i},
\end{equation*}
where index $i$ refers to multiple geofence constraints, that is, geofences with more complex geometry.
These functions can be combined into a single CBF candidate and used to construct the CBF $h$ via backstepping.
This process takes multiple steps; the details are found in~\cite{TamasTCST23}.

\begin{figure*}[!t]
    \centering
    \includegraphics{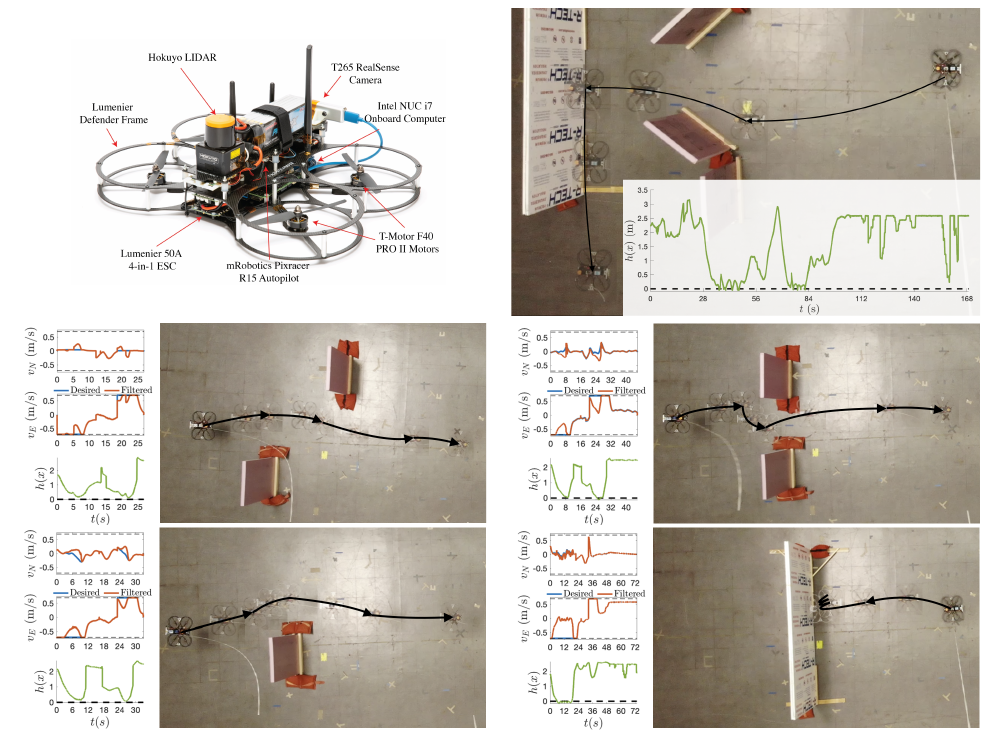}
    \caption{Safety-critical indoor flight tests with a quadrotor~\cite{DrewIROS21}.
    The quadrotor is controlled to traverse obstacle courses with various obstacle arrangements while maintaining a collision-free flight.
    The single integrator is used as ROM for the quadrotor's dynamics, while the distance from the obstacle is considered as the CBF.
    By incorporating these into a safety filter, safe velocity commands are computed, which are then tracked by the onboard flight controller.
    The end result is collision-free motion in each scenario.}
    \label{fig:drone}
\end{figure*}

The CBF can be used in the QP-based controller~\eqref{eq:safety-filter-qp-closed-form} to achieve safety-critical behavior.
The resulting motion is demonstrated in Fig.~\ref{fig:airplane} by the simulation of simultaneous collision avoidance and geofencing scenario.
The controlled aircraft seeks to track a straight trajectory, and its run-time assurance system intervenes to guarantee safety.
The aircraft first accelerates, pitches up, and turns left to avoid collision with the other aircraft, and then it is forced to turn right to avoid crossing the two geofence boundaries.
This behavior is generated by the backstepping-based CBF $h$, which was kept nonnegative throughout the motion.
As a result, the three position-based CBF candidates $h_{0,i}$ are also kept nonnegative, which indicates that the underlying maneuvers are executed with guaranteed safety.

\subsection{Safety-critical Control of Quadrotors}

Next, we illustrate safe behavior on another important class of aircraft: quadrotors.
We revisit the results of~\cite{DrewIROS21}, where the techniques discussed in Sec.~\ref{sec:tracking} were first demonstrated by hardware experiments on drones.
The quadrotor shown in Fig.~\ref{fig:drone} was utilized in indoor flight tests to traverse obstacle fields with various obstacle arrangements (see bottom panels).
In each scenario, the drone used an onboard flight controller to track velocity commands.
To obtain these commands, first, a desired velocity was provided by a high-level desired controller.
Then, using a single integrator as a ROM of the full quadrotor dynamics, a safety filter modified the desired velocity to a safe velocity command.
The CBF underlying this safety filter was the distance between the quadrotor and the obstacle.
Tracking of the resulting velocity resulted in collision-free flight, as the theory in Section~\ref{sec:tracking} suggests.

Importantly, safety filters can also be implemented to prevent a human pilot from crashing a drone.
The flight tests in~\cite{DrewIROS21} also demonstrated a case where a human was piloting the drone manually.
These experimental results are shown in the top right panel of Fig.~\ref{fig:drone}.
Here, a human pilot provides the desired velocity commands for traversing the field such that the drone is actively driven {\em towards} the obstacles.
Yet, even when the human pilot intends to hit the obstacles, the safety filter intervenes and prevents a collision.
As such, human pilots usually provide high-level commands for robotic systems like this drone, hence a high-level safety filter -- operating based on ROMs and CBFs -- is suitable for keeping the system safe.

\subsection{Safe Flying, Legged and Wheeled Robots}

\begin{figure*}[!t]
    \centering
    \includegraphics{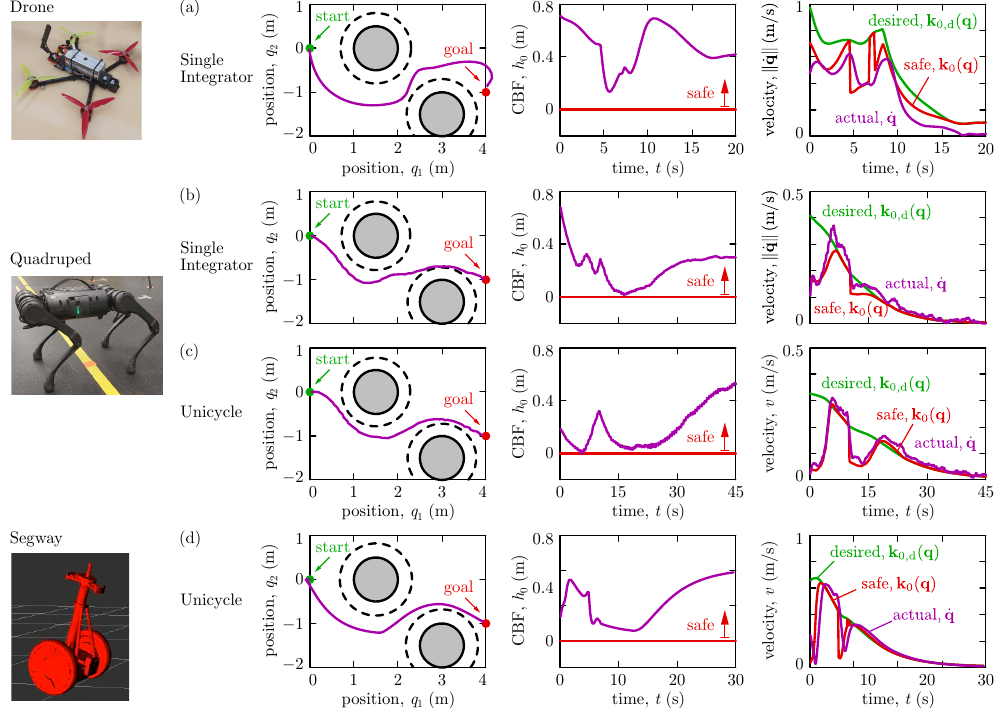}
    \caption{Illustration of the model-free safety-critical control paradigm from~\cite{TamasRAL22}.
    An obstacle avoidance task is executed on three fundamentally different systems: flying, legged, and wheeled robots.
    Each robot is controlled safely based on reduced-order (i.e., single integrator or unicycle) kinematics, by calculating safe velocity commands using CBFs and tracking these commands using on-board flight, walking, and driving controllers.
    (a) Hardware experiments on Drone, (b,c) hardware experiments on Quadruped, (d) high-fidelity simulations on Segway.}
    \label{fig:modelfree}
\end{figure*}

The control strategy discussed for quadrotors can be extended to a wide range of robotic systems.
We demonstrate this by revisiting the results from~\cite{TamasRAL22} where flying, legged, and wheeled robots were controlled via the same approach: stable tracking of safe ROMs.
This approach leverages the fact that many robotic systems have multi-layer structures in their dynamics, where the top layer captures the relationship between the configuration and velocity of robots while the bottom layer relates velocities to forces or torques.
As such, the top-level dynamics can be viewed as ROMs describing the evolution of the configuration.
If safety is captured by a set $\C_0$ in the configuration space (that is the case e.g.~for collision avoidance), then CBFs for these ROMs can be used to find safe velocity commands, which can be tracked by existing on-board controllers that make the robot fly, walk or drive.
This yields a simple method to guarantee safety of various robots.
Moreover, as was highlighted for quadrotors, the ROMs are often trivial equations with no parameters, like the single integrator in~\eqref{eq:robotic-ROM}.
Such ROMs lead to simple geometric expressions 
for the safe velocity, regardless of how complex the full model is.
We refer to this approach as {\em model-free} safety-critical control.

\begin{figure*}[!t]
    \centering
    \includegraphics{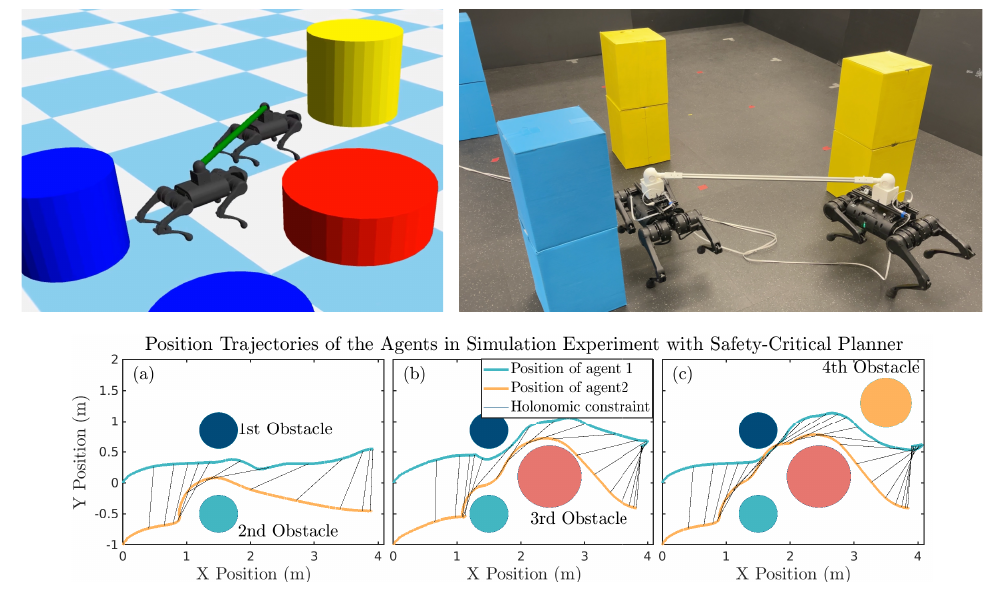}
    \caption{Simulation and hardware results corresponding the to collaborative locomotion case study, originally reported in~\cite{Jeeseop-Multi-Agent}.}
    \label{fig:quadruped_collaboration}
\end{figure*}

The model-free safety-critical control paradigm is illustrated in Fig.~\ref{fig:modelfree}.
Three fundamentally different robots -- a custom-built racing drone, a  Unitree A1 quadruped, and a Ninebot E+ Segway -- are controlled with the model-free approach to accomplish a reach-avoid task similar to that in Fig.~\ref{fig:double_int_backstepping}.
Using single integrator or unicycle reduced-order kinematics, CBF-based safe velocity expressions 
are computed for each robot, which are commanded as a reference signal to be tracked by the controller that flies the drone (established in~\cite{singletary2021onboard}), locomotes the quadruped (developed in~\cite{ubellacker2021verifying}) and drives the Segway (described in~\cite{AmesIEEEA20, TamasRAL22}), respectively.
The velocity tracking error, observed in the right panels, satisfies the bound~\eqref{eq:tracking-error-bound}, thus safety can be established according to Theorem~\ref{theorem:cbf-tracking-nonsmooth}.
Indeed, safe behavior was observed in hardware experiments (drone and quadruped) and high-fidelity simulations (segway), as indicated by the positive value of the CBF $h_0$ of the reduced-order kinematics.
Note that these results from \cite{TamasRAL22} did not include the robustness term with $\varepsilon$ in~\eqref{eq:top-layer-issf} (i.e., ${\varepsilon \to \infty}$ was taken), hence a different variant of Theorem~\ref{theorem:cbf-tracking-nonsmooth} with more restrictive assumptions was required to prove safety.
We will highlight the relevance of robustness terms in the upcoming subsections where CBFs are used on industrial manipulators and heavy-duty vehicles.

\subsection{CBFs in Collaborative Robotics}
In the previous case study \cite{TamasRAL22}, we demonstrated how ROMs may be used to develop safety-critical controllers for a variety of robotic systems, including legged robots. In the context of safe legged locomotion, this approach leveraged the system's existing control architecture, developed in \cite{ubellacker2021verifying}, and allowed to control a rather complex robotic system by simply passing safe reference commands, generated by models such as a single integrator or unicycle, to the existing architecture. In the present case study, we further explore how CBFs may integrate into a system's overall autonomy stack in the context of collaborative legged locomotion \cite{Jeeseop-Multi-Agent} as portrayed in Fig. \ref{fig:quadruped_collaboration}. 

Here, the objective is for a team of holonomically constrained robots, in this case, a team of quadrupeds, to collaborate and safely navigate around obstacles before arriving at a goal location. These holonomic constraints could represent, for example, a payload that these robots seek to transport, which constrains the team's overall formation. To complete this task, the control architecture is broken down into three layers, each leveraging a more detailed model of the interconnected robotic system. The top layer represents each quadruped as a double integrator and leverages CBFs to simultaneously enforce the holonomic constraints and obstacle avoidance. The outputs of the top layer are thus safe position and velocity trajectories that also respect the holonomic constraints imposed on the full-order dynamics. The middle layer seeks to bridge the gap between these reduced-order trajectories and the full-order dynamics by representing the robotic team as an interconnection of single rigid bodies (SRBs). At this level, the outputs of the top layer are used as reference commands for the center of mass of each SRB, which are tracked by a model predictive controller that outputs ground reaction forces (GRFs). These GRFs are input to the bottom layer, which leverages a high-fidelity model of each quadruped and a virtual constraint-based QP controller \cite{HamedRAL20,JeeseopTRO23} to generate torque inputs that impose the commanded GRFs and track the safe position and velocity trajectories generated by higher layers. 

The control architecture outlined above was implemented on a pair of Unitree A1 quadrupeds in both simulation and experimentally \cite{Jeeseop-Multi-Agent}, where the objective is for a pair of interconnected quadrupeds to navigate around obstacles to a goal location. As shown in Fig. \ref{fig:quadruped_collaboration}, in both simulation and hardware, the interconnected robotic system successfully navigates through simple (Fig. \ref{fig:quadruped_collaboration}a) and cluttered environments (Fig. \ref{fig:quadruped_collaboration}c). This is achieved by decomposing the control architecture into multiple layers and reasoning about both the system's holonomic constraints -- representing the interconnection of the robots -- and safety constraints at each layer using different model representations. Ultimately, this decomposition enables the implementation of safe and real-time collaborative locomotion. 

\subsection{Collision-free Food Preparation with Manipulators}

\begin{figure*}
    \centering
    \includegraphics[width=170mm]{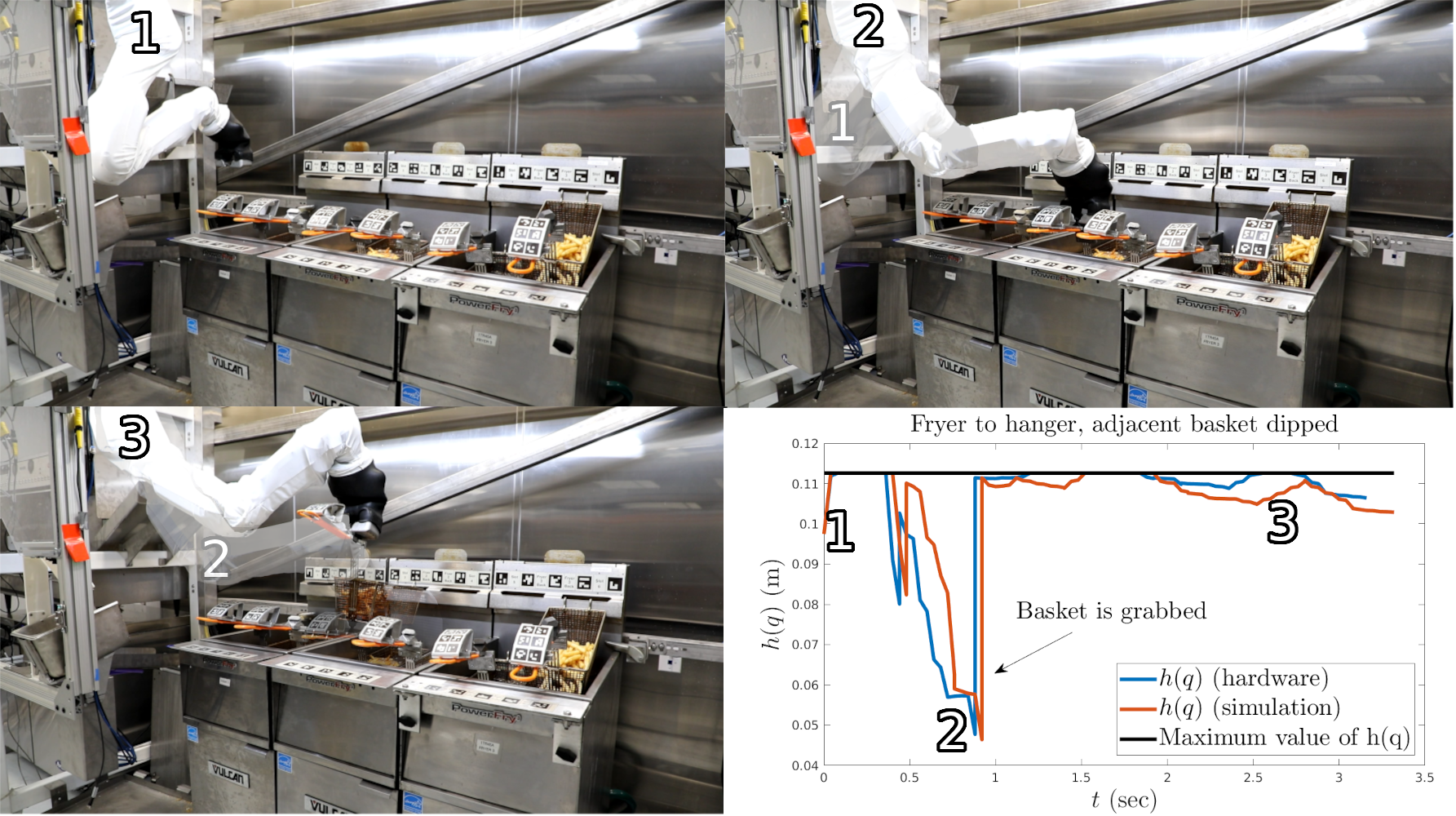}
    \caption{Collision-free food preparation with a Flippy2 robot, with results from~\cite{DrewRAL22}.
    Nominal motion plans that manipulate baskets of food are minimally modified using CBFs, in order to avoid collision between the robot and the kitchen equipment.
    Specifically, the reduced-order kinematics of the robot are used to synthesize a safe velocity using CBFs, which then were tracked by industrial low-level controllers.}
    \label{fig:cooking}
\end{figure*}

Next, we showcase the efficacy of utilizing CBFs and ROMs in the context of safe robotic manipulation.
In particular, we present a real-world industrial application, reported in~\cite{DrewRAL22}, wherein a manipulator is employed in a kitchen for automated food preparation that must be executed in a collision-free manner.
The manipulator, shown in Fig.~\ref{fig:cooking}, is a Miso Robotics Flippy2 robot.
This robot is intended to manipulate kitchen equipment in order to pick
up, deep fry, and dispense food while avoiding collision with its environment.
Executing such behaviors requires sophisticated motion plans, which are computed for various environmental factors and initial conditions.
Many of the required motion plans are similar trajectories with only slight deviations, accounting for the fact that food baskets may move and deform slightly, workers may push the equipment, or the robot may have a slightly different initial configuration.
Therefore, rather than replanning a trajectory in each slightly different situation, it is more efficient to use a CBF-based safety filter to modify a nominal trajectory online and provide formal safety guarantees.

Importantly, the manipulator has an efficient low-level control system that enables the tracking of trajectories and, in particular, velocity commands.
Hence, this architecture is well-suited for utilizing the approach outlined in Section~\ref{sec:tracking}.
Specifically, the kinematic equations of the robot can be used as a ROM to design safe velocity commands via CBF-based safety filters, which can be tracked by the low-level controller.
Ensuring safety at the ROM level via velocity commands -- rather than for the full dynamics by filtering the low-level controller -- was also motivated by the fact that the details of the low-level controller were proprietary, and could not be modified.
At the same time, the industrial low-level controller is well-designed for velocity tracking and capable of keeping the tracking error bounded as in~\eqref{eq:tracking-error-bound}.
As established by Theorem~\ref{theorem:cbf-tracking-nonsmooth}, this enables safe behavior for the full dynamics by the appropriate choice of a ROM-based safety filter.

In particular, the work in~\cite{DrewRAL22} used the signed distance between the closest point of the robot and its environment as CBF candidate $h_0$, and implemented the safety filter: 
\begin{equation*}
\begin{aligned}
    \bk_0(\bq,t) = \underset{\bv \in \mathbb{R}^n}{\operatorname{argmin}} & ~  ~ {\| \bv - \bk_{0,{\rm d}}(\bq,t) \|}^2 \\
    \mathrm{s.t.} & ~  ~ \bn(\bq)^\top \bJ(\bq) \bv \geq - \alpha h_0(\bq) + 2 J_{\max}  \dot{q}_{\max} ,
\end{aligned}
\end{equation*}
that minimally modifies a desired velocity $\bk_{0,{\rm d}}(\bq,t)$ given by a nominal motion plan to a safe velocity $\bk_0(\bq,t)$.
Here, safety is achieved by enforcing a CBF-based inequality constraint analogous to~\eqref{eq:top-layer-issf}.
The term on the left-hand side of this constraint is an approximation of the derivative of function $h_0$ along the kinematic ROM (with the Jacobian $\bJ$ and a normal vector $\bn$), while the last term on the right-hand side is intended to provide robustness against disturbances and approximation errors (with the bounds $J_{\rm max}$ and $\dot{q}_{\rm max}$ on Jacobian and velocity norms).
The resulting safe velocity was finally tracked by the robot's low-level controller to execute collision-free cooking.

The performance of the manipulator employing this control architecture is illustrated by hardware experiments in Fig.~\ref{fig:cooking}.
The objective of the robot is to pick up a food basket that has finished cooking and move it from the fryer to a hanger, allowing the oil to drip off the basket before serving.
Throughout this motion, the robot needs to operate in a dense workspace, where collision must be avoided with food baskets, fryers, the hood vent over the fryers, and a glass pane separating the manipulator from humans, leading to 36 collision objects in total.
Although the manipulation is done in a tight space with a few centimeters of clearance between the robot and the surrounding environment, the manipulator manages to accomplish the task without collision, thanks to the use of a safety filter at the reduced-order kinematics level.
This can be confirmed by the value of the underlying CBF candidate $h_0$, highlighted at the bottom right of Fig.~\ref{fig:cooking}, which is positive during the motion while its maximum value is only $11$ centimeters.
Importantly, the resulting behavior is reproducible:~\cite{DrewRAL22} reported that the use of CBFs led to collision-free behavior consistently in 100 subsequent test cases.

\begin{figure*}[!t]
    \centering
    \includegraphics{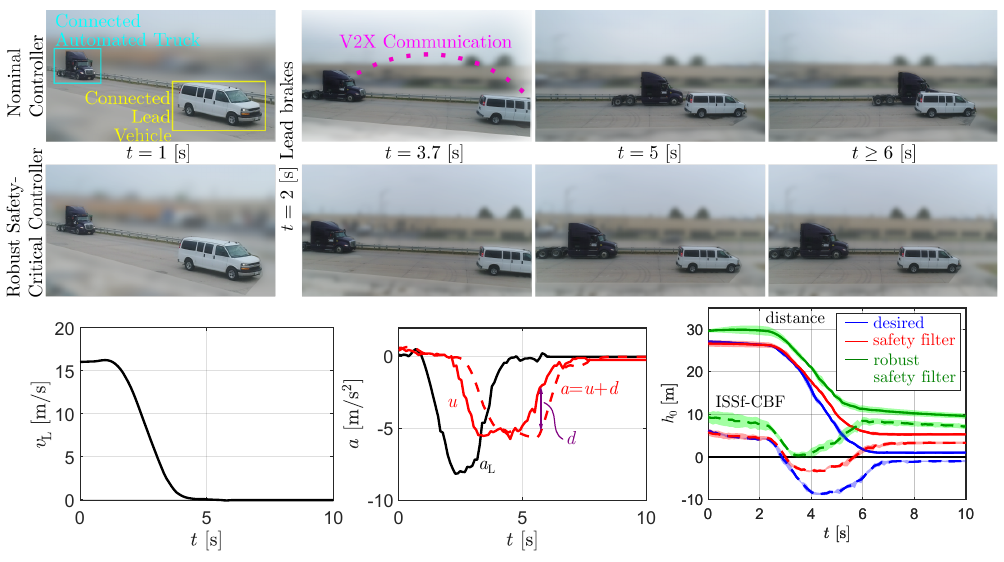}
    \caption{Input-to-state safety on heavy-duty trucks in emergency braking.
    A connected automated truck is controlled to track acceleration commands designed in a safety-critical fashion using a double integrator as ROM.
    The tracking errors act as a significant disturbance, hence robust safety-critical controllers are required to guarantee safe behavior.
    By utilizing tunable input-to-state CBFs, proposed in~\cite{AnilLCSS22}, for robust safety-critical control design, the truck safely executes the emergency braking maneuver without maintaining an overly conservative distance.
    Remarkably, this was not possible by traditional CBFs without added robustness.
    These results and figures have been adapted from~\cite{AnilTCST23}.}
    \label{fig:truck}
\end{figure*}

\subsection{Input-to-state Safety on Connected Automated Trucks}

Finally, we demonstrate safety-critical control of heavy-duty vehicles as 
originally reported in~\cite{AnilTCST23}.
Consider the connected automated truck in Fig.~\ref{fig:truck} that is controlled longitudinally to follow another vehicle on a straight road.
Throughout the motion, the truck must maintain a safe distance to avoid front-end collision, which may be crucial in situations like emergency braking.

The truck is equipped with a low-level control system discussed in~\cite{HeACC20} that regulates gas, brake pressure, and gear shifts to track acceleration commands.
Thus, the truck's desired acceleration is viewed as a high-level control input, and double integrator models (or variants thereof, involving resistance terms and other physical effects) can be used as ROMs to control the truck's motion.
For example, the following ROM was employed in~\cite{AnilTCST23}:
\begin{equation*}
\begin{aligned}
    \dot{D} = v_{\rm L} - v, \\
    \dot{v} = u + d, \\
    \dot{v}_{\rm L} = a_{\rm L},
\end{aligned}
\end{equation*}
where ${D \in \R}$ is the distance of the vehicles, ${v \in \R}$ is the speed of the truck, ${u \in \R}$ is its desired acceleration, ${d \in \R}$ is a disturbance, ${v_{\rm L} \in \R}$ is the speed of the lead vehicle, and ${a_{\rm L} \in \R}$ is its acceleration.
Furthermore, we have ${\bq=(D,v,v_{\rm L})}$ and ${\xi = u}$ with our previous notations.
Using the ROM, longitudinal car-following controllers can be designed at the acceleration level by measuring $D$, $v$, $v_{\rm L}$ and $a_{\rm L}$ using on-board range sensors like radar, as well as GPS and vehicle-to-vehicle connectivity.

With the estimated states, a desired {\em connected cruise controller}~\cite{ZhangITS16} can be utilized to execute car following:
\begin{equation*}
    k_{0,{\rm d}}(\bq) = A (V(D) - v) + B (W(v_{\rm L}) - v),
\end{equation*}
where ${A, B \in \R_{>0}}$ are control gains, ${V\,:\,\R\rightarrow\R}$ is the range policy that provides a desired velocity based on the distance, and ${W\,:\,\R\rightarrow\R}$ is the speed policy that takes the speed limit into account.
This desired controller can be incorporated into a CBF-based safety filter, where the CBF of the ROM:
\begin{equation*}
    h_0(\bq) = D - \rho(v,v_{\rm L})
\end{equation*}
involves a safe distance expression that depends on the speeds as given by ${\rho\,:\,\R^2\rightarrow\R_{\geq0}}$.
The corresponding safety filter generates safe acceleration commands, that can ultimately be tracked by the truck in order to maintain a safe distance.
If the tracking error is bounded, this leads to safe behavior as highlighted by Theorem~\ref{theorem:cbf-tracking-nonsmooth}.

Importantly, accurate tracking of accelerations is challenging on heavy-duty trucks, since they have large inertia and response time, as well as complicated underlying dynamics in the engine, powertrain and brake systems.
As a result, significant tracking errors inevitably occur that propagate as disturbance $d$ to the ROM.
This necessitates the use of safety-critical controllers that are robust to disturbances.
Specifically,~\cite{AnilTCST23} leveraged the concept of {\em tunable input-to-state safety} proposed in~\cite{AnilLCSS22}, and enforced:
\begin{equation}\label{eq:top-layer-tissf}
    L_{\bf_0}h_0(\bq) \!+\! L_{\bg_0}h_0(\bq)\bk_0(\bq) \geq - \alpha h_0(\bq) \!+\! \frac{\|L_{\bg_0}h_0(\bq)\|^2}{\varepsilon(h_0(\bq))},
\end{equation}
as a constraint in QP-based safety filters.
This constraint is a {\em tunable} counterpart of~\eqref{eq:top-layer-issf}, where $\varepsilon\,:\,\R\rightarrow\R_{>0}$ is a tunable function of $h_0$ to provide robustness near the boundary of the safe set only (while being less robust to disturbances when safety is not in danger of violation).
The tunability facilitates reducing the conservativeness of the controller, to allow the truck to keep shorter distances.

The end result is shown in Fig.~\ref{fig:truck}, which presents emergency braking experiments on a Navistar ProStar+ Class-8 truck as reported in~\cite{AnilTCST23}.
The lead vehicle brakes to a full stop (black lines), and the truck responds to this event with various controllers (colored lines).
The desired controller is unsafe during such a harsh maneuver (blue lines).
Similarly, a safety filter that enforces~\eqref{eq:backstepping-strict-inequality} without a robustness term (i.e., without the term of $\varepsilon$), although performing better, still cannot maintain safety (red lines).
This is due to the fact that the tracking of acceleration commands is imperfect and a significant disturbance arises (see purple arrow), while the underlying controller is not robust to disturbances.
The robust safety-critical controller that enforces~\eqref{eq:top-layer-tissf}, on the other hand, successfully guarantees safety.
This demonstrates the power of CBFs and ROMs in guaranteeing safe behavior on real-world systems and highlights that robustness against discrepancies between the ROM and the full system is crucial to achieving safety in practice.

\section{Discussion and Conclusions}\label{sec:conclusions}
Inspired by the success of reduced-order models in robotics, and the need for constructive techniques for CBFs, this paper presented a tutorial on using reduced-order models for safety-critical control. The core idea behind this methodology is to extend a CBF for a relatively simple system to a CBF for a complex system whose behavior, at a high level, is captured by its corresponding reduced-order model. We demonstrated different techniques, such as backstepping and Lyapunov-certified tracking, for constructing CBFs for relevant classes of control systems whose dynamics admit a particular layered structure. These systems include but are not limited to those encountered in robotics such as wheeled, legged, and flying robots. The central ideas of this approach were illustrated through theoretical results, numerical examples, and case studies that demonstrated the successful application of the ideas presented herein across various domains. 

Although the methods covered in this tutorial provide a fairly general way to construct CBFs for relevant classes of systems, they also possess several limitations that should be investigated in future research. Perhaps the greatest limitation the approaches presented herein is that CBFs were synthesized under the assumption of unlimited control authority. In reality, any physical system will posses actuator limits and designing CBFs that take into account such limits is of paramount importance. Popular approaches to constructing CBFs that account for actuation limits include backup CBFs \cite{GurrietICCPS18,YuxiaoCDC21}, input-constrained CBFs \cite{DevCDC21}, and integral CBFs \cite{AmesLCSS21}, among others. It may be possible to unite the ideas presented herein with such methods to systematically synthesize CBFs for high-dimensional systems with actuation limits. Initial steps towards this unification have been presented in \cite{TamasACC23} wherein the methods introduced in Sec.~\ref{sec:tracking} were combined with backup CBFs to develop safety-critical controllers based on reduced-order models that also account for actuation limits. Alternative approaches to accounting for actuation limits may involve the interplay between planning and control within a multi-rate framework \cite{NoelCDC22} in which trajectories of the reduced-order model are designed to be compatible with a lower-level controller with limited actuation authority.

Another question raised by the developments in this tutorial is: how does one choose a suitable reduced-order model? The results in Sec.~\ref{sec:backstepping} and Sec.~\ref{sec:robotic-cbfs} (with the exception of Sec.~\ref{sec:underactuated}) effectively require the full-order dynamics to be fully actuated, and demonstrate that, in such a situation, one may simply take the reduced-order model as a single integrator. The procedure in Sec.~\ref{sec:underactuated} demonstrates how CBFs may be constructed for underactuated systems under a certain set of assumptions, but falls far short of a complete characterization of synthesizing CBFs for underactuated systems. The challenges presented by underactuated systems are implicitly bypassed in Sec. \ref{sec:tracking} by assuming the existence of a low-level controller that tracks commands generated by a reduced-order model. However, the ability to construct such a controller will inevitably depend heavily on both the actuation capability of the system and on the richness of the reduced-order model. Fully characterizing when a reduced-order model is ``good" in the sense that its behavior may be roughly replicated by the full-order dynamics is an important open question that deserves a more thorough investigation. We believe classical tools from nonlinear control theory \cite{Isidori} such as the zero dynamics \cite{IsidoriEJC13}, virtual constraints \cite{Grizzle,MaggioreTAC13,AmesTCST20}, and output regulation \cite{IsidoriTAC90,GrizzleTAC94-model-matching,GrizzleTAC94-necessary} may play an important role in answering such questions.

While there are important theoretical questions that remain unanswered, the case studies presented in Sec.~\ref{sec:case-studies} indicate that the methods outlined in this tutorial tend to perform well in practice (i.e., when deployed on hardware) even when many of our standing assumptions, such as unlimited actuation capability, are violated. Ultimately, we believe developing principled approaches to handle such situations will only further improve the performance of the methods presented herein and facilitate their applications to a broader set of autonomous systems.

\section*{Acknowledgements}
This work was funded in part by the National Science Foundation under grant number 1932091. We thank Pio Ong for the helpful discussions and Jeeseop Kim and Anil Alan for kindly providing the figures from \cite{Jeeseop-Multi-Agent} and \cite{AnilTCST23}. 

\bibliographystyle{ieeetr}
\bibliography{biblio.bib}

\appendix

\section{Proofs}

\begin{proof}[of Theorem \ref{theorem:cbf-backstepping-mixed}]
    We leverage Lemma \ref{lemma:Lgh=0} to show that $h$ as in \eqref{eq:h-backstepping-mixed} is a CBF for the corresponding control affine representation \eqref{eq:two-layer-dyn-mixed-control-affine} of the mixed relative degree system \eqref{eq:two-layer-dyn-mixed}. We begin by computing the gradient of $h$ as:
    \begin{equation*}
        \nabla h(\bx) = 
        \begin{bmatrix}
            \nabla h_0(\bq) + \tfrac{1}{\mu}\pdv{\bk_0^{\bxi}}{\bq}(\bq)\T(\bxi-\bk_0^{\bxi}(\bq)) \\ 
            -\frac{1}{\mu}(\bxi-\bk_0^{\bxi}(\bq)).
        \end{bmatrix}
    \end{equation*}
    Thus, the Lie derivative of $h$ along $\bg$ as in \eqref{eq:two-layer-dyn-mixed-control-affine} is:
    \begin{equation*}
        L_{\bg}h(\bx)\T = 
        \begin{bmatrix}
            L_{\bg_0^{\bu}}h_0(\bq) + \tfrac{1}{\mu}\pdv{\bk_0^{\bxi}}{\bq}(\bq)\T(\bxi-\bk_0^{\bxi}(\bq))\bg_0^{\bu}(\bq) \\
            -\frac{1}{\mu}(\bxi - \bk_0^{\bxi}(\bq))\bg_{1}^{\bu}(\bq,\bxi).
        \end{bmatrix} 
    \end{equation*}
    We now analyze the behavior of $\dot{h}$ when:
    \begin{equation*}
        \begin{bmatrix}
            L_{\bg_0^{\bu}}h_0(\bq) + \tfrac{1}{\mu}\pdv{\bk_0^{\bxi}}{\bq}(\bq)\T(\bxi-\bk_0^{\bxi}(\bq))\bg_0^{\bu}(\bq) \\
            -\frac{1}{\mu}(\bxi - \bk_0^{\bxi}(\bq))\bg_{1}^{\bu}(\bq,\bxi)
        \end{bmatrix} =\begin{bmatrix}
            \bzero \\ \bzero
        \end{bmatrix}.
    \end{equation*}
    It thus follows from the assumption that $\bg_1^{\bu}$ is pseudo-invertible and the second equation in the above system that, when $L_{\bg}h(\bx)=\bzero$, we must have $\bxi - \bk_0^{\bxi}(\bq)=\bzero$. It then follows from the first equation of the above system that, when $L_{\bg}h(\bx)=\bzero$, we must also have $L_{\bg_0^{\bu}}h_0(\bq)=\bzero$. Now, computing the Lie derivative of $h$ along $\bf$ as in \eqref{eq:two-layer-dyn-mixed-control-affine} when $L_{\bg}h(\bx)=\bzero$, we have:
    \begin{equation*}
        \begin{aligned}
            L_{\bf}h(\bx) = & \begin{bmatrix}
                \nabla h_0(\bq) & \bzero
            \end{bmatrix}
            \begin{bmatrix}
                \bf_0(\bq) + \bg_{0}^{\bxi}(\bq)\bxi \\  \bf_1(\bq,\bxi)
            \end{bmatrix} \\
            = & L_{\bf_0}h_0(\bq) + L_{\bg_{0}^{\bxi}}h_0(\bq)\bxi \\
            = & L_{\bf_0}h_0(\bq) + L_{\bg_{0}^{\bxi}}h_0(\bq)\bk_0^{\bxi}(\bq) \\
            > & -\alpha(h_0(\bq)) - L_{\bg_{0}^{\bu}}h_0(\bq)\bk_0^{\bu}(\bq) \\
            = & -\alpha(h_0(\bq)) \\ 
            = & -\alpha(h(\bx)),
        \end{aligned}
    \end{equation*}
    where the third line follows from $\bxi=\bk_0^{\bxi}(\bq)$, the fourth from \eqref{eq:backstepping-strict-inequality-mixed}, the fifth from $L_{\bg_0^{\bu}}h_0(\bq)=\bzero$, and the sixth from $h_0(\bq)=h(\bx)$ (provided $L_{\bg}h(\bx)=\bzero$). It follows from Lemma \ref{lemma:Lgh=0} that $h$ is a CBF for \eqref{eq:two-layer-dyn-mixed-control-affine} on $\C$ as in \eqref{eq:C-backstepping}. \qed
\end{proof}

\begin{proof}[of Theorem \ref{theorem:ECBF}]
    We establish this result by showing that the function $h\,:\,\TQ\rightarrow\R$ as defined in \eqref{eq:ECBF} satisfies the barrier-like inequality $\dot{h}(\bq,\dot{\bq}) \geq - \alpha(h(\bq,\dot{\bq}))$ for the closed-loop system, allowing one to invoke the comparison lemma \cite[Lemma 3.4]{Khalil} to establish forward invariance of $\C$. To do so, we compute:
    \begin{equation*}
        \begin{aligned}
            \dot{h}(\bq,\dot{\bq}) = & \dot{h}_0(\bq,\dot{\bq}) - \frac{1}{\mu} \dot{V}(\bq,\dot{\bq}),
        \end{aligned}
    \end{equation*}
    noting that $\dot{h}_0(\bq,\dot{\bq}) =  \nabla h_0(\bq)\cdot \dot{\bq}$ and:
    \begin{equation*}
        \begin{aligned}
            \dot{V}(\bq,\dot{\bq}) = & (\dot{\bq} - \bk_0(\bq))\T\Big[\bD(\bq)\ddot{\bq} - \bD(\bq)\pdv{\bk_0}{\bq}(\bq)\dot{\bq}\Big] \\
            & + \frac{1}{2}(\dot{\bq} - \bk_0(\bq))\T\dot{\bD}(\bq,\dot{\bq})(\dot{\bq} - \bk_0(\bq)) \\
            = & -(\dot{\bq} - \bk_0(\bq))\T\Big[\bD(\bq)\pdv{\bk_0}{\bq}(\bq)\dot{\bq} + \bC(\bq,\dot{\bq})\dot{\bq} \\ & + \bG(\bq) - \bB\bk(\bq,\dot{\bq})\Big] \\ 
            & + \frac{1}{2}(\dot{\bq} - \bk_0(\bq))\T\dot{\bD}(\bq,\dot{\bq})(\dot{\bq} - \bk_0(\bq)) \\
            = & -(\dot{\bq} - \bk_0(\bq))\T\Big[\bD(\bq)\pdv{\bk_0}{\bq}(\bq)\dot{\bq} + \bC(\bq,\dot{\bq})\bk_0(\bq) \\ & + \bG(\bq) - \bB\bk(\bq,\dot{\bq})\Big], \\ 
        \end{aligned}
    \end{equation*}
    where the second equality follows from substituting in the dynamics \eqref{eq:robot-dyn} and the third from Property \ref{property:skew-symmetric}. Hence, $\dot{h}$ may be expressed as:
    \begin{equation*}
        \begin{aligned}
            \dot{h}(\bq,\dot{\bq}) = & \nabla h_0(\bq)\cdot \dot{\bq} + \frac{1}{\mu} (\dot{\bq} - \bk_0(\bq))\T\Big[\bD(\bq)\pdv{\bk_0}{\bq}(\bq)\dot{\bq} \\ &+ \bC(\bq,\dot{\bq})\bk_0(\bq) + \bG(\bq) - \bB\bk(\bq,\dot{\bq})\Big]  \\
            \geq & - \alpha(h(\bq,\dot{\bq})),
        \end{aligned}
    \end{equation*}
    where the inequality follows from \eqref{eq:ECBF-inequality}. It then follows from the comparison lemma that $h(\bq(t),\dot{\bq}(t))\geq h(\bq_0,\dot{\bq}_0)$ for all $t\in I(\bq_0,\dot{\bq}_0)$ so that if the system's initial condition satisfies $(\bq_0,\dot{\bq}_0)\in\C$, then $h(\bq(t),\dot{\bq}(t))\geq0$ for all $t\in I(\bq_0,\dot{\bq}_0)$, implying the forward invariance of $\C$. \qed
\end{proof}

\begin{proof}[of Theorem \ref{theorem:CBF-underactuated-1}]
    We use an argument similar to Lemma \ref{lemma:Lgh=0} to show that $h$ as defined in \eqref{eq:h1-robotic} is a CBF. We begin by computing the time derivative of $h$ to obtain:
    \begin{equation*}
        \begin{aligned}
            \dot{h}(\bx,\bu) = & \nabla h_{0,1}(\bq_1) \cdot \dot{\bq}_1 \!+\! \frac{1}{\mu}(\dot{\bq}_1 - \bk_{0,1}(\bq_1))\T\bar{\bD}_{1}(\bq)\pdv{\bk_{0,1}}{\bq_1}\dot{\bq}_1  \\ &- \frac{1}{\mu}(\dot{\bq}_1 - \bk_{0,1}(\bq_1))\T\bar{\bD}_{1}(\bq)\ddot{\bq}_1 \\
            & - \frac{1}{2\mu}(\dot{\bq}_1 - \bk_{0,1}(\bq_1))\T\dot{\bar{\bD}}_1(\bq,\dot{\bq})(\dot{\bq}_1 - \bk_{0,1}(\bq_1)) \\ 
            = & \nabla h_{0,1}(\bq_1) \cdot \dot{\bq}_1 \!+\! \frac{1}{\mu}(\dot{\bq}_1 - \bk_{0,1}(\bq_1))\T\bar{\bD}_{1}(\bq)\pdv{\bk_{0,1}}{\bq_1}\dot{\bq}_1  \\ &- \frac{1}{\mu}(\dot{\bq}_1 - \bk_{0,1}(\bq_1))\T\bB_1\bu \\ &+ \frac{1}{\mu}(\dot{\bq}_1 - \bk_{0,1}(\bq_1))\T\bar{\bH}_1(\bq,\dot{\bq}) \\
            & - \frac{1}{2\mu}(\dot{\bq}_1 - \bk_{0,1}(\bq_1))\T\dot{\bar{\bD}}_1(\bq,\dot{\bq})(\dot{\bq}_1 - \bk_{0,1}(\bq_1)). \\ 
        \end{aligned}
    \end{equation*}
    Collecting various terms in the above, we see that:
    \begin{equation*}
    \begin{aligned}
        L_{\bf}h(\bx) = & \nabla h_{0,1}(\bq_1) \cdot \dot{\bq}_1 + \frac{1}{\mu}(\dot{\bq}_1 - \bk_{0,1}(\bq_1))\T\bar{\bH}_1(\bq,\dot{\bq})  \\ & + \frac{1}{\mu}(\dot{\bq}_1 - \bk_{0,1}(\bq_1))\T\bar{\bD}_{1}(\bq)\pdv{\bk_{0,1}}{\bq_1}\dot{\bq}_1 \\ 
        & - \frac{1}{2\mu}(\dot{\bq}_1 - \bk_{0,1}(\bq_1))\T\dot{\bar{\bD}}_1(\bq,\dot{\bq})(\dot{\bq}_1 - \bk_{0,1}(\bq_1)) \\ 
        L_{\bg}h(\bx) = & - \frac{1}{\mu}(\dot{\bq}_1 - \bk_{0,1}(\bq_1))\T\bB_1,
    \end{aligned}
    \end{equation*}
    where $\bx=(\bq,\dot{\bq})$ and $\bf$ and $\bg$ are as in \eqref{eq:robot-dyn-control-affine}. Now, since $\bB_1$ is pseudo-invertible, we have:
    \begin{equation*}
        \begin{aligned}
            L_{\bg}h(\bx) = \bzero \iff & (\dot{\bq}_1 - \bk_{0,1}(\bq_1))\T\bB_1 =\bzero \\ \iff & \dot{\bq}_1 = \bk_{0,1}(\bq_1).
        \end{aligned}
    \end{equation*}
    Hence, when $L_{\bg}h(\bx)=\bzero$, we have:
    \begin{equation*}
        \begin{aligned}
            L_{\bf}h(\bx) = & \nabla h_{0,1}(\bq_1)\cdot \bk_{0,1}(\bq_1) \\ &> - \alpha(h_{0,1}(\bq_1))\\ & = -\alpha(h(\bq,\dot{\bq})),
        \end{aligned}
    \end{equation*}
    which implies that $h$ is a CBF for \eqref{eq:robot-dyn-control-affine}. \qed
\end{proof}

\begin{proof}[of Theorem \ref{theorem:cbf-tracking}]
    Computing the time derivative of $h$ yields:
    \begin{equation*}
        \begin{aligned}
            \dot{h}(\bq,\bxi) = & \dot{h}_0(\bq,\bxi) - \frac{1}{\mu \gamma_1}\dot{V}(\bq,\bxi) \\
            = & L_{\bf_0}h_0(\bq) + L_{\bg_0}h_0(\bq)\bxi - \frac{1}{\mu \gamma_1}\dot{V}(\bq,\bxi) \\
            = & L_{\bf_0}h_0(\bq) + L_{\bg_0}h_0(\bq)(\bk_0(\bq) + \bd) - \frac{1}{\mu \gamma_1}\dot{V}(\bq,\bxi) \\
            \geq & L_{\bf_0}h_0(\bq) + L_{\bg_0}h_0(\bq)(\bk_0(\bq) + \bd) + \frac{\gamma}{\mu \gamma_1}V(\bq,\bxi) \\
            \geq & - \alpha h_0(\bq) + \frac{1}{\varepsilon}\|L_{\bg_0}h_0(\bq)\|^2 \\ &- \|L_{\bg_0}h_0(\bq)\|\|\bd\| + \frac{\gamma}{\mu \gamma_1}V(\bq,\bxi),
        \end{aligned}
    \end{equation*}
    where the first inequality follows from \eqref{eq:tracking-Lyap-2} and the second from \eqref{eq:top-layer-issf}. After completing squares and further bounding $\dot{h}$, we have:
    \begin{equation*}
        \begin{aligned}
            \dot{h}(\bq,\bxi) \geq & - \alpha h_0(\bq) - \frac{\varepsilon}{4}\|\bd\|^2 + \frac{\gamma}{\mu \gamma_1}V(\bq,\bxi) \\
            \geq & - \alpha h_0(\bq) - \frac{\varepsilon}{4\gamma_1}V(\bq,\bxi) + \frac{\gamma}{\mu \gamma_1}V(\bq,\bxi) \\
            = & - \alpha h(\bq,\bxi) + \frac{1}{\mu \gamma_1}\Big(\gamma - \alpha - \frac{\varepsilon \mu}{4} \Big)V(\bq,\bxi),
        \end{aligned}
    \end{equation*}
    where the second inequality follows from \eqref{eq:tracking-Lyap-1} and the final equality from \eqref{eq:h-Lyap}. Hence, provided \eqref{eq:Lyap-safety-conditions} holds, then:
    \begin{equation*}
        \dot{h}(\bq,\bxi) \geq - \alpha h(\bq,\bxi),
    \end{equation*}
    implying $h$ is a barrier function for \eqref{eq:two-layer-dyn-control-affine} with $\bu=\bk(\bx)$ on $\C$ as in \eqref{eq:C-Lyap-tracking}, which implies that $\C$ is forward invariant for the closed-loop system by Theorem \ref{theorem:barrier}. \qed
\end{proof}

\begin{proof}[of Theorem \ref{theorem:cbf-tracking-nonsmooth}]
    Taking the time derivative of $h$ from \eqref{eq:h-time-varying} yields:
    \begin{equation*}
        \begin{aligned}
            \dot{h}(\bq,\bxi,t) = & \dot{h}_0(\bq,\bxi) + \frac{\gamma M}{\mu}e^{-\gamma t} \\
            = & L_{\bf_0}h_0(\bq) + L_{\bg_0}h_0(\bq)\bxi + \frac{\gamma M}{\mu}e^{-\gamma t} \\
            = & L_{\bf_0}h_0(\bq) + L_{\bg_0}h_0(\bq)(\bk_0(\bq) + \bd) + \frac{\gamma M}{\mu}e^{-\gamma t}. \\
        \end{aligned}
    \end{equation*}
    Lower bounding the above using \eqref{eq:top-layer-issf} yields:
    \begin{equation*}
        \begin{aligned}
            \dot{h}(\bq,\bxi,t) \geq & - \alpha h_0(\bq) + \frac{1}{\varepsilon}\|L_{\bg_0}h_0(\bq)\|^2 - \|L_{\bg_0}h_0(\bq)\|\|\bd\| \\ & + \frac{\gamma M}{\mu}e^{-\gamma t},
        \end{aligned}
    \end{equation*}
    which, after completing squares, may be further bounded as:
    \begin{equation*}
        \begin{aligned}
            \dot{h}(\bq,\bxi,t) \geq & - \alpha h_0(\bq) - \frac{\varepsilon}{4}\|\bd\|^2 + \frac{\gamma M}{\mu}e^{-\gamma t}. \\
        \end{aligned}
    \end{equation*}
    It then follows from the above and the bound on $\bd$ from \eqref{eq:tracking-error-bound} that:
    \begin{equation*}
        \begin{aligned}
            \dot{h}(\bq,\bxi,t) \geq & - \alpha h_0(\bq) - \frac{\varepsilon M}{4}e^{-\gamma t} - \frac{\varepsilon \delta}{4} + \frac{\gamma M}{\mu}e^{-\gamma t} \\
            = & - \alpha h_0(\bq) +\frac{M}{\mu}\Big(\gamma - \frac{\varepsilon\mu}{4} \Big)e^{-\gamma t} - \frac{\varepsilon \delta}{4}.
        \end{aligned}
    \end{equation*}
    Using the definition of $h$ from \eqref{eq:h-time-varying}, we then have:
    \begin{equation*}
        \begin{aligned}
            \dot{h}(\bq,\bxi,t) \geq & -\alpha h(\bq,\bxi,t) + \frac{M}{\mu}\Big(\gamma - \alpha - \frac{\varepsilon \mu}{4} \Big)e^{-\gamma t}. \\ 
        \end{aligned}
    \end{equation*}
    Thus, provided \eqref{eq:Lyap-safety-conditions} holds, then:
    \begin{equation*}
        \dot{h}(\bq,\bxi,t) \geq - \alpha h(\bq,\bxi,t).
    \end{equation*}
    It then follows from the comparison lemma that $h(\bq(t),\bxi(t),t)\geq h(\bq_0,\bxi_0, 0)$ for all $t\in I(\bq_0,\bxi_0)$ so that if the system's initial condition satisfies $(\bq_0,\bxi_0)\in\C(0)$, then $h(\bq(t),\bxi(t),t)\geq0$ for all $t\in I(\bq_0,\bxi_0)$, implying the forward invariance of $\C(t)$.\qed
\end{proof}

\end{document}